\message{STBASIC.TEX TeX Macro Library}
\message{ }








\def\df{\leaders\hbox to 0.6em{\hss.}\hfill}


\def\section#1{\bigbreak\medskip\centerline{#1}\par\nobreak\medskip\markpage}

\def\subsection#1#2{\bigbreak\noindent{\bf#1\hskip 0.9em\relax#2}\par
   \nobreak\medskip\markpage}

\def\subsubsection#1#2{\medbreak\noindent{\sl#1\hskip 0.60em\relax#2}\par
   \nobreak\medskip\markpage}

\def\today{\advance\year by -1900 
   \number\month/\number\day/\number\year}
\def\yearmonthday{\number\year\space
   \ifcase\month\or January\or February\or March\or April\or May\or June\or
   July\or August\or September\or October\or November\or December\fi
   \space\number\day}

\newcount\num

\def\nextnum{\global\advance \num by 1 \number\num}
\def\nextitem{\leavevmode
   \hbox{\ifnum\num>8 \kern-0.43em\fi \nextnum.\kern0.60em}}
\def\bfnextitem{\leavevmode
   \hbox{\ifnum\num>8 \kern-0.43em\fi \bf\nextnum.\kern0.60em}}

\newcount\colnum

\def\nextcolnum{\global\advance \colnum by 1 \number\colnum}
\def\nextcolumn{\leavevmode
   \hbox{{\it \ifnum\colnum<9 \phantom{1}\fi Column \nextcolnum:}\kern0.60em}}

\newcount\fig

\def\nextfig{\global\advance \fig by 1 \number\fig}

\newcount\cap

\def\nextcap{\global\advance \cap by 1 \number\cap}

\newcount\letter

\def\nextlet{\global\advance \letter by 1
   \ifcase\letter\or A\or B\or C\or D\or E\or F\or G\or H\or I\or
   J\or K\or L\or M\or N\or O\or P\or Q\or R\or S\or T\or U\or V\or W\or X\or
   Y\or Z\fi}

\newdimen\bigindent \bigindent=3.5in
\def\letterhead{\hsize=6in\interlinepenalty=2000\parskip=6pt minus 3pt
  \pretolerance=750
  \def\topline##1{\hbox to\hsize{\hfil##1\hskip\rightskip}}
  \footline={\ifnum\pageno=1
    \hss\hbox{\vrule height 0.4in width 0pt}
    \eightrm Operated by the Association of Universities for Research in 
    Astronomy, Inc., for the National Aeronautics and Space Administration\hss
    \else\hfil\fi}
  \null
  \vskip-0.2in
  {\advance\rightskip by -0.75in
    \topline{3700 San Martin Drive}
    \topline{Baltimore, MD 21218}
    \topline{(301) 338-4718}\par}
  \vskip30pt minus 15pt
  {\leftskip=\bigindent\yearmonthday\par}}

\def\arpanetletterhead{\hsize=6in\interlinepenalty=2000\parskip=6pt minus 3pt
  \pretolerance=750
  \def\topline##1{\hbox to\hsize{\hfil##1\hskip\rightskip}}
  \footline={\ifnum\pageno=1
    \hss\hbox{\vrule height 0.4in width 0pt}
    \eightrm Operated by the Association of Universities for Research in 
    Astronomy, Inc., for the National Aeronautics and Space Administration\hss
    \else\hfil\fi}
  \null
  \vskip-0.2in\vskip-3\baselineskip
  {\advance\rightskip by -0.75in
    \topline{3700 San Martin Drive}
    \topline{Baltimore, MD 21218}
    \topline{(301) 338-4718}
    \topline{{\elevenrm BITNET:} \tt golombek@stsci}
    \topline{\elevenrm SPAN: \tt SCIVAX::GOLOMBEK}
    \topline{{\elevenrm ARPANET:} \tt golombek@scivax.arpa}\par}
  \vskip30pt minus 15pt
  {\leftskip=\bigindent\yearmonthday\par}}

\def\gosbletterhead{\hsize=6in\interlinepenalty=2000\parskip=6pt minus 3pt
  \pretolerance=750
  \def\topline##1{\hbox to\hsize{\hfil##1\hskip\rightskip}}
  \footline={\ifnum\pageno=1
    \hss\hbox{\vrule height 0.4in width 0pt}
    \eightrm Operated by the Association of Universities for Research in 
    Astronomy, Inc., for the National Aeronautics and Space Administration\hss
    \else\hfil\fi}
  \null
  \vskip-0.375in
  {\advance\rightskip by -0.75in
    \topline{General Observer Support Branch}
    \topline{3700 San Martin Drive}
    \topline{Baltimore, MD 21218}
    \topline{(301) 338-4996}\par}
  \vskip30pt minus 15pt
  {\leftskip=\bigindent\yearmonthday\par}}



\def\indentleft{\advance\leftskip by 50pt\interlinepenalty=750}
\def\inndentleft{\advance\leftskip by 78pt\interlinepenalty=750}
\def\narrower{\advance\leftskip by 0.42in\advance\rightskip by 0.42in
  \interlinepenalty=750}
\def\nnarrower{\advance\leftskip by 50pt\advance\rightskip by 45pt
  \interlinepenalty=750}

\def\checkbox{\nnarrower\parindent=0pt\itemitem{\vbox{\hrule height.7pt
  \hbox{\vrule width.7pt height6pt \kern6pt \vrule width.7pt}
  \hrule height.7pt}$\,$}}  


%
%
\newcount\index \index=100
\def\markpage{\advance\index by 1 \count\index=\pageno}
\def\begintableofcontents{\begingroup
  \index=100 \frenchspacing\interlinepenalty=750
  \parskip=0.1pt plus 1pt minus 0.1pt \parindent=0.3in
  \def\dfi{\advance\index by 1 \df\number\count\index}
  \def\in{\par\hskip-0.2in\indent \hangindent2\parindent \textindent}    
  \def\inin{\par\hskip0.32in\indent \hangindent3\parindent \textindent}
  \def\ininin{\par\hskip0.95in\indent \hangindent4\parindent \textindent}}



{\obeylines\gdef\startdisplay#1
  {\catcode`\^^M=5$$#1\halign\bgroup\indent##\hfil&&\qquad##\hfil\cr}}
\outer\def\enddisplay{\crcr\egroup$$}

\chardef\other=12

{\obeyspaces\gdef {\ }} 

  \font\twentyfourrm=cmr10 scaled 2488
  \font\twentyfouri=cmmi10 scaled 2074   
  \font\twentyfoursy=cmsy10 scaled 2074
  \font\twentyrm=cmr10 scaled 2074      
  \font\twentyi=cmmi10 scaled 2074   
  \font\twentysy=cmsy10 scaled 2074
  \font\eighteenrm=cmr10 scaled 1728
  \font\eighteeni=cmmi10 scaled 1728 \font\eighteensy=cmsy10 scaled 1728
  \font\fourteenrm=cmr10 scaled 1440
  \font\fourteeni=cmmi10 scaled 1440 \font\fourteensy=cmsy10 scaled 1440
  \font\twelverm=cmr12
                
  \font\twelvei=cmmi12               \font\twelvesy=cmsy10 scaled 1200
  \font\elevenrm=cmr10 scaled 1095
    
  \font\eleveni=cmmi10 scaled 1095   \font\elevensy=cmsy10 scaled 1095
  \font\tenrm=cmr10
                   
  \font\teni=cmmi10  \font\tensy=cmsy10  
  \font\ninerm=cmr9

  \font\ninei=cmmi9                  \font\ninesy=cmsy9
  \font\eightrm=cmr8
  \font\seveni=cmmi7 \font\sevensy=cmsy7

\def\commonstuff{
  \parindent=0.42in       
  \def\skipline{\vskip\baselineskip}
  \hyphenpenalty=200\pretolerance=300\tolerance=600 
  \interlinepenalty=100\clubpenalty=500\widowpenalty=500
  \nonfrenchspacing\singlespace\rm}

\def\twelvepoint{
  \font\bf=cmbx12
  \font\it=cmti12
  \font\sl=cmsl12
  \font\tb=cmtt10 scaled 1200 
  \font\tt=cmtt8 scaled 1440
  \textfont0=\twelverm \scriptfont0=\tenrm     
    \scriptscriptfont0=\sevenrm                 
  \def\rm{\fam0 \twelverm}   
  \textfont1=\twelvei  \scriptfont1=\teni  
    \scriptscriptfont1=\seveni                  
  \def\mit{\fam1 } \def\oldstyle{\fam1 \twelvei}
  \textfont2=\twelvesy \scriptfont2=\tensy 
    \scriptscriptfont2=\sevensy                 
  \def\singlespace{\baselineskip=13.5pt\lineskiplimit=-5pt
    \lineskip=0pt
    \parskip=1.25pt plus 1.5pt minus 0.25pt}  
  \def\oneandahalfspace{\baselineskip=18pt\parskip=0pt plus 1pt}
  \def\doublespace{\baselineskip=24pt\parskip=0pt plus 0.5pt}
  \footline={\ifnum\pageno=1 \hfil
             \else\hss\twelverm-- \folio\ --\hss\fi} 
  \def\pagenumbers{\footline={\hss\twelverm-- \folio\ --\hss}}  
  \def\romanpagenumbers{\footline={\hss\twelverm-- \romannumeral\folio\ --\hss}}
  \commonstuff}

\def\tenpoint{
  \font\it=cmti10
  \font\sl=cmsl10
  \font\bf=cmb10
  \textfont0=\tenrm \scriptfont0=\sevenrm     
    \scriptscriptfont0=\fiverm                 
  \def\rm{\fam0 \tenrm}   
  \textfont1=\teni  \scriptfont1=\seveni  
    \scriptscriptfont1=\fivei                  
  \def\mit{\fam1 } \def\oldstyle{\fam1 \teni}
  \textfont2=\tensy \scriptfont2=\sevensy 
    \scriptscriptfont2=\fivesy                 
  \def\singlespace{\baselineskip=12pt\lineskiplimit=0pt
    \lineskip=-0.5mm       
    \parskip=2pt plus 1pt minus 1pt}  
  \footline={\ifnum\pageno=1 \hfil
             \else\hss\tenrm-- \folio\ --\hss\fi} 
  \def\oneandahalfspace{\baselineskip=18pt\parskip=0pt plus 1pt}
  \def\doublespace{\baselineskip=24pt\parskip=0pt plus 1 pt}
  \def\pagenumbers{\footline={\hss\tenrm-- \folio\ --\hss}}  
  \def\romanpagenumbers{\footline={\hss\tenrm-- \romannumeral\folio\ --\hss}}
  \commonstuff}

\def\elevenpoint{
  \font\it=cmti10 scaled 1095
  \font\sl=cmsl10 scaled 1095
  \font\bf=cmb10 scaled 1095 
  \font\tt=cmtt10 scaled 1095
  \textfont0=\elevenrm \scriptfont0=\tenrm     
    \scriptscriptfont0=\ninerm                 
  \def\rm{\fam0 \elevenrm}   
  \textfont1=\eleveni  \scriptfont1=\teni  
    \scriptscriptfont1=\ninei                  
  \def\mit{\fam1 } \def\oldstyle{\fam1 \eleveni}
  \textfont2=\elevensy \scriptfont2=\tensy 
    \scriptscriptfont2=\ninesy                 
  \def\singlespace{\baselineskip=13pt\lineskiplimit=-5pt
    \lineskip=0mm       
    \parskip=2pt plus 1pt minus 1pt}  
  \footline={\ifnum\pageno=1 \hfil
             \else\hss\elevenrm-- \folio\ --\hss\fi} 
  \def\oneandahalfspace{\baselineskip=19pt\parskip=0pt plus 1pt}
  \def\doublespace{\baselineskip=26pt\parskip=0pt plus 1 pt}
  \def\pagenumbers{\footline={\hss\elevenrm-- \folio\ --\hss}}  
  \def\romanpagenumbers{\footline={\hss\tenrm-- \romannumeral\folio\ --\hss}}
  \commonstuff}

\def\eighteenpoint{           
  \font\bf=cmbx10 scaled 1728
  \font\it=cmti10 scaled 1728
  \font\sl=cmsl10 scaled 1728
  \font\tb=cmtt10 scaled 1728
  \font\tt=cmtt10 scaled 1728
  \textfont0=\eighteenrm \scriptfont0=\fourteenrm
    \scriptscriptfont0=\twelverm                 
  \def\rm{\fam0 \eighteenrm}   
  \textfont1=\eighteeni  \scriptfont1=\fourteeni  
    \scriptscriptfont1=\twelvei                  
  \def\mit{\fam1 } \def\oldstyle{\fam1 \eighteeni}
  \textfont2=\eighteensy \scriptfont2=\fourteensy 
    \scriptscriptfont2=\twelvesy                 
  \def\singlespace{\baselineskip=21pt\lineskiplimit=-5pt
    \lineskip=0pt
    \parskip=4pt plus 1pt minus 1pt}  
  \def\oneandahalfspace{\baselineskip=30pt\parskip=0pt plus 1pt}
  \def\doublespace{\baselineskip=40pt\parskip=0pt plus 1pt}
  \footline={\ifnum\pageno=1 \hfil
             \else\hss\eighteenrm-- \folio\ --\hss\fi} 
  \def\pagenumbers{\footline={\hss\eighteenrm-- \folio\ --\hss}}  
  \commonstuff}

\def\twentypoint{
  \font\bf=cmbx10 scaled 2074
  \font\it=cmti10 scaled 2074
  \font\sl=cmsl10 scaled 2074
  \font\tb=cmtt10 scaled 2074
  \font\tt=cmtt10 scaled 2074
  \textfont0=\twentyrm \scriptfont0=\eighteenrm     
    \scriptscriptfont0=\fourteenrm                 
  \def\rm{\fam0 \twentyrm}   
  \textfont1=\twentyi  \scriptfont1=\eighteeni  
    \scriptscriptfont1=\fourteeni                  
  \def\mit{\fam1 } \def\oldstyle{\fam1 \twentyi}
  \textfont2=\twentysy \scriptfont2=\eighteensy 
    \scriptscriptfont2=\fourteensy                 
  \def\singlespace{\baselineskip=24pt\lineskiplimit=-5pt
    \lineskip=0pt
    \parskip=5pt plus 1.5pt minus 1.5pt}  
  \def\oneandahalfspace{\baselineskip=33pt\parskip=0pt plus 1pt}
  \def\doublespace{\baselineskip=44pt\parskip=0pt plus 0.5pt}
  \footline={\ifnum\pageno=1 \hfil
             \else\hss\twentyrm-- \folio\ --\hss\fi} 
  \def\pagenumbers{\footline={\hss\twentyrm-- \folio\ --\hss}}  
  \def\romanpagenumbers{\footline={\hss\twentyrm-- \romannumeral\folio\ --\hss}}
  \commonstuff}

\def\twentyfourpoint{
  \font\bf=cmbx10 scaled 2488
  \font\it=cmti10 scaled 2488
  \font\sl=cmsl10 scaled 2488
  \font\tb=cmtt10 scaled 2488
  \font\tt=cmtt10 scaled 2488
  \textfont0=\twentyfourrm \scriptfont0=\twentyrm     
    \scriptscriptfont0=\eighteenrm                 
  \def\rm{\fam0 \twentyfourrm}   
  \textfont1=\twentyfouri  \scriptfont1=\twentyi  
    \scriptscriptfont1=\eighteeni                  
  \def\mit{\fam1 } \def\oldstyle{\fam1 \twentyfouri}
  \textfont2=\twentyfoursy \scriptfont2=\twentysy 
    \scriptscriptfont2=\eighteensy                 
  \def\singlespace{\baselineskip=28pt\lineskiplimit=-5pt
    \lineskip=0pt
    \parskip=5pt plus 1.5pt minus 1.5pt}  
  \def\oneandahalfspace{\baselineskip=42pt\parskip=0pt plus 1pt}
  \def\doublespace{\baselineskip=56pt\parskip=0pt plus 0.5pt}
  \footline={\ifnum\pageno=1 \hfil
             \else\hss\twentyfourrm-- \folio\ --\hss\fi} 
  \def\pagenumbers{\footline={\hss\twentyfourrm-- \folio\ --\hss}}  
  \def\romanpagenumbers{\footline={\hss\twentyfourrm-- \romannumeral\folio\ --\hss}}
  \commonstuff}

\def\spose#1{\hbox to 0pt{#1\hss}}
\def\lta{\mathrel{\spose{\lower 3pt\hbox{$\mathchar"218$}}
     \raise 2.0pt\hbox{$\mathchar"13C$}}}
\def\gta{\mathrel{\spose{\lower 3pt\hbox{$\mathchar"218$}}
     \raise 2.0pt\hbox{$\mathchar"13E$}}}

\def\in{\indent}
\def\inin{\in{\in}
\def\ininin{\inin{\in}}}

\hsize = 6.9in

\null
\vskip 200pt
\centerline{\bf Matching the Cosmic Star Formation History}
\centerline{\bf to the Local Galaxy Population}
\vskip 16pt
\centerline{Neil Trentham}
\centerline{Institute of Astronomy}
\centerline{University of Cambridge}
\centerline{Madingley Road}
\centerline{Cambridge CB3 0HA}
\centerline{United Kingdom}
\vfil \eject

\noindent
{\bf ABSTRACT}
\vskip 5pt

\noindent
In this review I will describe a number of recent advances in extragalactic 
astronomy.  First of all I will 
describe our current best estimates of the star formation history of the 
Universe.  Then I will describe measurements of local galaxies and their 
stellar populations, concentrating on measurements of the luminosity 
functions and stellar population compositions 
of the different kinds of galaxies.  Finally, I 
will investigate the relationship between these two sets of results.
The ultimate aim 
is to tell at what stage in the history of the Universe the different stars 
seen in the local galaxies formed.
At present much is known but there are significant uncertainties and
I will highlight some prospects for the future.

\vskip 50pt
\noindent
{\bf 1 INTRODUCTION}
\vskip 5pt

\noindent
Extragalactic astronomy 
is a particularly active area of scientific research.  Of particular
importance is the recognition that the formation and evolution of
galaxies are essentially {\it cosmological} processes and that 
extragalactic astronomy is intimately related to cosmology. 

There are essentially two kinds of matter in galaxies: dark matter
and luminous matter.  The dark matter is probably cold, meaning that
it does not diffuse out of gravitational condensations [1,2].  
Its formation is tied up with the 
physics of the early Universe and its assembly into discrete
galaxies is caused by the gravitational growth of small perturbations
present at very early times [3].  Most luminous
matter is in the form of stars, and it is the distribution of
these stars into populations in different kinds of galaxies and
the formation of these stellar populations that is the subject of this review.

One area of study of considerable current interest is the determination of
the star-formation history of the Universe i.e.~determining when in
the past the stars seen now in galaxies formed.
Classically, this was determined from optical surveys of field galaxies
[4]. 
But we know that the far-infrared background measured by $COBE$
[5,6,7] 
is high, suggesting that much of
cosmic star-formation could be dust-enshrouded and missing from optical
surveys.
Submillimetre surveys (e.~g.~ref.~8) then confirmed a great
deal of star-formation was indeed happening in infrared galaxies that
are optically faint.  These are separate galaxies from 
those seen in optical surveys, so we are not talking about large amounts
of obscured star formation happening within optically-identified galaxies  
of known redshift, rather a new population of galaxies altogether.
The submillimetre surveys showed that the galaxies exist in enough
number to explain the far-infrared background and possibly dominate the 
cosmic star-formation
history.
Unfortunately the large beamsize of the SCUBA instrument used in these
surveys means that we cannot identify the infrared galaxies individually
so that we cannot study them and determine their redshifts.
In an important recent development [9,10,11],  
four such
infrared galaxies have been identified by virtue of hosting gamma-ray
bursts (a phenomenon which seems to be intimately linked with ongoing
star formation).  These four galaxies are exactly what we expected
them to be:
the have high ($>$ 100 M$_{\odot}$ yr$^{-1}$) star-formation rates inferred from
submillimetre and/or optical measurements, but low optical fluxes,
presumably due to internal obscuration.  So while we are still some
way from determining the redshift distribution of these infrared
galaxies (only four are known), we are now beginning to compile a
sample which can be studied in some detail.  

The accurate determination of the
galaxy luminosity function $\phi (L)$, 
defined as the number density of
galaxies per unit luminosity $L$, has  
been another active area of study.  The general form of the
total galaxy luminosity function is well described by the
Schechter function [12]:
$$\phi (L) = \phi^* \exp(-{L\over{L^*}}) 
\left({L\over{L^*}}\right)^{\alpha}\,{1\over{L^*}},\eqno(1)$$
where $\phi^*$ is a characteristic density,
$L^*$ is a characteristic luminosity,
and $\alpha$ is the faint-end slope. 
This function provides an acceptable fit to the total
galaxy luminosity function in both clusters and the field,
although the contributions to the total 
luminosity function from different galaxy types is somewhat
different in the two environments (e.g.~ref.~13):
in clusters ellipticals and lenticulars dominate at the bright end and 
dwarf ellipticals at the faint end whereas in the field
spirals dominate at the bright end and
dwarf irregulars and dwarf ellipticals 
exist in equal number at the faint end.

Two lines of current research have been relevant in this context.
Firstly, the normalization of the galaxy luminosity function, and
consequently the luminosity density of
the Universe, has
been determined from large redshift surveys like SDSS
[14] and 2dF (see ref.~15, where this redshift survey is considered in
conjunction with the 2MASS near-infrared survey).
Secondly, the contribution from galaxies with extremely low
surface brightnesses has been shown to be small, from deep optical
surveys (e.~g.~ref.~16).
 
It has been conventional to assume that almost all the gas which
was converted into stars within galaxies exists in the stellar populations
that we see today.  The possibily that this view may be incorrect has recently been
suggested by the result of the MACHO gravitational
microlensing project [17]: 
as much as a few percent of galactic halos may be made up of
objects with masses of about 0.5 M$_{\odot}$.   
Two possibilities are that these are
stellar remnants like cold white dwarfs [18; however see ref.~19] 
or low-mass stars that 
failed to initiate nuclear burning for some reason.
Either of these scenarios would require that most of the star formation
that happened over the history of the Universe occurred in galaxy
{\it halos}, which is very much in contradiction to the traditional view.

In this article we review both measurements of the cosmic star formation
history and of the distribution of stars in galaxies in the local Universe.
We then review the techiques that can be employed to match the two sets of
observations so that we know at what
point in the history of the Universe the stars
that we see in the galaxies around us formed.
Prospects for the future are highlighted.
While we will concentrate on the traditional view that most stars formed
in high-density environments well inside dark-matter halos and all but
the highest-mass ones (whose lifetime plus the age of the Universe at the
redshift of formation is less than a Hubble time) exist in the visible parts
of galaxies today, we will also discuss the implications of the assertions
outlined in the previous paragraph.

Earlier it was remarked that extragalactic astronomy is closely
related to cosmology, and consequently the values of many
quantities depend on the cosmological model chosen.  
Masses derived from gravitational motions are proportional to the luminosity
distance $d_L$ and luminosities derived from fluxes scale as 
$d_L^2$.  Number densities scale as $V^{-1}$ so mass densities
scale as $d_L V^{-1}$ and luminosity densities as $d_L^2 V^{-1}$, where
$V$ is the cosmological volume element.  
The star-formation rate density at any redshift $z$ is usually derived
from a luminosity density and similarly depends on $d_L (z)^2 V(z)^{-1}$;  
for a given cosmological model the star-formation rate density
is therefore proportional to the Hubble
constant $h$ since $d_L \propto h^{-1}$ and $V \propto h^{-3}$.  
The current density in stars $\rho_*$ is proportional to the-integral of this star-formation rate density and so independent of
$h$ (and is weakly dependent on the cosmology).  Expressed in units of the
critical density $\Omega_* = \rho_* / \rho_c$, then $\Omega_* \sim h^{-2}$,
since $\rho_c = 3 H_0^2 / 8 \, \pi G \sim h^2$.
These sscaling relations assume that the luminosity-to-star-formation-rate
conversion factor (this comes from a Galactic calibration) does not depend on the
Hubble constant.

Stellar densities derived from the local luminosity function
do not depend on the cosmology and have a different
dependence on the Hubble constant.
Luminosities scale as $h^{-2}$ and volumes as $h^{-3}$ so that the
luminosity density of the Universe scales as $h$.  Therefore
$\rho_* \sim h$ and $\Omega_* \sim h^{-1}$.
The above scaling relations assume that the stellar mass-to-light ratios of
galaxies (normally these come from population
synthesis models) do not depend on the Hubble constant.

Throughout this work we will 
assume a model with a non-zero cosmological
constant: $h=0.65$ ($h$ is the Hubble constant $H_0$ in
units of 100 km s$^{-1}$ Mpc$^{-1}$), $\Omega_{\Lambda}=0.7$, 
$\Omega_{\rm matter}=0.3$.  
The luminosity-distance for this model [20] is 
$d_L(z) = {{c}\over{H_0}}\,
(1+z) \, \int_0^{z} {{1}\over
{\sqrt{(1+z^{\prime})^2 (1+\Omega_{\rm matter} z^{\prime}) - 
z^{\prime} (2+z^{\prime}) \Omega_{\Lambda}}}}
\, {\rm d}z^{\prime}.$ 

\vskip 20pt

\noindent
{\bf 2 THE COSMIC STAR FORMATION HISTORY}

\vskip 10pt

\noindent
{\bf 2.1 The Madau Plot}
\vskip 5pt

\noindent
The star formation history of the Universe is normally presented 
as the ``Madau'' or ``Madau-Lilly'' Plot
[4,21].
The ordinate axis of this plot is redshift $z$, which is related directly 
to cosmic time, assuming some cosmological model (see Table 1, which is 
appropriate for a model with a cosmological constant; e.g.~ref.~22). 
The abscissa axis of this plot is the 
total star-formation rate happening within some 
average volume element that is comoving with the Hubble flow.  
At the present time ($z=0$) this volume element represents some typical 
Mpc$^3$ in the Local Universe.

Most contemporary observations suggest that the Madau Plot is best considered
in two parts--an optical part and an infrared part.
The optical part is very well determined from a large number of redshift surveys currently being
performed with large telescopes (see ref.~23 for a compilation).
The idea here is to construct an (optical) flux-limited sample of galaxies and
measure redshifts for those galaxies.

\vfil \eject

\centerline{\bf Table 1}
\vskip 8pt
\centerline{\bf Age of Universe for $h=0.65$, $\Omega_{\Lambda}=0.7$, $\Omega_{\rm matter}=0.3$
cosmology$^{*}$}
\vskip 8pt
{$$\vbox{
\halign {\hfil #\hfil && \quad \hfil #\hfil \cr

Redshift $z$ & Age/Gyr &\cr 
\noalign{\smallskip} \noalign{\smallskip}
\cr
\noalign{\smallskip}
0 & 14.5  &\cr
1 &  6.2  &\cr
2 &  3.5  &\cr
3 &  2.3  &\cr
4 &  1.6  &\cr
5 &  1.2  &\cr
10 &  0.5  &\cr
    &       &\cr
\noalign{\smallskip}\cr}}$$}
$^{*}$Note that age $t(z) = {{c}\over{H_0}}\,
\int_z^{\infty} {{1}\over{1+z^{\prime}}} {{1}\over
{\sqrt{(1+z^{\prime})^2 (1+\Omega_{\rm matter} z^{\prime}) - z^{\prime} (2+z^{\prime}) \Omega_{\Lambda}}}}
\, {\rm d}z^{\prime}$
[ref.~20].

\vskip 12pt
The infrared part is less well-determined.  
The galaxies that are important here are those whose 
optical and ultraviolet light, which comes from young OB stars and
traces ongoing star formation directly, has been 
extinguished by dust.  This dust is made of graphite and 
silicates produced within the star-forming galaxies in red giant atmospheres and supernovae
[24].
These galaxies are faint at optical wavelengths 
and are consequently missing from the redshift 
surveys described above. 
Even if the {\it galaxy} is present in optical surveys due to some
residual optical light that escapes or is generated at low skin-depth,
the vast majority of the star formation is unaccounted for if optical observations
alone are available.
We know that large numbers of these 
infrared galaxies exist from the high infrared background 
[5,6,7] and from submillimetre galaxy surveys 
(ref.~25 and references therein), but the galaxies cannot generally be identified on an 
individual basis so that we cannot determine their redshifts and therefore we cannot place them 
on the Madau Plot. 

What is becoming increasingly clear is that the 
populations of optical and infrared galaxies that contribute to
the two parts of the Madau Plot are basically disjoint.  
This statement is not 100\% true, but provides a useful starting
position to adopt.
Infrared galaxies have low optical fluxes due to internal dust
obscuration and optical galaxies, like high-redshift Lyman-break galaxies, 
tend to have low infrared and submillimetre fluxes [26] 
because their total star formation rates and energy 
outputs tend to be small in comparison with infrared galaxies.  A small number of
galaxies exist which have extremely high infrared {\it and} 
optical luminosities (the ``Class-2'' sources of Smail et al.[25], but
these are rare.  They can, however, be studied in detail by virtue of their
optical identification - the prototype
is SMM J02399$-$0136 [27].

It is worth explaining in this context exactly what is meant by ``infrared" as
opposed to ``optical" galaxy.
The important discriminant is whether or not a reasonable fraction (say 50\% or
so) of the star formation in a galaxy can be inferred from optical observations.
Most ``optical" galaxies, like the Milky Way and late-type spiral galaxies, emit much 
of their energy at far-infrared wavelengths so that much of the star
formation accounted for in the ``optical" part of the Madau Plot includes a contribution
from star formation 
absorbed and reradiated by dust.  But the crucial thing to note in this
context is that these
galaxies also  emit a reasonable proportion of their energy at rest-frame
optical and ultraviolet wavelengths (see Figure 2 of ref.~28).  
``Infrared" galaxies do not do this and we  
only have a realistic measure of their energy output and star-formation rate from
far-infrared/submillimetre (or radio continuum)
observations.  We will see that a good prototype of an ``infrared"
galaxy is the host galaxy of GRB 980703, where the optically-derived star formation
rate is 10 -- 30 M$_{\odot}$ yr$^{-1}$ [29] but the total 
(infrared + optical) star formation rate is more like 500 M$_{\odot}$ yr$^{-1}$ 
[9].  The fraction of the star-formation in this galaxy determined
by optical measurements is therefore only 
4\%, much
less than the 30 -- 70\% that is typical for local late-type star-forming galaxies
[28].

The two types of star-forming galaxies could well have
different redshift distributions.
We cannot therefore obtain the infrared Madau Plot simply by multaplicative
scaling of the optical Madau Plot.

The proportion of the
total star formation happening in the Universe in infrared galaxies
$f_{\rm IR}$ is probably greater than 50\% [30]. 
One plausible scenario consistent with
current observation is presented in Figure 1.
The optical Madau plot, as indicated by the dashed line, is
very well determined by a large number of observations
(the filled circles) but the infrared part
(the solid line minus the dashed line) is poorly constrained
and is model-dependant, and we only really know its
integral over redshift (and even then with considerable uncertainty).

\includegraphics{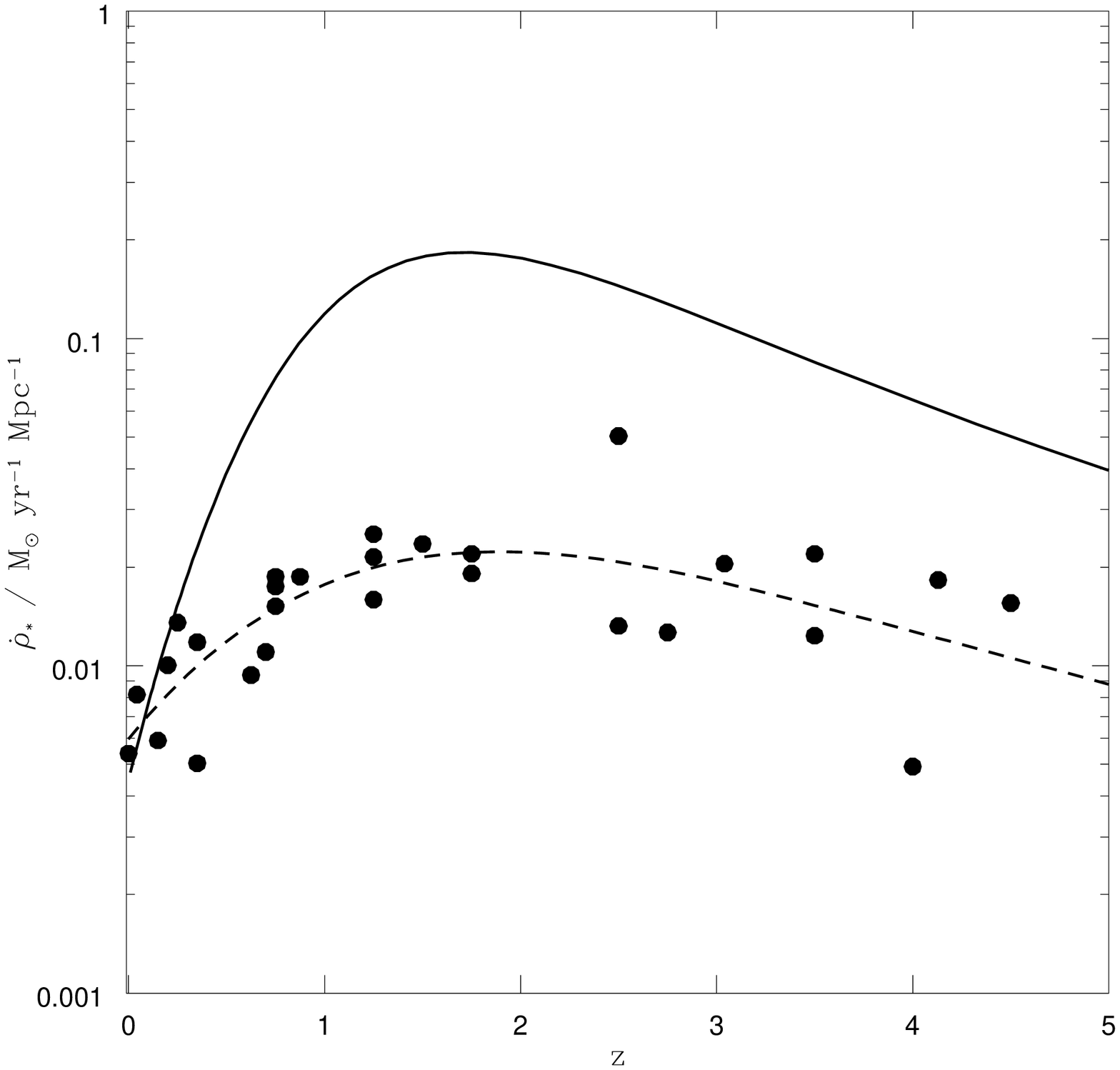}

\vskip 300pt
\noindent
{\bf Figure 1:} The comoving star-formation rate of the Universe as a function of redshift,
from ref.~30. 
The points represent the optical data compiled by Somerville 
et al.~(ref.~23) with no corrections for dust extinction; the dashed line
represents a 4th-order polynomial fit to this data -- the optical Madau
plot. The solid line represents an estimate of
the total (optical and infrared) Madau plot, from the
models of Blain [31,32,33], assuming no AGN
contribution to the far-infrared and submillimetre backgrounds and counts. The
stellar IMF of Kroupa, Tout \& Gilmore [34] is assumed.

\vskip 20pt
\noindent
{\bf 2.2 The Extragalactic Background Light}
\vskip 5pt

\noindent
The extragalactic background radiation emerging at different wavelengths is
shown in Figure 2.  The total extragalactic background light (EBL) is
given by the integral over wavelength of the line in Figure 2 and is
approximately equal to 55 nW m$^{-2}$ sr$^{-1}$ (ref.~35). 
 
The EBL shows two peaks: an optical/near-infrared one at about 1 $\mu$m and
a far-infrared one at about 100 $\mu$m.  The optical-near-infrared EBL is
derived by summing galaxy counts from field surveys [35,36]. 
The far-infrared background comes from $COBE$ DIRBE [5,6] 
and FIRAS [7] 
measurements.

The total energy 
in the optical regime of the EBL is approximately equal to the total
energy in the far-infrared regimes.  A useful exercise is to
attempt to translate this
observation into a value of $f_{\rm IR}$.  In order to do this
we need to consider two further issues. 

Firstly, we must establish which parts of the EBL come from physical
processes other than star formation.
Radiation from old, evolved, stars is certainly a contributor to the EBL, but
this radiation is unlikely to be mistaken for radiation originating from
young newly-formed OB stars since old stars have spectral energy distributions
(SEDs) peaked in the near-infrared, whereas young stars have SEDs peaked at
optical or ultraviolet wavelenghs and any enshrouding dust has an SED peaked at
far-infrared wavelengths.
The contribution from active galactic nuclei (AGNs) is potentially more
difficult to correct for, since these (and their associated dust shrouds)
can in principle 
radiate energy at any wavelength.  Indeed, if quasars form via the
evolutionary sequence [37,38] 
cold ultraluminous infrared galaxy (ULIG; e.g.~Arp 220) $\rightarrow$ warm ULIG
(e.~g.~Markarian 231) $\rightarrow$ infrared quasar (e.~g.~I Zw 1)
$\rightarrow$ optical quasar  
(e.~g.~3C 273; these could be radio-loud late in their
evolution), then we might expect AGNs to contribute at some level
to both the far-infrared (when they are cold ULIGs) and optical (when they
are quasars) backgrounds.
However, we do not expect the contribution to the EBL 
from AGNs to be large, else
the local density of supermassive black holes in the centers of nearby
galaxies, the end products of AGN activity, would be larger than that
observed [39].  The total contribution to the
EBL from AGNs (assumed to happen at mean redshift $<z_{\rm AGN}>$ and efficiency
$\eta$, which is 0.057 for disk accretion onto a Schwarzchild
black hole) is
$I_{\rm AGN} =
5 \, \left({{h}\over{0.65}}\right) \, \left({{\eta}\over{0.057}}\right) \,
\left({{3}\over{1+<z_{\rm AGN}>}}\right) \, {\rm nW}\,{\rm m}^{-2}\,{\rm sr}^{-1}$, which is
about 10\% of the total EBL [35].  This turns out to be smaller than the other
uncertainties, like those described in the next paragraph.

\includegraphics{fig2.ps}

\null
\vskip 320pt
\noindent
{\bf Figure 2:} The extragalactic background light,
from ref.~40. 
This figure was compiled using many literature sources, represented
by the numbers.  The reader is referred to ref.~40 for
the original sources.
The cosmic microwave background radiation spectrum is also shown.

\vskip 20pt

Secondly, the average redshifts of the galaxies that contribute to the optical
and infrared parts of the Madau Plot may be different.  The extragalactic
background intensity $I_* \propto \dot{\rho_*} \, ({1+<z>})^{-1}$, where 
$<z>$ is the mean redshift of the sources (assumed to form stars at an
average rate $\dot{\rho_*}$) contributing to $I$.  Consequently if
the infrared sources are at significantly higher redshift than the optical ones,
then they may trace the formation of a larger mass of stars for a given amount of
background radiation generated.

We can now estimate $f_{\rm IR}$ from the extragalactic background light
as follows. 
In addition to the simplifications described in Section 2.1, let us consider
approximating the background radiation by two parts (representing the two regimes in
Figure 2): an optical part $I_{\rm opt}$ and an infrared part $I_{\rm IR}$.  Let us
assume that the infrared galaxies are at an average redshift $<z_{\rm IR}>$ and
that the optical galaxies are at an average redshift $<z_{\rm opt}>$
and further that the optical galaxies emit an average fraction $R$ of their energy in
the infrared ($R \sim 0.5$ for local galaxies [28] but might be
significantly different at higher redshifts).
Then all other things (like the stellar initial mass function and the 
bolometric-luminosity to star-formation-rate conversion factor) being equal, 
$$f_{\rm IR} =   
{{1}\over{1+{{({1+<z_{\rm opt}>})}\over{({1+<z_{\rm IR}>})}} \,  
\left( (1-R) {{I_{\rm IR}}\over{I_{\rm opt}}} - R \right)^{-1}}}.\eqno(2)$$ 
Although there are many assumptions inherent in applying this equation it
is useful to note the important result:   
the greater $<z_{\rm IR}>$ for a given $<z_{\rm opt}>$, the higher
$f_{\rm IR}$ for a given ratio of
$I_{\rm IR}$ and $I_{\rm opt}$.

\vskip 20pt
\noindent
{\bf 2.3 The Optical Madau Plot} 
\vskip 5pt

\noindent
For an optically-selected sample of galaxies, if redshifts and distances are
available for all the galaxies, then if we are able to estimate star-formation
rates for the galaxies from ultraviolet or H$\alpha$ luminosities, we 
can place them on the Madau Plot.  For any complete
sample, the optical Madau Plot can then be derived.
This completeness has to be present both in terms of 
\vskip 1pt \noindent
(i) luminosity: 
this is achieved if the faintest galaxies in
the sample have low enough luminosity and are far enough down the Schechter (1976)
luminosity function that the total luminosity density in fainter galaxies is
small; if this is not true then a correction to the cosmic star formation rate
can be made to take into account the contribution from these low-luminosity galaxies,
assuming that star-formation rate scales with luminosity;
\vskip 1pt \noindent
(ii) surface-brightness: galaxies with surface-brightnesses significantly below the
sky will be missing from optically-selected  
samples, but corrections due to this effect are normally assumed negligible since
the fraction of the luminosity density in low-surface-brightness
star-forming galaxies is small.

At low redshifts $z<1$ spectroscopic redshifts may be obtained for optically
complete (in a flux-limited sense) samples and the cosmic star formation rate derived from
measurements of the H$\alpha$ 656.3 nm [41,42] or
[OII] 372.7 nm [43,44] lines.  The cosmic star formation
rate at low $z$ is presented in those papers (see also ref.~45). 
For redshifts $1<z<3$ it
is not practical to obtain spectroscopic redshifts for complete 
samples so that photometric redshifts (e.g.~ref.~46)
are often used
(the deepest
surveys, like the Caltech Faint Galaxy Redshift Survey are highly complete only to
$z \sim 1.5$ [47]).  Additionally, for $z>1$, the H$\alpha$ line is
redshifted from the optical to the near-infrared, so that H$\alpha$ luminosities
are difficult to obtain for complete samples -- ultraviolet/optical luminosities
then become the most direct method of estimating star formation rates. 
At higher redshifts $3<z<5$ galaxies can be selected from optical
surveys using the Lyman-break technique [48].  Spectroscopic
redshifts can then be obtained for these Lyman-break galaxies and star-formation rates
derived from their rest-frame ultraviolet luminosities.  The optical Madau
Plot at $3<z<5$ can then be constructed from the samples of
Lyman-break galaxies [49].  

Much of the energy produced by young OB stars in these optical galaxies is absorbed
by dust and reradiated at far-infrared wavelengths (the SED of
many optical galaxies has a peak in the far-infrared).  
This is what we parameterized by the variable $R$ in the previous paragraph.
Each measured point on the optical Madau Plot must be therefore multiplied by a factor $P
\approx (1-R)^{-1}$.
The value of $P$ depends on the redshift and on the nature of the tracer used to
derive the star formation rate.  At $z \sim 0$, the star-formation tracer is the H$\alpha$ 
(656.3 nm) luminosity and  $P \sim 2$ (ref.~28).  
An interesting estimate of $P$ is provided by $ISO$ mid-infrared 15 $\mu$m observations
[50].  This work suggests that $P \sim 3$ at a mean redshift 0.6.   
At high redshifts $z \sim 4$, the star-formation
tracer is the ultraviolet luminosity at about 150 nm, which is very susceptible to
attenuation by dust [51,52].  Consequently a higher
value of $P \sim 5$ is required [49; however see the discussion in ref.~53 abour the
concordance between H$\beta$ and ultraviolet star-formation rates in Lyman-break
galaxies].

At very high $z>5$ we expect the cosmic star formation rate to be dominated by optical,
not infrared, galaxies since the Universe has not had enough time to generate the large
amounts of dust required to produce infrared galaxies.  
Break techniques can certainly be used to find $z>5$ galaxies (e.g.~refs.~54 and 55),
but at the highest redshifts the most profitable technique is likely
to be narrowband imaging searches for Ly$\alpha$ emitters [56].
These high $z$ emission-line searches are helped by the redshift-dependence
of equivalent width EW $\sim$ $(1+z)$,  
and the technique has been successful at finding galaxies up to a redshift
$z=6.56$ [57].
Converting Ly$\alpha$ emitter space densities into a cosmic star formation rate is
difficult since it is not possible to establish what fraction of optical galaxies of
a given redshift {\it are} Ly$\alpha$ emitters (above some luminosity threshold).
This fraction might be quite low since Ly$\alpha$ is a resonant line and photons may be
subject to multiple scatterings and have a low escape fraction; many galaxies with
high star-formation rates may have low Ly$\alpha$ luminosities.
From measurements of Ly$\alpha$ emitters,
Hu et al.~[56] estimate a lower limit on the cosmic star formation rate
(adapted to our cosmology and stellar IMF -- see Section 1 and Figure 2) 
of about 0.003 -- 0.006 M$_{\odot}$
yr$^{-1}$ Mpc$^{-3}$ for redshifts $3<z<6$,
which is less than the value of $0.09 \, \left({{ P}/{5}}\right)$ 
 M$_{\odot}$
yr$^{-1}$ Mpc$^{-3}$ measured at $z \sim 4$ from Lyman-break surveys 
[49]. 
A somewhat higher range of 0.07  -- 1.4 M$_{\odot}$
yr$^{-1}$ Mpc$^{-3}$ has been estimated at $z=6.6$ [57]. 

It is important to note that the only measurable quantity in all
the surveys described above is a luminosity density, and that if an
appreciable fraction of cosmic star formation happened in environments where
this luminosity density is low, much could be missed due to $(1+z)^{-4}$
cosmological surface-brightness dimming [58].  The
determinations of the cosmic star-formation rate at any redshift from these
surveys should therefore be regarded as lower limits, more so at the highest
redshifts. 

\vskip 20pt
\noindent
{\bf 2.4 The Infrared  Madau Plot}
\vskip 5pt

\noindent
The infrared Madau Plot represents the total star formation so
obscured that it is very much under-represented the optical (rest-frame
ultraviolet at high redshift) surveys described in the previous section.
It is {\it not} merely the absorbed and reradiated light from optical
galaxies (parameterized by $P$ and $R$); this reradiated energy is not
sufficient to explain the high infrared background.

At $z=0$, ultraluminous infrared galaxies (ULIGs; ref.~59) like
Arp 220 are the types of galaxies with these properties -- unremarkable at
optical wavelengths but very luminous in the infrared. 
But these contribute negligibly to the global $z=0$ star-formation rate.
Furthermore, were they to exist at high redshift, ULIGs could not have produced
the local galaxy population. 
This is because the molecular gas density in the central regions generating most of
the energy in ULIGs is high ($>$ 100 M$_{\odot}$ pc$^{-3}$), whereas a negligible
fraction ($\sim$ 1\%) of the stars in the local Universe -- elliptical galaxy
centres -- exist at such high
densities [60].  
Semi-analytic models  
[61] also predict that the gas in galaxies generating the infrared Madau Plot must be somewhat
more extended than in local ULIGs.
Large reservoirs of extremely dense cold gas at high redshift are certainly
known to exist (e.~g.~around APM 08279+5255 at $z=3.9$ [62])
so it is quite possible that these exist and are associated with much of the
cosmic star formation.

It is interesting to note that the star-formation rate 
at $z=0.6$ measured from a 15-$\mu$m
sample observed with $ISO$ [50] corresponds exactly to
the star-formation rate determined from optical measurements 
[21] with $P=3$.  This implies either at $z=0.6$ infrared galaxies
contribute negligibly to the Madau Plot or the dust temperatures in the
infrared galaxies are so low that their SEDs peak at long
enough wavelengths that their 15-$\mu$m fluxes are small.  This latter possibility
is what is suggested by the submillimetre observations described in the next
section.

\vskip 20pt
\noindent
{\bf 2.4.1 Constraints from submillimetre surveys}
\vskip 5pt

\noindent
Submillimetre surveys are one way to find high-redshift infrared galaxies.  Although
submillimetre measurements probe well down the Rayleigh-Jeans long-wavelength tail
of the SED of most infrared galaxies, current submillimetre cameras, like the
SCUBA bolometer array on the JCMT in Hawaii [63]
operating at 850 $\mu$m and
450 $\mu$m, are only sensitive enough to find the
most luminous of these galaxies.
The main issue preventing very deep surveys is source confusion [64].
Observing gravitationally lensed sources behind massive galaxy clusters
[8] increases the sensitivity of SCUBA by factors equal to
the magnifications of the background sources, typically 2 -- 5 [25].

The submillimetre number counts are presented in Figure 3.  It is immediately
obvious from this figure that the submillimetre counts probe to a far deeper
equivalent flux limits than $ISO$ infrared surveys.  The $ISO$ surveys probe closer
to the SED peaks of the majority of the infrared galaxies but the
telescope + detector efficiency is lower.

The exact form of the 850-$\mu$m counts is only well constrained at high fluxes.  Between
2 mJy and 10 mJy the differential number counts are approximately
$n(S) \approx 30000 \left(0.7 + S \right)^{-3.2}$, where $S$ is the flux density in mJy
[65]. Fainter than about 1 mJy, the counts are constrained primarily by the
extragalactic background light $\int S \, n(S) \, {\rm d}S$, which at 850 $\mu$m is about
0.55 nW m$^{-2}$ sr$^{-1}$ [7].  Combining these results, Barger et al.~[65]
show that the bulk of the 850-$\mu$m background light comes from sources with fluxes of about
1 mJy and inferred star formation rates of about
200 M$_{\odot}$ yr$^{-1}$.

\vfil \eject 

\includegraphics{fig3.ps}

\null
\vskip 320pt
\noindent
{\bf Figure 3:} Galaxy number counts in the far-infrared, submillimetre,
and millimetre regimes, 
from ref.~40.
The data is compiled from many sources and the reader is referred to ref.~40 for
the original sources.
The light points represent 850-$\mu$m SCUBA counts, the dark points on the right represent
95-$\mu$m $ISO$ data, and the dark square 
on the left represents the 2.8-mm BIMA limit of Wilner \& Wright [66].
The lines represent two models from the literature.

\vskip 20pt 

Aside from verifying the fact that the infrared galaxies  
exist in the right number to produce the extragalactic background light, the
main information to be gained from the submillimetre observations are the luminosities
and temperatures of the galaxies.
Luminosities (and hence star formation rates)
can be obtained from fluxes even if we do not know the
redshifts because infrared galaxies have negative $K$ corrections at 850 $\mu$m:
$f_{\nu} = {L_{\nu (1+z)}} (1+z)/{4 \pi d_L(z)^2 }$ and infrared galaxies have SEDs
where an increase in redshift causes ${L_{\nu (1+z)}}$ to increase by an amount
approximately equal to the increase in ${d_L(z)^2/(1+z)}$ at 850 $\mu$m for
$1<z<4$.  The total luminosity $L = \int_0^{\infty} L_{\nu} {\rm d}{\nu}$,
where $L_{\nu}$ at any frequency can be derived from its value at 850 $\mu$m if
the dust temperatures are known.
Temperatures are known to be close to 37 K due to joint consideration of
850-$\mu$m and 450-$\mu$m counts and modeling [32,67].
Further information is difficult to obtain because of the large SCUBA beamsize
FWHM $\sim$ 14 arcseconds.  Normally we cannot tell which of the large number of
optical sources (if any) within the SCUBA beam is the submillimetre source and
this lack of optical identification means we cannot obtain the redshift. 

A number of submillimetre galaxies do have optical or near-infrared counterparts
but these are rare, particularly at the low 850-$\mu$m fluxes responsible for
generating most of the background.  Examples include the Class-2 SCUBA galaxies
[25],
like SMM J02399$-$0136 ($z=2.80$ [27]),
SMM J14022+2512 ($z=2.56$) and
SMM J02399$-$0134 ($z=1.06$).
These are lightly obscured star-forming galaxies with huge optical/UV and
H$\alpha$ star-formation rates, so high that the reradiated energy from the
stars that {\it have} been obscured is enough to make these into
luminous submillimetre galaxies.
Other examples are extremely red objects (EROs, or Class-1 SCUBA
Galaxies [68,69]), perhaps 
similar to the well-studied galaxy HR10 at $z=1.4$ [70].
 
The next generation of submillimetre cameras like SCUBA-2 and eventually
ALMA will provide information, in particular positions,
 at a far greater level of detail than the
results presented here.  The redshift distribution of submillimetre sources
may well turn out to be the most direct measure of the infrared Madau Plot.

\vskip 20pt
\noindent
{\bf 2.4.1.1 Constraints from Radio Followup of Submillimetre Sources}
\vskip 5pt

\noindent
Infrared galaxies that are luminous at submillimetre wavelengths
have radio fluxes that are faint but detectable [71,72]. 
This means that accurate positions can be obtained which in turn means
redshifts can be obtained.  If this can be done for a complete
sample of submillimetre sources down to very faint flux levels, we can
then construct the infrared Madau Plot.
To date this has been done for a sample of 8 mJy SCUBA sources
[73], but these do not contribute significantly to the
infrared background.  A sample of more typical 1 mJy sources would be
required; with SCUBA this can only be achieved with the aid of
gravitational lensing by imaging sources behind massive clusters.

\vskip 20pt
\noindent
{\bf 2.4.2 Constraints from infrared background and source counts}
\vskip 5pt

\noindent
Far-infrared surveys (which need to be undertaken from space, since the
earth's atmosphere is opaque for most of 40 -- 1000 $\mu$m)
offer another direct method of finding infrared galaxies.
To date, sensitivity considerations (see the $ISO$ points on Figure 3) have  
resulted in these surveys finding only the most luminous sources.

The next generation of instruments on infrared satellites should be at
least an order of magnitude more sensitive.
Consider, for example, a ``typical" infrared galaxy with an 850-$\mu$m flux of
1 mJy (see section 2.4.1).  The flux density $S_{\nu} \sim L_{\nu} \sim
{\nu}^{3.5}$ a modified Rayleigh-Jeans spectrum for typical dust [31]. 
Therefore the 160-$\mu$m flux is 350 mJy 
(independent of redshift).   
These sources should be easily detectable with $SIRTF$/MIPS, which has a
5$\sigma$ sensitivity at 160 $\mu$m of 50 mJy for a 500 second exposure
(this is confusion-limited [74]). 
The poor resolution
($\lambda / {2 D}$ = 19$^{\prime \prime}$ at 160 $\mu$m for the 85 cm
telescope) and  
large pixel size (16$^{\prime \prime}$) may, however, prevent straightforward
identifications of the infrared sources.
Redshift determinations will then be
difficult and it will not be possible to construct the infrared Madau
Plot from these observations alone.  Observations with $SIRTF$/MIPS
at 24 $\mu$m and 70 $\mu$m could help in this regard (the pixel size at
these wavelengths is smaller and the resolution is better) 
but sources at high redshift and/or with
only very cold dust may well go undetected at these shorter wavelengths.

\vskip 20pt
\noindent
{\bf 2.4.3 GRB host galaxies} 
\vskip 5pt

\noindent
Gamma-ray bursts (at least the long-duration ones
which last longer than 2 seconds)
seem to come from massive collapsing
stars.  If the proportion of collapsing stars
that become GRBs is roughly constant from
one star-forming region to another (a major
assumption), then the
redshift distribution of GRBs tells us the star
formation history of the Universe.

This assertion is based on
various pieces of circumstantial evidence which
when combinrd seem to be reasonably compelling:

\vskip 5pt
\noindent
1) GRBs tend to happen 
in galaxies with high star formation rates.
Often these have been identified as
blue star-forming
galaxies at high-redshift, perhaps similar
to local HII galaxies [75].  Most compelling,
however, is that GRBs have now been
seen in four infrared galaxies with huge
star formation rates: GRB 980703, GRB
010222, GRB 000418, and GRB 000210
[10].

\vskip 5pt
\noindent
2) On theoretical grounds,  a natural
mechanism for producing a long-duration burst is the cataclysmic collapse
of a massive star leading to a hypernova-type explosion
[76].

\vskip 5pt
\noindent
3) On two occasions iron-line and edge features
have been identified which are associated with GRBs [77,78] 
suggesting a link with supernovae and therefore with collapsing massive stars.

\vskip 5pt
\noindent
4) On four occasions,
a red component
with an optical flux consistent with that of a supernova was discovered several
weeks after the burst (e.g.~ref.~79).

\vfil \eject

\noindent
One additional result that is extremely important is:

\vskip 5pt

\noindent
{\bf To date, three out of the four GRBs -- 980703, 010222 and 000418 -- that have happened 
in infrared-luminous galaxies have had bright optical afterglows}

\vskip 5pt
This means that these GRBs and their host galaxies can have their redshifts
measured -- for example, by the presence of absorption lines from the host
galaxy in the afterglow spectrum.  In turn, this means that we can 
measure
the redshift distribution of {\it all} gamma ray bursts with optical
afterglows, including the ones
in infrared as opposed to optical galaxies.  

We can directly compare the redshift distribution of GRBs to the predictions
from models of
the cosmic star formation history.  All the bursts with redshift
measurements to date are listed in Table 2 [80].
A comparison with current data is
shown in Figure 4.

\includegraphics{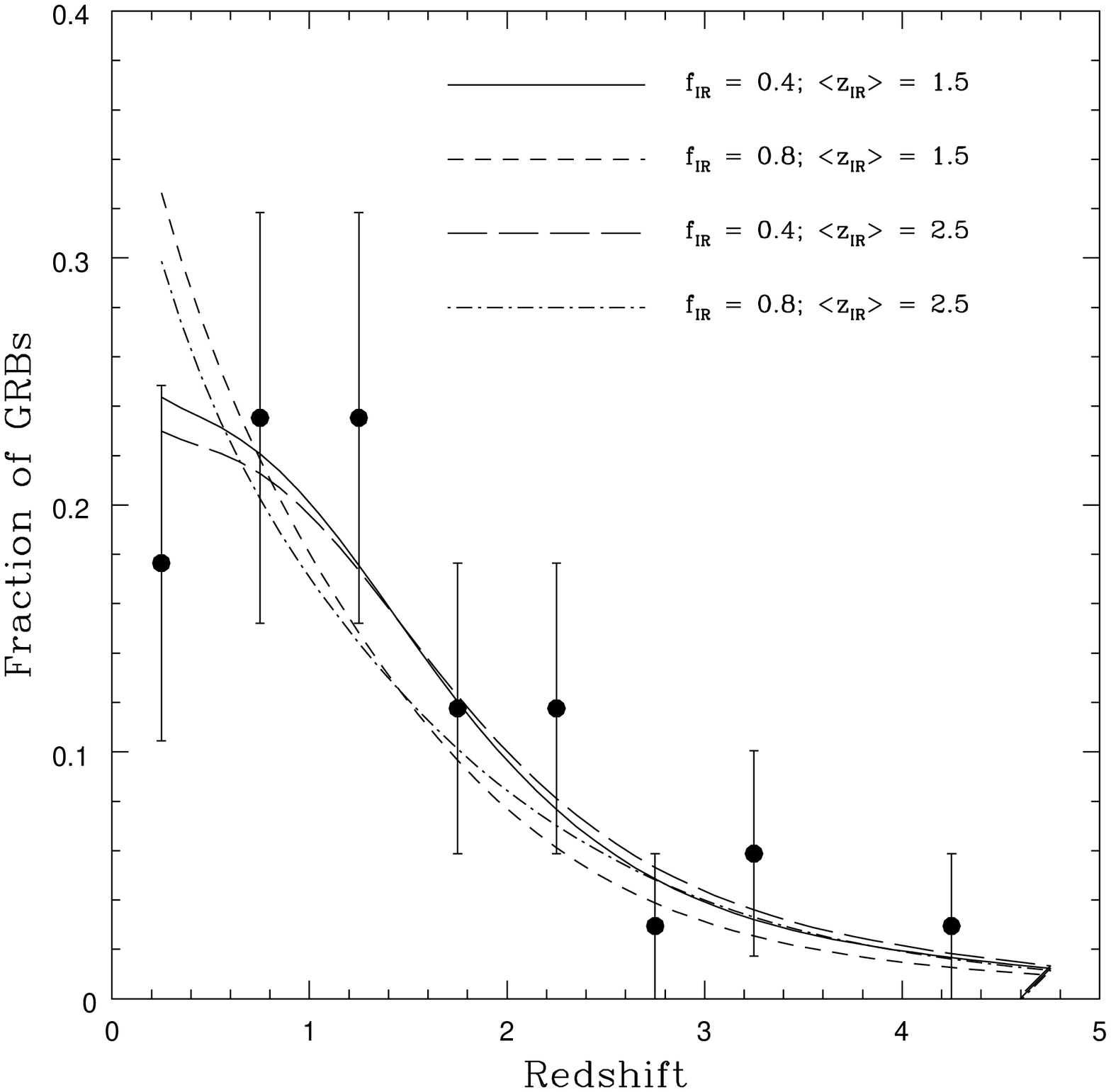}

\null
\vskip 300pt
\noindent
{\bf Figure 4:} The redshift distribution of GRBs compared with
predictions from different models of the cosmic star formation
history.  The data points represent the 34
bursts to date with redshifts known to better than 0.25.  The lines
represent different models of the cosmic star-formation history,
consisting of a contribution from ultraviolet-bright
galaxies (the optical Madau plot, from ref.~23) 
and a contribution from
infrared-bright galaxies.  The infrared bright galaxies are
assumed to have a gaussian redshift distribution with mean
$<z_{\rm IR}>$, variance $\sigma_z = 0.5$, and to contribute
a fraction $f_{\rm IR}$ to the total star formation.
The calculation of the lines further assumes that the number
of GRBs per unit mass converted into stars is not a strong
function of redshift and that GRBs have a luminosity function
$\phi(L)$ where $\int_{4 \pi d(z)^2 f_{\rm lim, det}}
^{\infty} \phi(L) {\rm d}L$ does not vary strongly with
redshift -- all the relevant GRBs were Beppo Sax or HETE-2 detections.

\vfil \eject

\vskip 35pt
\centerline{\bf Table 2}
\vskip 8pt
\centerline{\bf 
Gamma Ray Bursts with Redshift Measurements}
\vskip 8pt
{$$\vbox{
\halign {\hfil #\hfil && \quad \hfil #\hfil \cr

Burst & Redshift &\cr
\noalign{\smallskip} \noalign{\smallskip}
\cr
\noalign{\smallskip}
GRB 970228 & 0.695  &\cr
GRB 970508 & 0.835  &\cr
GRB 970828 & 0.96   &\cr
GRB 971214 & 3.42   &\cr 
GRB 980326 & $\sim 1$   &\cr
GRB 980329 & $<3.9$   &\cr
GRB 980425 & 0.0085 &\cr  
GRB 980703 & 0.97   &\cr 
GRB 990123 & 1.60   &\cr 
GRB 990506 & 1.30   &\cr 
GRB 990510 & 1.62   &\cr 
GRB 990705 & 0.86   &\cr 
GRB 990712 & 0.43   &\cr 
GRB 991208 & 0.71   &\cr 
GRB 991216 & 1.02   &\cr 
GRB 000131 & 4.5    &\cr 
GRB 000210 & 0.85   &\cr 
GRB 000214 & $0.37 < z < 0.47$ &\cr 
GRB 000301C & 2.03  &\cr 
GRB 000418 & 1.12   &\cr 
GRB 000911 & 1.06   &\cr 
GRB 000926 & 2.07   &\cr 
GRB 010222 & 1.48   &\cr 
GRB 010921 & 0.45   &\cr 
GRB 011121 & 0.36   &\cr 
GRB 011211 & 2.14   &\cr 
GRB 020405 & 0.69   &\cr 
GRB 020813 & 1.25   &\cr 
GRB 021004 & 2.3    &\cr 
GRB 021211 & 1.01   &\cr 
GRB 030226 & 1.98   &\cr
GRB 030323 & 3.37   &\cr
GRB 030328 & 1.52   &\cr
GRB 030329 & 0.168  &\cr
GRB 030429 & 2.65   &\cr
&       &\cr
\noalign{\smallskip}\cr}}$$}

For such a comparison to be meaningful we need to make some assumptions:
\vskip 2pt \noindent
1) we must assume some redshift-dependance on the fraction of massive stars
which eventually generate GRBs.  One extreme situation is that
this fraction is independent of redshift -- this may happen if the stellar
IMF is universal and if mass is the main stellar parameter in determining
whether or not a star becomes a GRB.  The other extreme would be
a situation where some redshift-dependant parameter (possibly metallicity) is
the most important issue in determining whether or not a star becomes a GRB.
\vskip 2pt \noindent
2) a GRB luminosity function must be assumed.  One extreme situation might 
be to assume a $\delta$-function at high luminosities -- then we observe all
GRBs, whatever their redshift.  The other extreme would be a luminosity
function that is steeply rising at low luminosities.  Then, any flux-limited
sample of GRBs would be dominated by nearby GRBs of low intrinsic
luminosity.

\vskip 2pt
It is clear from Figure 4 that the current data do not provide any
meaningful constraints on the models because the Poisson errors are far
too large.  Nevertheless, the fact that this method is sensitive to
star formation in both ultraviolet and infrared galaxies makes it very
attractive.

The areas of study required to turn this into a profitable method
are:

\vskip 4pt \noindent
$\bullet$ Very many more GRBs with redshifts need to be observed.  Several
hundred are required if we are to distinguish between the models in
Figure 4.  The $SWIFT$ satellite should provide this.

\vskip 4pt \noindent
$\bullet$ The requirements for generating a GRB need to be understood 
so that assumption 1) can be tested.  This will probably come from both
theoretical modeling and observations of the environments in which GRBs
occur.

\vskip 4pt \noindent
$\bullet$ The gamma-ray luminosity function of GRBs needs to be measured,
since in order to place specific models on Figure 4 we need to know this.
Estimates of the luminosity function can be made from the flux statistics
of bursts with measured afterglow redshifts [81].

\vskip 4pt \noindent
$\bullet$
Many GRB host galaxies should be seen at infrared
wavelengths by $SIRTF$.
This is be a useful test of the whole pictures.

\vskip 4pt
In summary, this seems to be a useful method although much
additional data is needed before it can provide an accurate
measurement of the cosmic star formation history.  The
optical part of the cosmic star formation history is already
reasonably well established.  The infrared part will
come from ALMA.  In the meantime, the GRB redshift
distribution should provide interesting insights.  In the
long term (post-ALMA) it should provide a useful consistency
check and will probably tell us something about the physics of
GRBs.

\vskip 20pt
\noindent
{\bf 2.5 Obscuration Extent} 
\vskip 5pt

\noindent
Given that most of the star formation in the Universe seems to be
taking place in infrared galaxies, it is worth addressing the issue of how
obscured this star formation is.  Figure 5 shows a rough analysis.

The symbol ``UV" represents the location of star-forming galaxies that
are bright at rest-frame ultraviolet wavelengths; these are what generate
the optical Madau Plot.  
At low redshift, the most luminous examples of these are late-type spiral
e.g.~Sc galaxies.
At high redshift, these are Lyman-break galaxies.
These objects have moderate or little obscuration ($A_V \leq 1$ mag typically),
moderate or low star-formation rates ($< 10$ M$_{\odot}$ yr$^{-1}$ normally,
although a few have star-formation rates greater than this), and consequently
low submillimetre fluxes (typically below 0.1 mJy).  Consequently
they occupy the bottom left corner of Figure 5. 

The symbol ``C2" represents the location of Class-2 SCUBA galaxies.
These are luminous submillimetre source with high star formation rates
that are also luminous at optical wavelengths.
The fact that the star formation rate derived from optical measurements is
comparable to the star formation rate derived from submillimetre measurements
means that the star formation in these galaxies is only lightly obscured
and hence they occupy the top left corner of Figure 5.

Most of the star formation probably happened in infrared galaxies, the
ones that dominate the infrared background (``IRB").  Suspected
examples are the three GRB host galaxies (``GRBh") described in the 
previous section.
That the afterglows were able to emerge from the dusty host galaxies follows
from the high specific intensity of the radiation source.
It also tells us that the obscuration in the host galaxies,
while wiping out most of the ultraviolet light, cannot be {\it too}
severe.  For example, were the hosts to be like local ultraluminous
galaxies such as Arp 220 (``ULIG") the afterglows would not have emerged -- there are
tens of magnitudes of visual extinction to the cores of these galaxies,
which is where all the energy is being generated.

The line-of-sight extinction to a GRB is, however, complicated by local
physical processes.  High-energy radiation from the GRB
may heat and sublime nearby dust [82,83,84,85],
but since the X-ray/UV flux $\sim d^{-2}$, dust
grains at $>>$ 10 pc from the GRB are unaffected [82], and 
it is unlikely that the GRB can affect the absorption properties
of dust on the scale of an entire galaxy.

So it appears that the infrared GRB hosts are quite unlike local
ultraluminous galaxies.  They are also different from Class 2-SCUBA
galaxies since their optical fluxes are low.
They are presumably some intermediate type of galaxy.

This is also what is suggested from evolutionary arguments
[60]: only about
1\% of the stars in local galaxies are found at densities above
100 M$_{\odot}$ Mpc$^{-3}$. 
This is gas density in the cores of local ultraluminous
galaxies like Arp 220 [86], which is where most
of the energy is generated.  

There are, however, three bursts which could have host galaxies
similar to local ULIGs.  These are the dark bursts GRB 000210,
GRB 970828 and
GRB 990506, which all had radio but not
optical afterglows.  In these cases the optical afterglow could have
been extinguished by a dense interstellar medium.  This possibility
is further suggested for GRB 970828 by the observation is a
merging/interacting system [87], like local ULIGs.  Additionally 
GRB 000210 is in an infrared-luminous galaxy that is not very
luminous at rest-frame optical wavelengths [88],
again like local ULIGs.

The bursts GRB 990308, GRB 980329 and GRB 980326 all had extremely faint
host galaxies but had optical afterglows [80] so these could perhaps be
some intermediate kind of galaxy between those described in the 
previous paragraph and infrared galaxies like the host galaxy of
GRB 980703.  
 
\includegraphics{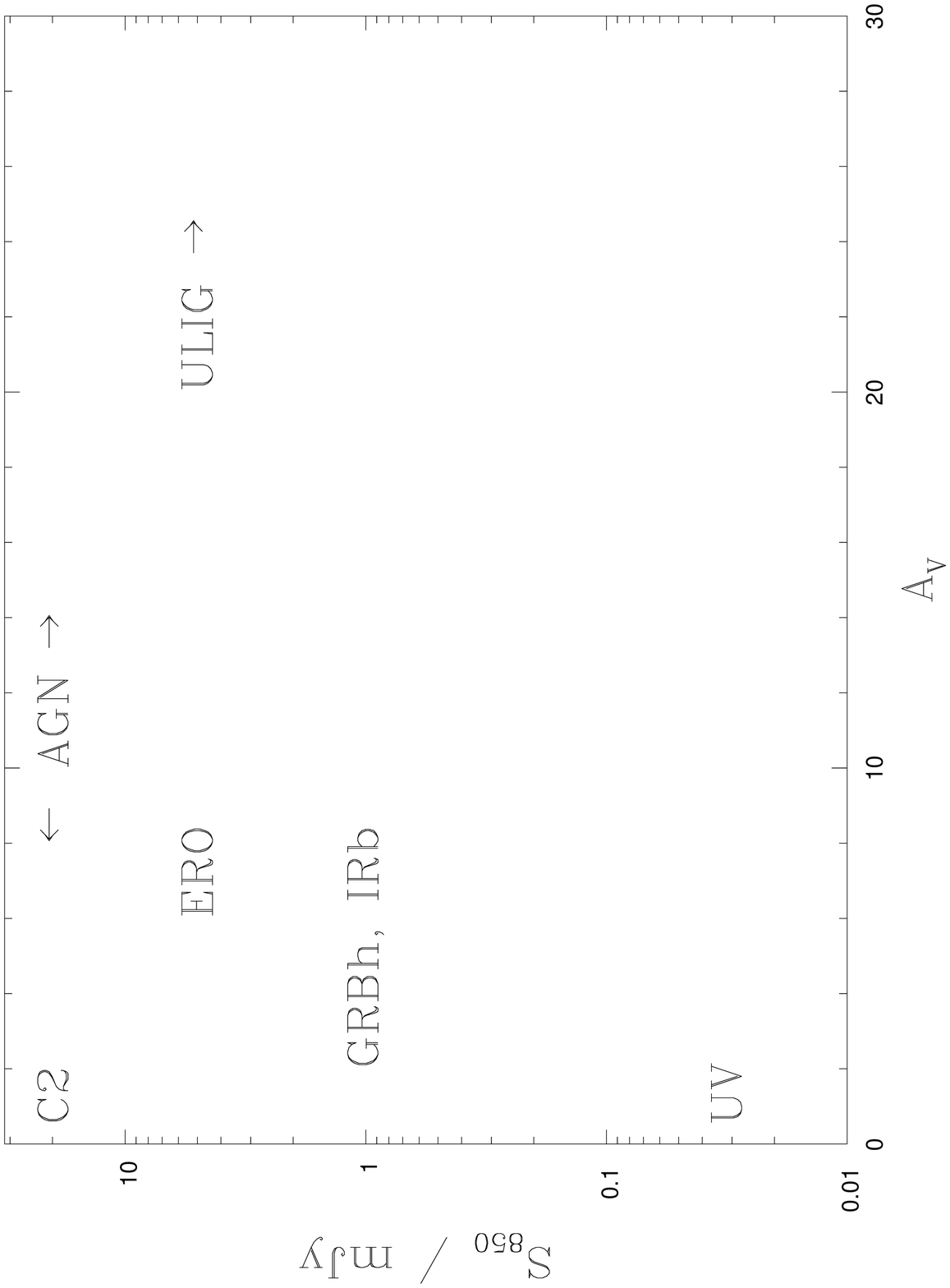}

\null
\vskip 346pt
\noindent
{\bf Figure 5:} Cartoon showing the total obscuration, parameterized
by the $V$-band extinction $A_V$ and the submillimetre flux
$S_{850}$ for various extragalactic objects.  See the text for
details.
\vskip 8pt

One more, independent, piece of evidence that suggests GRB host galaxies
exhibit only modest obscuration is the analysis of Galama \& Wijers
[83].  They show that the column densities of gas in GRB host
galaxies are typically $10^{22} - 10^{23}$ cm$^{-2}$, higher than in
spiral galaxies like 
the Milky Way, but lower than in the dense molecular cores of local
ULIGs like Arp 220.

In summary, there is considerable evidence that the star formation in
the infrared galaxies which dominate the infrared background and are likely
to produce most of the stars we see locally is only lightly obscured.
There must be some obscuration so as to (i) generate the high infrared
luminosities, (ii) extinguish sufficient rest-frame ultraviolet light
to prevent the galaxies from being luminous in optical samples, and
(iii) be consistent with column densities of $10^{22} - 10^{23}$ cm$^{-2}$.
There cannot be too much or (i) the optical afterglows would not emerge
and (ii) too many stars in local galaxies would reside at densities well in
excess of 100 M$_{\odot}$ pc$^{-3}$. 

\vskip 5pt

EROs are somewhat more obscured than these infrared galaxies since the
optical radiation (but not the near-infrared radiation) is completely
extinguished -- the GRB host galaxies described in the previous section
all had optical detections, although they were faint.   
They also have slightly higher submillimetre fluxes: the three
EROs discovered by Smail et al.~[68] and Gear et al.~[69] have
(unlensed) 850-$\mu$m fluxes of 8.6, 1.6, and 8.8 mJy, 
whereas the infrared background
is dominated by sources with fluxes closer to 1 mJy.  The EROs
therefore lie slightly upwards and to the right of the
GRBh/IRB point on Figure 5; presumably they are a somewhat more
obscured version of the more common type of infrared galaxy
indicated by the ``GRBh" symbol.  

Active galactic nuclei (AGNs) represent some of the most luminous
objects at optical [89], infrared 
[90], and submillimetre 
wavelengths.  Ultimately this is because accretion onto a black
hole is about 100 times more efficient at generating photons per
unit mass than star formation.
AGNs occupy the upper region of Figure 5
whether they are discovered by optical [91]
or submillimetre [92] measurements. 
It is possible that the horizontal position of a particular
object is determined by its evolutionary state.  In the
scheme of Sanders et al.~[37,38] quasars form by the
evolutionary sequence:
cold ULIG (e.~g.~Arp 220) $\rightarrow$
warm ULIG (e.~g.~Mrk 231) $\rightarrow$
infrared quasar (e.~g.~I Zw I) $\rightarrow$
optical quasar (e.~g.~3C 273; the most massive examples of
these may be radio-loud,  particularly in the late phases
of evolution).  On moving along this sequence, increasing
amounts of dust are blown away from the central regions
as the central black hole grows in size and the temperature
increases.  A given AGN therefore moves to the left on
Figure 5 as it evolves.  When the objects are very cold,
note that a starburst, not an AGN, is likely to be the
instantaneous power source (see ref.~93) 
and the models of
Rowan-Robinson \& Efstathiou
[94]. 

\vskip 20pt
\noindent
{\bf 2.6 Star Formation Rate Distribution Function}
\vskip 5pt

\noindent
A useful quantity to consider is the star
formation rate distribution function (SFRDF) in galaxies, defined
where ${\rm SFR}\,{\rm d SFR}$ is the number of galaxies with
star formation rates between SFR and SFR+dSFR.
The SFRDF tells us the sizes of the galaxies in which stars form.
This is proportional to the ultraviolet
luminosity function for optical galaxies
at low [14,15] and high [49] 
redshift.
For infrared galaxies, the SFRDF can be determined at low redshift
from the infrared luminosity function (e.~g.~ref.~95),
once the contribution for infrared emission from optical galaxies ($R>0$)
is subtracted.

The high infrared background means that the importance of the
infrared SFRDF relative to the optical SFRDF must evolve
strongly with redshift [32], but it is difficult to
quantify at redshifts $z>0.5$, which is where most cosmic star formation
is happening.
It is very probably top-heavy compared with the optical SFRDF, since
(i) most of the high infrared background is generated by
sources with $S_{850}$ = 1 mJy which have star formation
rates ($\sim$ 100 M$_{\odot}$ yr$^{-1}$) 
in excess of the most luminous optical galaxies, and (ii) the
four infrared galaxies discovered as GRB host galaxies
[10] all had star formation rates this high.

At the highest luminosities, many infrared sources may well be AGNs
so that establishing the SFRDF at the extremely-high SFR end may be
difficult.

This means that most of the stars that we see in local galaxies formed within big
galaxies.  The SFRDF is related to the local luminosity function, but
not directly, since mergers between stellar systems
produce final galaxies (which is what the
local luminosity function measures) bigger than the progenitors
(which is what the SFRDF measures).  The top-heaviness 
of the local luminosity function compared to the SFRDF is therefore an
indication of the role of stellar-kinematic mergers of existing systems
in producing local galaxies. 

\vskip 20pt
\noindent
{\bf 2.6.1 Absorption line systems}
\vskip 5pt

\noindent
Star-forming galaxies presumably form from HI galaxies, so one might expect
the mass function of HI clouds (seen in absorption in quasar spectrum) to be
similar to the SFRDF.
Lanzetta et al.~[58] demonstrates that for the two to be consistent, more 
star formation is required than is seen in optical surveys.  They suggest
that much of the star formation is missing from these surveys at high redshift due to
$(1+z)^4$ cosmological surface-brightness dimming.  Another possibility is
that much of the star formation that we might expect to occur from the HI
column density function is happening in infrared galaxies that are missing
from the optical surveys altogether. 

In any case, the concordance between the cosmological density 
$\Omega_*$ of stars at
$z=0$ and the integrated cosmological density
$\Omega_{\rm HI}$ in HI systems at all redshifts ($ \int_0^4 \Omega_{\rm HI}(z)
\, {{\rm d}z}  /   \int_0^4 \, {{\rm d}z} \approx \Omega_*$;
ref.~96) means that
all stars that we see locally are made of hydrogen that was part of an
HI cloud at some redshift, assuming that the atomic hydrogen got
converted to stars on short timescales.
The clouds at high redshift are 
simply acting as reservoirs.  

\vfil \eject

\null
\includegraphics{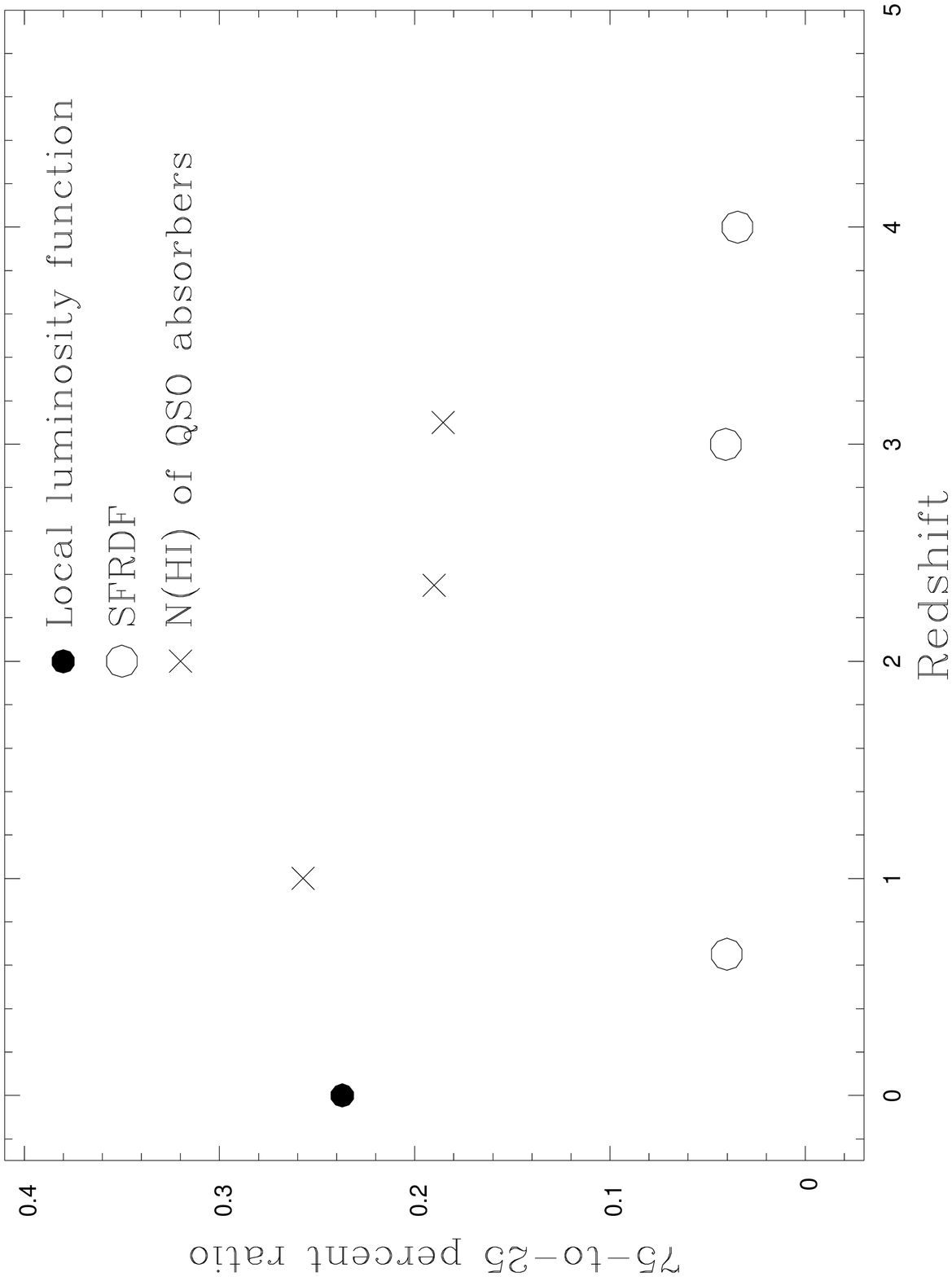}

\null
\vskip 336pt
\noindent
{\bf Figure 6:} The 75-to-25-percent ratio for the quantities
described in Section 2.6.  The $z=0$ galaxy luminosity
function is from the SDSS collaboration [14]. 
The SFRDF and the redshift evolution of the infrared galaxy
luminosity functiøn is from ref.~30.
The $z=3$ and $z=4$ optical galaxy luminosity functions are those of
Steidel et al.~[49] and at $z=0.65$ it is the 280-nm
luminosity function of Cohen [47].
We assumed further that the stellar IMF
is that of Kroupa et al.~[34], that the infrared 
galaxies have dust temperatures of 37 K [32], that
the infrared luminosity-to-star-formation-rate conversion
factor for our IMF is 1.2 $\times$ 10$^9$ L$_{\odot}$ M$_{\odot}$$^{-1}$
yr [98], and that for optical galaxies,
ultraviolet luminosities can be converted to star-formation rates
using Kennicutt's relation [99]:
SFR (M$_{\odot}$ yr$^{-1}$) = $1.4 \times 10^{-28} 
\,L_{\nu}$ (erg s$^{-1}$ Hz$^{-1}$) between 150 nm and 280 nm.
The SFRDF is then
SFRDF$_{\rm opt}$ + SFRDF$_{\rm IR}$ =
$0.00475  \, ({\rm SFR}/2.8)^{-1.0} \, \exp(-{\rm SFR}/2.8) \, 1/2.8 +
           0.000865 \, ({\rm SFR}/5.3)^{-0.1} \,
\exp(-0.95 \log_{10}(1 + {\rm SFR}/5.3)^2) \, 1/{\rm SFR}$ ($z=0.65$); 
$0.0016 \, ({\rm SFR}/6.1)^{-1.6} \, \exp(-{\rm SFR}/6.1) \, 1/6.1 + 
           0.00088 \, ({\rm SFR}/9.9)^{-0.1} \,
\exp(-0.95 \log_{10}(1 + {\rm SFR}/9.9)^2) \, 1/{\rm SFR}$ ($z=3$);
$0.0010 \, ({\rm SFR}/6.1)^{-1.6} \, \exp(-{\rm SFR}/6.1) \, 1/6.1 +
           0.00078 \, ({\rm SFR}/5.8)^{-0.1} \,
\exp(-0.95 \log_{10}(1 + {\rm SFR}/5.8)^2) \, 1/{\rm SFR}$ ($z=4$)
M$_{\odot}$ yr$^{-1}$ Mpc$^{-3}$, where
SFR is in units of M$_{\odot}$ yr$^{-1}$. 
Note that in all these functions SFRDF$_{\rm IR}$ is
dominated by higher-SFR galaxies than SFRDF$_{\rm opt}$.
The absorption-line system HI column density 
$N({\rm HI})$ distribution function is taken from the
measurements of P{\'{e}}roux et al.~[96].

The alternative view is
that clouds at $z=z_1$ simply evolve into clouds at $z=z_2<z_1$ and
the atomic hydrogen in the clouds was converted to stars
on long timescales, if at all (most hydrogen seen in HI systems at high
redshift might subsequently have been ionized and be part of the intergalactic medium
today).  If this alternative view is correct then it is plausible that most
hydrogen that is in stars at $z=0$ never resided in an HI system at any redshift
for any appreciable time.

The catchment volumes of the clouds at high redshift
could correspond to present-day galaxies.
A quantitative analysis is suggested.
A useful mathematical construction that describes the shape of a
function $f(x)$ of a measurable variable $x$ that is most reliably measured
at large $x$ is the
75-to-25-percent ratio
$$r={{x_{75}}\over{x_{25}}}$$
where
$$\int_{x_{75}}^{\infty} x \, f(x) \, {{\rm d}x} \,\,/
\,\, \int_{0}^{\infty} x \, f(x) \, {{\rm d}x} = 0.75$$
and
$$\int_{x_{25}}^{\infty} x \, f(x) \, {{\rm d}x} \,\,/
\,\, \int_{0}^{\infty} x \, f(x) \, {{\rm d}x} = 0.25.$$
A value $r \sim 1$ suggests $f(x)$ is top-heavy and high-$x$
biassed.  A value $r << 1$ suggests $f(x)$ is low-$x$
biassed.  Figure 6 shows $r$ for a number of related functions.

Figure 6 shows that many of the stars in local galaxies reside in
large systems and that much of the gas in high-redshift absorption-line
systems also exists in large systems.
Yet much of the star-formation in the Universe is not happening in
the galaxies with the highest individual star formation rates.  There are
many possible explanations for this interesting result.
Possibly the high-redshift absorption-line
systems represent the catchment areas of
systems that later become gravitationally
bound to form galaxies, yet the stars were formed in smaller
systems that later
merged to form the final galaxies that we see today.
This could be tested by studying the evolution of the correlation
function between infrared galaxies and evolved stellar
systems (seen in near-infrared surveys), the evolved stellar systems
being the end-products of infrared galaxies which have completed their
star formation.  The correlation function between infrared galaxies and
other infrared galaxies may be useful too but on the other hand may
be very small if infrared galaxies within a particular catchment
volume burst at different times
For example,  the luminous galaxy
SMM J14011+0252 [25] appears to be a many-component
system at optical wavelengths, and it is possible that one component
contains an obscured starburst whereas the others are evolved stellar
systems.

The idea that the HI clouds are systems that have yet to collapse and form
galaxies is also suggested by measurements of abundances in damped Ly$\alpha$
systems at high redshift [97].  The abundances in the clouds are
lower than those of Milky Way stars with ages comparable to the cosmological
ages of the clouds.  This suggests that at the cosmological ages in question
galaxies like the Milky Way, whose formation is traced by the SFRDF, have already
formed (presumably in smaller units), but the gas in the HI systems has not
yet been converted into stars in galaxies yet.

\vskip 20pt
\noindent
{\bf 2.7  Summary of Section 2}
\vskip 5pt

\noindent
The main result of this section is that most of the star formation in the
Universe is happening in infrared rather optical galaxies with
star formation rates of about 100 M$_{\odot}$ yr$^{-1}$.
Most of this star formation is unaccounted for in optical surveys because
of internal extinction, but this internal extinction is not large; a
number of independent arguments suggest that it is no more than a few
magnitudes in $A_V$.
The host galaxy of GRB 980703 appears to be a good prototype.
The starbursts in these galaxies are likely to be short-lived [100].

The precise fraction of star formation $f_{\rm IR}$
that happened in infrared galaxies is uncertain but it is probably
above 50 \% [30].

One of the main constraints on $f_{\rm IR}$ come from the spectrum of the extragalactic
background light.
In particular, our estimate of $f_{\rm IR}$ will be too high if infrared part of
the EBL spectrum is overestimated, perhaps due to contamination from a warm
Galactic component [101].

It will also be too high if the optical part of the the EBL spectrum is
underestimated, as would happen if there exists significant amounts of optical
flux originating from the outer parts of galaxies.  This would contribute to
the EBL [102], but would not be accounted for in determinations
of the EBL obtained by summing contributions from resolved sources.  It would
also be unaccounted for in determinations of the optical Madau Plot, particularly
if the star formation is happening at high redshift where $(1+z)^{-4}$ surface-brightness
dimming is important [58].
If this were the case, there would need to be very many stars or their remnants today in
the outer parts of galaxies and their halos, like the cold white dwarfs observed by
Oppenheimer et al.~[18].

Yet another issue which could lead to an overestimate of
$f_{\rm IR}$ is the temperature $T$ of dust-enshrouded starbursts.
If they are significantly lower than 40 K, then the star-formation rates implied
by radio measurements for galaxies like the host galaxy
of GRB 980703 will be too high.  The logic is as follows.
The temperatures are determined by modelling, number counts, the EBL, and
SEDs for a few luminous sources [32].  But if the bulk of the energy is
produced by sources of moderate luminosity (like the SCUBA 1 mJy sources)
that are colder
than the more luminous sources that contribute to the bright end of the
counts and have well-determined SEDs, then the bolometric luminosities
of the moderate-luminosity starbursts ($L_{\rm bol} \sim S_{850} \, T^{4.5}$)
and their star-formation rates (SFR $\sim L_{\rm bol}$) will be overestimated.
When we then determine SFR from a radio flux for any galaxy based on the
radio-submillimetre flux calibration [71,72], it will be too high.
When summed over the entire infrared poplation based on flux and/or
background measurements, the consequent value of $f_{\rm IR}$ will be an
overestimate.

\vskip 28pt

\noindent
{\bf 3 THE LOCAL GALAXY POPULATION}

\vskip 10pt 

\noindent
{\bf 3.1 Galaxy types and stellar populations}
\vskip 10pt 

\noindent
{\bf 3.1.1 Disks and Spheroids} 
\vskip 5pt

\noindent
Most familiar luminous galaxies are giant galaxies, either ellipticals or
spirals. 

Giant ellipticals have high
brightnesses at their centres and 
absolute $B$ magnitudes between about $-$25 and $-$15. 
Elliptical galaxies are featureless, with brightness profiles that
are high in the center and lower far away from the center.
They are red and metal-rich (typically 1 -- 2 Z$_{\odot}$), consisting mainly of old stars.
 
Spiral galaxies have absolute B magnitudes between about $-$24 and $-$18.  They
often look smooth near their centres (where the brightnesses are highest) and
have spiral arms at large radii from ther centres.  The spiral arms are often
irregular in form and one can see many condensations or knots, like HII regions
which are making stars.
Spiral galaxies can generally be separated into bulges and
disks [103,104].  The bulges are predominantly composed of
old, red Population II stars.  
The disks contain young, blue Population I stars, particularly in spiral
arms where there is considerable ongoing star formation.

The familiar Hubble sequence 
Sa$\rightarrow$Sab$\rightarrow$Sb$\rightarrow$Sbc$\rightarrow$Sc$\rightarrow$Scd$\rightarrow$Sd
is used for categorizing spiral galaxies.  The main parameter which varies 
along the sequence is the relative luminosities of the bulge and disk [105,106].
Galaxies classed Sa have substantial bulges.  Galaxies classed Sd
have no bulges.  Familiar examples are M81 (Sab), M31 (Sb), the Milky Way (Sbc), 
M51 (Sbc), M101 (a luminous Scd) and M33 (a low-luminosity Scd).

A less common type of giant galaxy is the peculiar galaxy.
They are often systems of interacting galaxies (e.g.~ref.~107)
or galaxies that have undergone a recent interaction.
These galaxies normally have SEDs which peak at infrared wavelengths.  Familiar
examples are the Antennae (NGC 4038/9), NGC 7252 and Arp 220.  

Galaxies with absolute blue magnitudes fainter than $-$18 are 
mostly dwarf galaxies.  
There are two kinds: red dwarf elliptical (dE) galaxies with old stars, smooth
morphologies and no detectable HI gas, and blue dwarf irregular (dIrr) galaxies which have
younger stars, complex morphologies (often they are star-forming condensations
embedded within a luminous matrix) and appreciable HI gas.
The dwarf irregulars are generally metal-poor ($<$ 0.1 Z$_{\odot}$) whereas most
dwarf ellipticals have higher metallicities (although still lower than those of
giant elliptical galaxies).
The dEs in the Local Group cover a large range in absolute magnitude, from
NGC 205 ($M_V=-15$) to Draco ($M_V=-8.5$).  There are also a number of
dIrr galaxies in the Local Group, like the Large and Small
Magellanic Clouds (LMC and SMC) and NGC 6822. 

If a galaxy has a bright quasar at the center, the quasar is usually so bright that
one cannot see the rest of the galaxy.  Such objects therefore look pointlike.

Finally, there do exist other, rarer, less classifiable galaxies, like giant 
low-surface-brightness galaxies that can barely be seen above the sky.
Malin 1 [108] is the prototype of this kind of galaxy.

\vskip 20pt
\noindent
{\bf 3.1.2 Halo Stellar Populations and MACHOs}
\vskip 5pt

\noindent
Galaxy halos mostly consist of dark matter, but there are small number of field
halo stars.  These are Population II stars, identified in the Milky
Way by their halo kinematics.
The fraction of the Galaxy's stellar mass that is in the halo is very
small (a few percent or less), but the exact mass of the halo stellar component
is known very imprecisely.  Freeman \& Bland-Hawthorn [109] suggest a total stellar halo mass of
about 10$^9$ M$_{\odot}$, but Binney \& Merrifield [110] point out that many of
the blue horizontal-branch stars used to estimate the halo mass might in fact be
thick disk members and that the field halo stellar mass may be only about 10$^8$ M$_{\odot}$.

Another stellar mass component of the Galactic Halo is the globular cluster (GC) population.
These are about 150 dense, very old Population II systems of characteristic masses
$10^5 - 10^6$ M$_{\odot}$.  Individual globular clusters are susceptible to tidal
disruption (e.~g.~Pal 5; ref.~111)
and much of the field halo could have been generated in this way, although
for most halo stars to have been part of a GC at some time in the past the GCs would
have to have been very much more numerous then than now.  

In an interesting recent development, Oppenheimer et al.~[18] have
discovered a number of cold (presumably old)
white dwarfs in the Milky Way halo.
They are almost dark: the prototype WD0346+246 has a mass of
about half a solar mass but a $V$ absolute magnitude 
of about 17 [112]. 
Oppenheimer et al.~[18] suggest that about 2\% of the halo dark matter could
be in the form of these white dwarfs.  If a contribution from helium white dwarfs
(which are not detected by Oppenheimer and collaborators) is included, this
fraction could be higher.  The total mass of the progenitors of
these stars, massive stars which would have completed their evolution and
become white dwarfs long ago and since cooled, is then comparable to the mass  
of stars in the rest of the galaxy.  The
mass in these remnants is far greater than that predicted from the
extrapolation of the normal field halo star mass function.

If the Milky Way is typical of other galaxies, then an appreciable fraction of the
star formation in the Universe would then have produced stars which reside in
galaxy {\it halos}.
This would be quite remarkable in light of our other knowledge of the galaxy
formation process so it is worth examining possible sources of systematic error
in the derivation of the 2\% fraction quoted above.
Possible systematic errors could arise from: 
\vskip 1pt \noindent 
(1) contamination from a thick disk [113];
\vskip 1pt \noindent
(2) contamination from stellar remnants ejected from the Galactic disk
by 3-body or multiple encounters [114], or the
remnants of donor stars from disk Type Ia supernovae [115];
\vskip 1pt \noindent
(3) the derivation of density in the V/V$_{\rm max}$ method used by
Oppenheimer et al.~[18] is sensitive to assumptions which are difficult to test about the joint
proper-motion--luminosity distribution function of the subsample used to derive the
density. 
\vskip 1pt \noindent
A second remarkable aspect of this result is that it would not be consistent
with our picture of Galactic or cosmological chemical evolution [19]. 
The problem is that the white dwarf progenitors would eject
a large mass of metal-rich gas.  This gas is not present in the Milky Way,
nor (should it have been ejected by a galactic wind) in Lyman-$\alpha$ forest clouds. 

Additionally, the results from the MACHO microlensing collaboration are now
available.  The purpose of this project was to look of lensing events by
dark objects in the outer Milky Way halo where the sources were LMC stars.  
The results are [116]: between 8 \% and 50 \% of the dark halo
consists of objects with characteristic masses between 0.15 M$_{\odot}$ and
0.9 M$_{\odot}$.
This mass may or may not be baryonic.  If it is, many of the baryons in
the Universe will be in the form of MACHOs.  Because the mass range above
corresponds to stellar masses, MACHO production, probably in galaxy
halos, would then dominate the cosmic
star formation history.  This would again be a remarkable result,
given what we know about star and galaxy formation.
 
The important issue then is to determine whether or not MACHO production
is or is not represented in the Madau Plot, and this depends on the
nature of individual MACHOs themselves.  Possible candidates include: 
\vskip 1pt \noindent 
(1) normal hydrogen-burning stars.  This would appear to be ruled out as the
MACHO results would overpredict the observed field halo star number density by
at least an order of magnitude; 
\vskip 1pt \noindent
(2) non-hydrogen-burning stars.
Brown dwarfs are stars too low in mass ($< 0.08 \, {\rm M}_{\odot}$) for
hydrogen to ignite, but the MACHOs are more massive than this.
If somehow stars with masses $\sim 0.5 \, {\rm M}_{\odot}$ can form but
fail to ignite hydrogen, these could be MACHOs if they exist
predominantly in galaxy halos.
Possible mechanisms for forming such objects are described by
Hansen [117] and Lynden-Bell \& Tout [118], based on
the collapse of thermal instabilities in primordial gas. 
In this instance the MACHOs would contribute significantly to the
baryon mass density of the Universe, but not to the Madau Plot
since they would never generate radiation at any wavelength; 
\vskip 1pt \noindent
(3) cold white dwarfs.  Although they have the right masses,
the mass density of cold white dwarfs
measured by Oppenheimer et al.~[18] appears too low for most MACHOs to
be this kind of object.  More generally, stellar remnants that
form as the end stages of stellar evolution are unlikely to be MACHOs if they
as numerous as implied by the microlensing results, because they would
then produce more helium and heavy elements than is consistent with
observation [19,119]; 
\vskip 1pt \noindent
(4) cold dark matter condensations. 
Most of the halo appears to be made of dark matter that is probably
cold.  The behaviour of instabilities in a dark matter fluid in which is
embedded large number of stars is complex, particularly if the stars
have a different velocity structure to the dark matter.  
If the instabilities grow to masses comparable to those of the stars,
they could be MACHOs.  Because the dark matter in this case would not
be baryonic, the formation of these MACHOs would not be represented on the
Madau Plot.  Note the importance of a field halo star population in
such a scenario.  In this scenario, galaxies lacking a substantial
field halo population (like Local Group dwarf ellipticals)
would not contain many MACHOs; 
\vskip 1pt \noindent
(5) quark nuggets [120] 
or other exotic forms of dark matter.  Again, if the
dark matter is of this type its formation would not be included in the
Madau Plot.  But there would certainly be an element of coincidence at
work here: it is not clear why the dark matter would clump on exactly the
same mass scales as normal stars and their remnants.  
\vskip 1pt \noindent
Finally, it is worth remarking that the
interpretation of the MACHO observations is subject to numerous
potential systematic errors (e.g.~anomalous LMC substructure, self-lensing);
so the numbers given in the previous paragraph could in principle be too
high.  The EROS 
microlensing project [121] has found less candidate events than the 
MACHO project and as more data is taken and analysed, potential systematic
errors associated with particular lines of sight will become better
understood.

\vskip 20pt
\noindent
{\bf 3.2 Luminosity Functions}
\vskip 10pt

\noindent
The total luminosity function of galaxies is well described by a Schechter
function (Equation (1).  One of the results from the large redshift surveys
that are currently in progress [14,15,47,122,123,124,125,126] 
is a precise measurement of the galaxy
luminosity function of the local Universe.  

\vskip 10pt
\noindent
{\bf 3.2.1 Type-specific Luminosity Functions} 
\vskip 5pt

\noindent
If the luminosity function is broken down into the contributions from 
different types of galaxies, more complex patterns emerge [13,127].  
The luminosity functions of E, S0,
Sa, Sb, Sc, and Sd galaxies are described approximately by Gaussian
functions, in order of decreasing mean luminosity.  The luminosity functions of
dIrr and dE galaxies are described by Schechter functions of comparable
characteristic luminosity.  It is further 
possible to perform bulge-disk decompositions and to obtain the
luminosity functions of different stellar populations, most significantly for
the bulges and disks in Sa and Sb galaxies (the star-forming disk component is
negligible in earlier-type S0 and E galaxies, and the bulge is
negligible in later-type Sc and Sd galaxies).  The luminosity functions
for each stellar population can be converted from optical to near-infrared 
$K$-band (2.2 $\mu$m) luminosity functions using 
optical--near-infrared colours (e.~g.~ref.~128). 
The $K$-band luminosity functions are particularly useful since
the $K$-band luminosities are not strongly affected by internal obscuration
from dust within the galaxies ($A_K = 0.1 A_V$).
 
\vskip 20pt
\noindent
{\bf 3.2.2 The Stellar Initial Mass Function}
\vskip 5pt

\noindent
The luminosities of stellar populations 
depend mainly on the contributions from high-mass stars.
In young Populations I these are blue OB stars.  In old
Populations II these are red giants and supergiants.   
The masses of the populations, however, depend primarily
on the contributions from low-mass stars, whose observational
signatures are weak.

A stellar initial mass function (IMF) $\xi (m)$ therefore needs to
be {\it assumed} when comparing observational quantities which
depend on luminosities to theoretical quantities which depend on
masses.  Population masses, mass-to-light ratios, and star
formation rates all depend on the mass integral
$\Xi_M = \int_0^{\infty} \, m \, \xi(m) \, {\rm d}m$.

Luminosities, on the other hand, are closely related to
the same integral performed above some threshold mass
$m_l$: $\Xi_L = \int_{m_l}^{\infty} \, m \, \xi(m) \, {\rm d}m$. 

The general form of the IMF is a steep power
law at high masses which flattens at 0.5 -- 1 M$_{\odot}$.
The IMF does not appear to vary significantly from place to
place [129].
The most simple approximation to the IMF is a single power-law
$\xi (m) \propto m^{\alpha}$ [130,131].
More complex functions that are derived from direct fits to
data are the IMFs of Gould et al.~[132] and Kroupa
et al.~[34].
The Kroupa et al.~[34] IMF has the advantage of a simple
analytic form: 
$$\xi(m) = \cases{1.87 \, \xi (1) \, m^{-1.3} & 0.08 $\leq$ m $<$ 0.5, \cr
                  \xi(1) \, m^{-2.2} & 0.5 $\leq$ m $<$ 1.0, \cr
                  \xi(1) \, m^{-2.7} & 1.0 $\leq$  m, \cr
}\eqno(3)$$
where $m$ is the stellar mass in units of the solar mass.

A commonly used IMF is the Salpeter [130] IMF with $\alpha=-2.35$ and a 
lower-mass cutoff of $m_l$ = 0.1 M$_{\odot}$.
Let us call the mass-integral of this IMF $\Xi_0$.
If then lower-mass cutoff is raised to 0.15  M$_{\odot}$, then
$\Xi = 0.87 \, \Xi_0$.  For a Gould, Bahcall \& Flynn [132] IMF below
1 M$_{\odot}$ and a Salpeter one above it,
$\Xi = 0.60 \, \Xi_0$ (see ref.~106).
For the Kroupa, Tout \& Gilmore [34] IMF with a lower-mass cutoff of
0.1 M$_{\odot}$, $\Xi = 0.45 \, \Xi_0$.  This is the IMF we assume elsewhere
in this review.
The reason that the mass integral is so much (about a factor of two) lower
than the Salpeter one is the flattening of the IMF at low masses and the
consequent absence of low-mass stars, the ones that dominate the mass
integral.

\vskip 20pt
\noindent
{\bf 3.2.3 Type-specific Mass Functions}
\vskip 5pt

\noindent
The mass function of entire stellar populations may be derived by
combining galaxy luminosity functions, bulge-to-disk ratios and 
population mass-to-light ratios $\Gamma$.  This is done in
Figure 7.  The quantity $\Gamma \propto \Xi$, so each of the
curves in Figure 7 scales with the value of $\Xi$ appropriate to
that population.  If the IMF is universal, all curves scale by the same
amount.  

\includegraphics{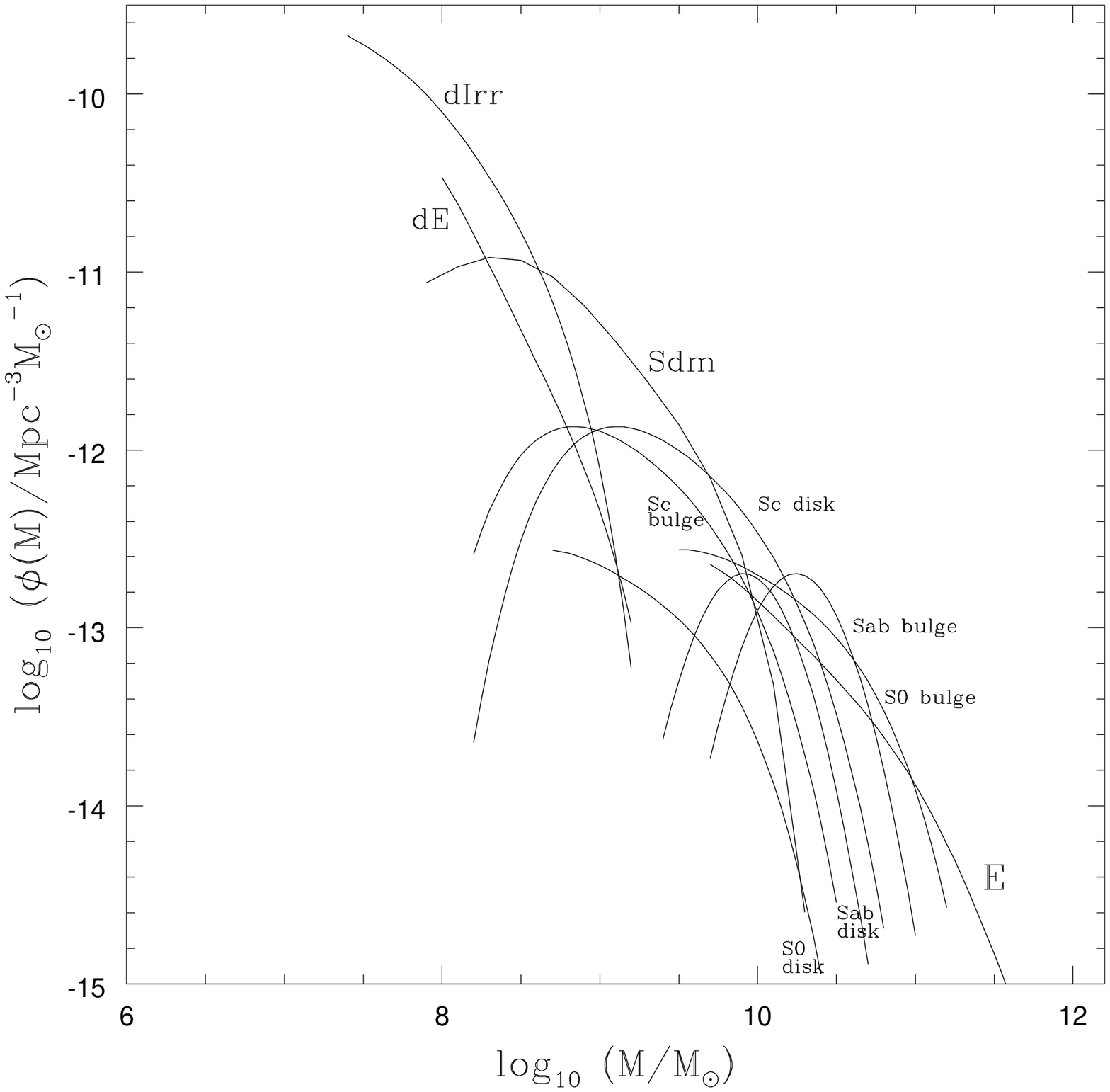}

\vskip 300pt
\noindent
{\bf Figure 7:} The galaxy mass functions $\phi(M)$ for elliptical
(E), lenticular (S0), spiral (Sab, Sc, Sdm), dwarf irregular
(dIrr) and dwarf elliptical (dE) galaxies.  The bulge and disk
component of the spiral and lenticular galaxies are considered
separately.  This figure is based on the luminosity functions of
Binggeli et al.~[13] and the mass-to-light ratios, bulge-to-disk
ratios and cosmological luminosity density of Fukugita et al.~[106].
The mass-to-light ratios used here are average values for classes of
stellar populations (e.g.~spheroids or disks).  A more rigorous
analysis would include the luminosity dependance of the stellar
mass-to-light ratios
($M/L \propto L^{0.35}$; ref.~133, 
which follows from the fact that lower luminosity
galaxies have lower metallicities [134]. 

\vskip 20pt

\noindent
Note that these mass functions apply to the stellar populations
alone and are independent of the presence or absence of dark matter.
Furthermore, for stellar populations the mass-to-light ratios are
increasing functions of luminosity, but when the 
dark matter is included
they become {\it decreasing} functions of luminosity [135] i.~e.~lower-luminosity
galaxies are increasingly dark-matter dominated.  

Figure 7 shows that the most massive systems are luminous elliptical
galaxies.  No spiral bulges or disks are as luminous as the most
massive ellipticals.
These massive systems probably turned their gas into stars very early
in the history of the Universe.  They have likely cleared out their
catchment volume early on and have little or no gas
left to form disks and/or turn into stars.  This follows from the facts that
(1) the earliest-forming systems have had the most time to convert gas into
stars, and (2) the most massive systems have large gravitational potential
wells and so are particularly efficient at collecting gas, compressing it
to high densities, and turning it into stars. 

In clusters of galaxies (note that Figure 7 is appropriate for the
field and is not relevant in this context), 
the supergiant cD galaxies are even more luminous
than the most luminous field elliptical galaxies.  
These are probably formed by special cluster processes linked to the
tidal disruption of individual cluster galaxies in the presence of the
massive cluster dark-matter halo.  This is inferred by: 
\vskip 1pt \noindent
1) the positions of cD galaxies in the precise centres of galaxy clusters; 
\vskip 1pt \noindent
2) the huge extended stellar halos around cD galaxies;  
\vskip 1pt \noindent
3) the high stellar velocity dispersions in the outer parts of cD galaxies [136],
implying the presence of dark matter in amounts characteristic of galaxy
clusters rather than individual galaxies. 

Finally note that spheroid populations (ellipticals and bulges) have higher
mass-to-light ratios than galaxy disks, by a factor of about 4.  This is why there are
comparable masses of bulge and disk in Sb and some Sc galaxies, even though the disks
dominate the optical luminosities.

\vskip 20pt
\noindent
{\bf 3.2.4 Low-luminosity and Low-surface-brightness Galaxies}
\vskip 5pt 

\noindent
One long-standing concern in extragalactic astronomy is that
galaxies with extremely low
surface brightnesses could dominate the luminosity density of
the Universe [137].
These would missing from optical surveys
because they are never visible above the night sky.
Recent measurements have shown, however, 
that the contribution from galaxies with extremely low
surface brightnesses has been shown to be small.  These conclusions
follow from deep optical
surveys (the survey of Trentham \& Tully [16] reaches nearly
29 $R$ mag arcsec$^{-2}$).  The results are particularly
important at low luminosities, where most galaxies, like the
Local Group dwarf spheroidals, are known to have low
surface brightnesses [138]. 
These low-luminosity, low-surface-brightness dwarfs, like Draco [139], have
high dark-matter fractions but their number densities are not high enough that their
dark matter dominates the total mass density. 

In the Local Group, incompleteness in the luminosity function due to our
missing very low surface brightness objects is perhaps more of a concern since
they cover such large areas of sky.  Around M31 the census [140] appears to 
almost complete.  Around Milky Way, new dwarfs like
Cetus [141] are still being discovered, but these 
have absolute magnitudes fainter than $M_B = -11$, the magnitude limit of the
luminosity function discussed in this section). 

\vskip 20pt
\noindent
{\bf 3.2.5 Environmental Effects}
\vskip 5pt

\noindent
There is increasing evidence that the galaxy luminosity
function is somewhat different in high density environments like
galaxy clusters than in low density environments (``the field").  There
appear to be two types of galaxy luminosity function -- one for
evolved regions (where the elliptical galaxy fraction is high, the
galaxy density is high and the crossing time is short) like the Virgo
Cluster and one for unevolved regions like the
Ursa Major Cluster (an accumulation of about 80 spiral 
and lenticular galaxies; ref.~142) and the Local Group.  The Virgo and Ursa Major LFs
represent natural prototypes for the two kinds of luminosity function (see
Figure 8) due to the small Poisson counting errors in the functions
(ref.~143 for Ursa Major, ref.~144 for
Virgo). 

The main differences between the two are that (1) amongst the more luminous galaxies, 
the elliptical galaxy
fraction is higher in clusters than in the field and (2) in clusters there are more low-luminosity
dwarf galaxies per giant galaxy, although {\it all} environments are deficient in
dwarf galaxies compared to the predictions from cold dark matter theory, which is
otherwise very successful at explaining the properties of galaxies and their large
scale distribution.  

The preponderance of giant
elliptical galaxies in clusters is probably due to ram-pressure stripping of
gas in cluster galaxies by the hot intracluster medium and by the numerous 
galaxy-galaxy interactions in the history of cluster galaxies, particularly in the
early stages of cluster assembly, before the galaxies acquired high velocities
(due to the high cluster velocity dispersion).

The preponderance of dwarf
elliptical galaxies in clusters is probably due to the fact that the stars in
the dark halos surrounding dwarf galaxies formed at an earlier time than the
stars in field galaxies.  This is borne out by the red colours of cluster dwarfs
compared to field dwarfs.  
The higher 
dwarf-to-giant ratios in clusters may be due in part to the fact that cluster
dwarfs, unlike field dwarfs, formed prior to reionization [145],
although it seems likely that this is not the only important physical effect
[146].  More generally, the conditions under which stars formed
in galaxies in dense environments (that would later become galaxy clusters)
in the early Universe is very likely different 
from the way stars formed in galaxies in normal, diffuse environments at
late times.

In the context of the current discussion we are interested in the {\it field}
luminosity function, not the cluster one.  The fraction of the mass in the
Universe in dense elliptical-rich clusters is small, perhaps not more than a few percent.
But the first galaxies in the Universe to form probably exist in such
environments today.  These tend to be the easiest galaxies to study, particularly
at high redshift, so care must be taken not to draw conclusions about cosmology
from the properties of small numbers of extreme objects.

\includegraphics{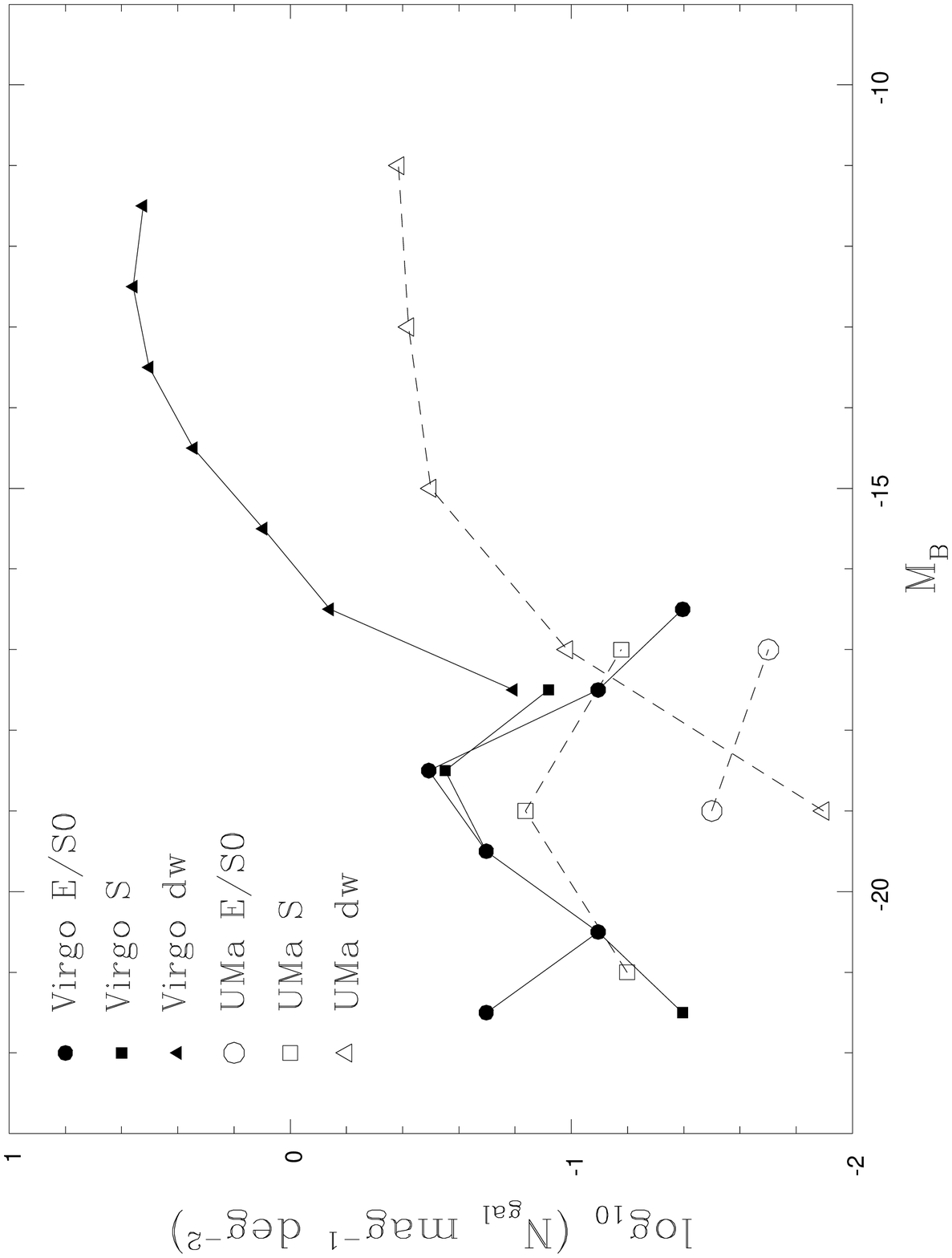}

\null
\vskip 346pt
\noindent
{\bf Figure 8:} The galaxy luminosity function for the
Virgo (filled symbols; ref.~144) and
Ursa Major (open symbols; ref.~143) Clusters for
elliptical/lenticular (E/S0), spiral (S), and
dwarf (dw) galaxies. 

\vskip 20pt
\noindent
{\bf 3.3 Structural and Kinematic Parameters}
\vskip 5pt

\noindent
Giant elliptical galaxies   
have brightness profiles well fit by a de Vaucouleurs $r^{{1}\over{4}}$ law;
$$I(r) = I_e \, e^{\, -7.67 \, 
\left({\left({{r}\over{r_e}}\right)^{{1}\over{4}}
-1}\right)},\eqno(4)$$
where $I_e$ and $r_e$ are the effective 
brightness and radius.
The structural parameters -- the absolute magnitude, effective 
surface brightness, central velocity distribution, effective radius 
radius -- of elliptical galaxies are described by
the fundamental plane [147].
Importantly, as the luminosity of elliptical galaxies increases, their central
surface brightness {\it decreases} (see Figure 9).  Luminous elliptical galaxies like
M87 are diffuse in their centre and are extremely large.
Small elliptical galaxies like M32 are dense and bright in their centres and
are extremely compact.
The most luminous ellipticals also tend to       
have dense stellar cores.  These are rarely isothermal
[148] and have light profiles
that are flatter than an  extrapolation      
of an $r^{{1}\over{4}}$ law to small radii.
Lower luminosity ellpticals tend not to                                
have these cores;                            
rather they have power-law brightness profiles that extend all the way to their centres
[149].
More luminous ellipticals also tend to have isophotes that are boxy,                              
lower-luminosity ellipticals that are disky [150].

The most luminous ellipticals are flattened by velocity dispersion anisotropy,
lower luminosity ellipticals by rotation [151].
The cores in the luminous ellipticals are often                                  
kinematically decoupled from the rest                            
of the galaxy (e.~g.~ref.~152)     
and can even be counter-rotating.
Kinematic measurements [39,153] 
also infer the presence of supermassive black              
holes in the centres             
of elliptical galaxies, probably the remnants of dead quasars and AGNs.
The black holes have masses roughly in proportion with the stellar masses of
the {\it entire} galaxy: $M_{\rm bh} \approx 0.005 - 0.010 \, M_*$ for our adopted
(low-$\Xi$) IMF and cosmology.

\null
\vskip 45pt

\includegraphics{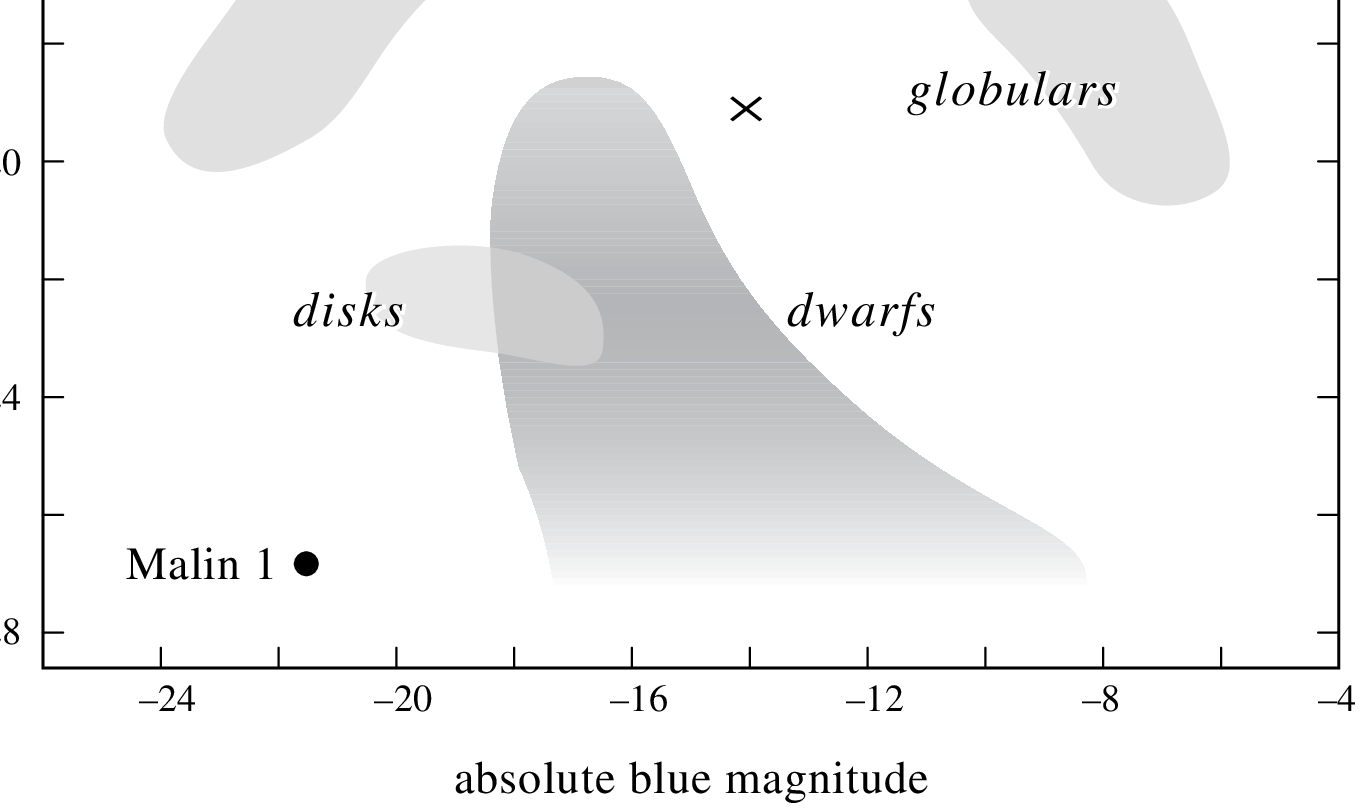}

\null
\vskip 276pt
\noindent
{\bf Figure 9:} The absolute magnitude vs.~central surface brightness
relationship for various stellar systems, adapted from
Binggeli [138].
Stellar systems do not normally occupy the regions between the shaded
areas, notable exceptions being Malin 1 and M32.  Another possible group of
exceptions are star-forming compact galaxies, like
Markarian 1460 [154]. 
For other examples, see the samples studied by Doublier [155]
and Drinkwater [156,157].

\vskip 20pt
\noindent

Bulges of spiral galaxies are similar in many ways to low-luminosity
elliptical galaxies.  They have brightness profiles that
follow a $r^{{1}\over{4}}$ law, lie on the same fundamental plane as
elliptical galaxies, are flattened by rotation 
as opposed to velocity dispersion anisotropy, and have black holes whose
masses follow the same black-hole-mass vs.~stellar mass  
correlation as elliptical galaxies.

The disks of spiral galaxies have lower surface brightnesses than elliptical
galaxies and have brightness profiles  which follow an exponential law:
$$I(r) = I_0 \, e^{-{{r}\over{h}}},\eqno(5)$$
where $I_0$ and $h$ are the central surface brightness 
and scale length. 
Many disks have central surface brightnesses
close to precisely 21.6 $B$ mag arcsec$^{-2}$ [158].

Dwarf galaxies also have exponential light
profiles.
They have decreasing surface brightness as they get lower in luminosity, a correlation
in the opposite sense to                                        
that for elliptical galaxies.
The faintest dwarf galaxies, Draco and Ursa Minor ($M_V=-9$), have very faint surface brightnesses
indeed and are barely visible above the sky.  
Both types of dwarf galaxy have similar correlations of
magnitude  with surface brightness
[159] and appreciable dark matter halos
(e.~g.~ref.~139).

Globular clusters have light profiles that are well-fit by King models [160].
Like dwarf galaxies, they have decreasing surface brightness as they decrease in luminosity,
but they do not have substantial dark
matter halos and have much higher surface brightnesses than
dwarf galaxies.  Consequently they are 
easy to distinguish from dwarf galaxies although they have comparable
optical luminosity. 
Globular clusters are spherical in shape and
normally found in the halos of giant galaxies, sometimes far out in the outer
halo e.g.~the Pal Clusters and NGC 2419 [161] around the Milky Way.
The formation mechanism for globular clusters is poorly understood [162] 
but there is evidence of a supermassive black hole in the centre of the
Milky Way globular cluster M15 [163,164] 
and this may provide some important clues. 
Globular clusters that are unambiguously intergalactic have yet to be found, 
even in galaxy clusters [165].

\vskip 20pt
\noindent
{\bf 3.4 Colour-magnitude Diagrams and Star Formation Histories of Local Group Galaxies}
\vskip 5pt

\noindent
Detailed star formation histories are available for Local Group galaxies
because individual stars can be resolved, even at the low-mass end of the main
sequence.  Some general results are summarized by Hodge [166].
Of particular note are
\vskip 1pt \noindent
1) M31 and the Milky Way have quite different star formation histories.  In M31 almost
all the stars were made at early times.  In the Milky Way
many stars were made early on, but the star formation has continued
to the present day, the current rate being about 1 M$_{\odot}$ yr$^{-1}$;
\vskip 1pt \noindent
2) the star formation in the Magellanic Clouds is extremely sporadic, with
bursts happening on scales of 1 kpc separated by times of about 10$^8$ yr.
When we look at most dIrr galaxies (like the Magellanic Clouds), we typically
see a small number of extremely blue regions of enhanced star formation  
embedded in a blue luminous matrix of typical dimension 1 kpc.
In order to generate this instantaneous view, the bursts must have lifetimes
comparable to 10$^8$ yr and the young stars responsible for the blue light
must form with such a high velocity that many are able to leave their
birthsites before they die.  There must also be a plentiful gas supply all over
the galaxy to fuel this star formation.  Note that dIrr galaxies are
observed to be HI-rich;
\vskip 1pt \noindent
3) Local Group dwarf elliptical galaxies have a wide variety of star formation
histories.  For example, in Ursa Minor all the stars are very old and metal-poor --
indeed the colour-magnitude of this galaxy [167,168] looks similar to that of a Milky
Way globular cluster.  On the other
hand, in Carina (e.~g.~ref.~169) most of the 
stars formed several Gyr after 
star formation began. 

In the Local Group, M31 and the Milky Way are the only two
luminous giant galaxies of type that dominate the local cosmic
star formation rate.
So we need to look beyond the Local Group if we wish to get colour-magnitude diagrams
of typical luminous star-forming galaxies.
The problem is that, for such galaxies, 
we can only resolve the most luminous stars (this is increasingly
true with increasing distance and consequently with increasing sample size).
In general, we must 
model star formation histories based on stellar evolutionary theory (that is
well-tested in the Milky Way and Local Group),
colour-magnitude diagrams of the most luminous stars, broadband colours of the galaxies,
and various spectroscopic indicators [170,171,172,173,174,175].
 
These models work well but uniqueness is a problem, especially as to determining
precisely when unresolved very old stars (the ones which dominate the stellar light in most
massive galaxies) form -- this is because observables which are sensitive to the
presence of these stars, like near-infrared colours, vary only slightly with
age when the stars get old.
The situation will improve significantly with the advent 
of the {\it Next Generation Space Telescope}, when we will be able to
probe much further down (in terms of stellar luminosity) the colour-magnitude
diagram for galaxies beyond the Local Group.

\vskip 20pt
\noindent
{\bf 3.5 Calculation of $\Omega_*$}
\vskip 5pt

\noindent
Fukugita et al.~(ref.~106, hereafter FHP98) sum over the optical ($B$-band) galaxy
luminosity function   to determine
the luminosity density of the Universe and then compute 
$\Omega_*$, the cosmological density in stars in units of the
critical density.
The principal uncertainties in this analysis are the normalization of
the galaxy luminosity function (and consequently the galaxy luminosity density)
and the stellar initial mass function.
They find
$$\Omega_* = 0.0024 \, h^{-2} \, \left({{\cal{L}_B}\over{2.0 \times 10^8 \, h \, {\rm L_{\odot}
Mpc^{-3}}}}\right) \left({{\Xi}\over{{\Xi}({\rm GBF} )}}  
\right),\eqno(6)$$
where ${\cal{L}_B}$ is the local luminosity density of the Universe and
${\Xi}({\rm GBF} )$ is the mass-integral of the stellar IMF appropriate to the
IMF of Gould et al.~[132].
About 3/4 of these stars are in spheroids (ellipticals and bulges of spirals) and 1/4 in disks.
Adopting $h=0.65$, and 
using the value of ${\Xi}$ implicit elsewhere in this article (from Kroupa
et al.~[34], hereafter ${\Xi}({\rm KTG} )$), Equation (6)
implies $\Omega_* = 0.0036$. 
The uncertainty in this value is large and depends on a number of factors:
\vskip 1pt \noindent
1) Uncertainty in ${\cal L}_B$.  This is probably about 10\% of ${\cal L}_B$, 
estimated by FHP98 based on an intercomparison of the results of various
redshift surveys;
\vskip 1pt \noindent
2) Uncertainty in the mass integral of the stellar IMF $\Xi$.  This is difficult to
estimate because the contribution to the total stellar mass from low-mass stars is poorly
determined in any galaxies other than the Milky Way.  A plausible estimate of the
uncertainty can be obtained by comparing the values of $\Xi$ from different
Galactic surveys: $\sigma (\Xi) / \Xi \approx [{\Xi}({\rm KTG} ) - {\Xi}({\rm GBF} )] /
{\Xi}({\rm KTG} ) = 0.33$      
\vskip 1pt \noindent
3) Uncertainty in the global mass-to-light ratio $<\Gamma>$ averaged over all
stellar populations, which can be further divided
into two:
\vskip 1pt \noindent
3a) Uncertainty in the conversion between mass and luminosity for a given population with
known IMF.  This is due to uncertainty in the models described in the previous section,
which are due to uncertainties in radiation transfer modelling in stars, 
in modelling stellar evolution (which is sensitive to stellar ages), and in the
contributions from stellar remnants of different types.  It is further complicated
by our lack of knowledge about the precise nature of internal extinction.
FHP98 use the models of Charlot et al.~[176] to determine $\Gamma = 5.4 - 8.3$ for the
$B$-band, corresponding to a fractional uncertainty of 0.42.
\vskip 1pt \noindent
3b) Uncertainty in the proportions of different stellar populations in different
galaxies.  For example, Simien \& de Vaucouleurs [105] determine that only 61 \% of the
light in elliptical galaxies comes from old stars with high mass-to-light ratios, 
not the 100 \% conventionally assumed (e.~g.~FHP98 and references therein).  As Population II
stars typically have $\Gamma$ about 4 times higher than Population I stars, this
uncertainty leads to a fractional uncertainty in $\Omega_*$ of about
$(4 - 4 \times 0.61 - 0.39)/4 = 0.29$ for the elliptical galaxy contribution.  For other
galaxy types, this source of error is less, and the luminosity-weighted fractional uncertainty
averaged over all galaxy types is closer to 0.2.
\vskip 1pt \noindent
Combining all these uncertainties suggests a value $\Omega_* = 0.0036 \pm 0.020$. 

A similar analysis done in $K$-band also yields $\Omega_* = 0.0036$, with comparable
error.  The motivation was to use data where the effects of internal extinction 
are small and we are able to observe the stellar populations directly.
Although the method was similar,
in deriving this number I tried as far as possible to use different datasets:
the luminosity density was drawn from the $K$-band luminosity function of
Szokoly et al.~[177], near-infrared mass-to-light ratios from Thronson \& Greenhouse
[178], and near-infrared colours from de Jong [179] and Huang et al.~[128].
It is therefore reassuring that the results are so similar.

In summary I conclude 
$$\Omega_* = 0.0036 \pm 0.020.$$ 
If a Salpeter, not Kroupa et al.~IMF [34] is used in deriving $\Xi$, a value of $\Omega_*$
a factor of two higher.
   
These values of $\Omega_*$ are much lower than the values of
$\Omega_{\rm baryon} = 0.052 \pm 0.008 \, (h/0.65)^{-2}$ inferred from
big-bang nucleosynthesis (refs.~180 and 181;
however note the discrepant results in  the deuterium abundance derived by
Songaila et al.~[182] and O'Meara et al.~[183])
and microwave background anisotropies [184,185]. 
Most of these missing baryons are not associated with galaxies at all
($\Omega_{\rm gas \,\, in \,\, galaxies} \sim 0.0006$; FHP98).  Most are in ionizied
gas in small groups.  Theoretical arguments [186] based on
the broadening of Ly$\alpha$ forest lines in quasar absorption spectra suggest that
most baryons exist in the intergalactic medium (IGM) at $z \sim 2$.
The same seems to be true at $z=0$ although the growth of structure in the
Universe between $z=2$ and $z=0$ means that this gas is now associated more
with diffuse galaxy groups (FHP98) than in a genuine, smooth IGM.
 
So a picture is emerging where the mass content of the Universe [185] 
is about 
65 \% dark energy (perhaps a cosmological constant),
30 \% dark matter (probably cold) which has a pressureless equation of state and 
clumps gravitationally, and
5 \% baryons of which about one-tenth is in stars.
It is interesting that
this inventory, suggested mainly from observations of the cosmic microwave
background in conjunction with studies of high-redshift supernovae
[187], is consistent with the distribution of
material in rich galaxy clusters [188].
These clusters are bound systems that assemble at late times but represent high-$\sigma$ peaks in
the mass distribution, so it is likely that the star formation in the galaxies within
them was complete at early times and that the cluster material distribution  
is close to the ``end-state" distribution for the Universe.
This notion is consistent with what we saw in Figure 1 -- the star formation rate at the
present time is very much lower than what it was in the past.

A question that follows from the previous paragraph: is the cold dark
matter associated with individual galaxies containing stellar populations?
Giant galaxies do not possess sufficient dark matter for this to be the case.
For example, M31 has a total (including dark matter) mass-to-light ratio of about
60 [189] and a stellar mass-to-light ratio of about 5.
Therefore about 1 part in 12 by mass is in stars, much higher than the ratio
0.5/30 implied in the previous paragraph.  Dwarf galaxies are also not a
candidate.  Galaxies like Draco possess a great deal of dark matter, but not enough
to explain the difference between the observed field galaxy luminosity function
and the cold dark matter mass function (Figure 10).  Therefore most cold dark matter is 
likely to reside in individual dark halos containing no stars [190] 
or to be smoothly distributed.

\includegraphics{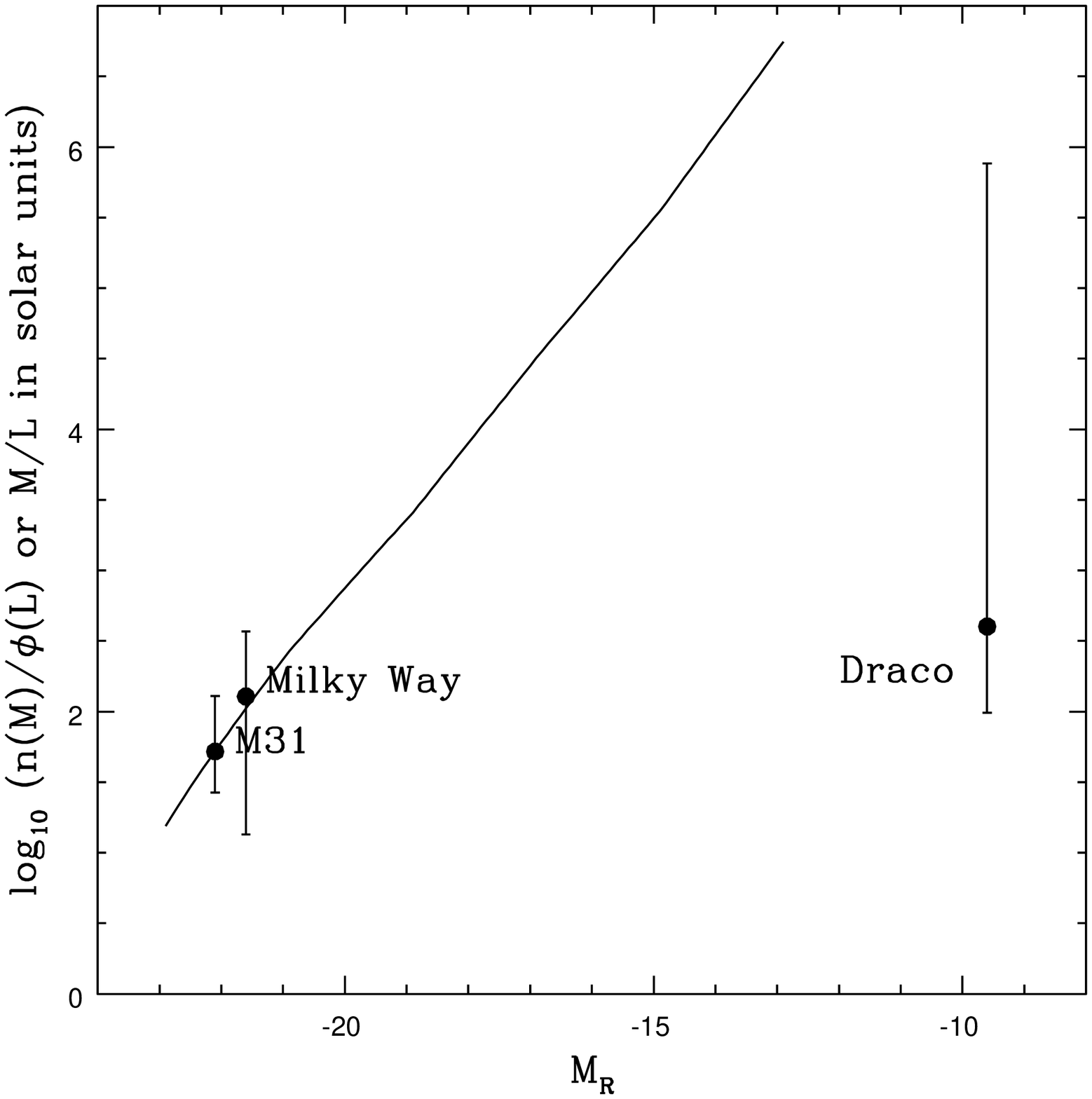}

\null
\vskip 300pt
\noindent
{\bf Figure 10:} The ratio of the galaxy mass function to the field luminosity function, 
normalized so that a galaxy having $M_R = -22.1$
has a mass-to-light ratio of 60, as appopriate for M31, given the
analysis of Evans \& Wilkinson [189].  The mass function is
computed from Press \& Schechter [191] theory, assuming the CDM
power spectrum of Efstathiou, Bond \& White [192] with a shape parameter
$\Gamma = 0.3$ and a normalization $\sigma_8 = 1$.  The luminosity
function is the Ursa Major luminosity function [143], 
assumed to be representative of
the field.
The three points show the mass-to-light ratios of M31
[189], the Milky Way [193] and
Draco [194 with an upper limit from ref.~195],
using absolute magnitude from ref.~196, adjusted to the $R$-band assuming $V-R=1$.
 
\vskip 20pt
\noindent
{\bf 4 THE MATCH BETWEEN THE COSMIC STAR FORMATION HISTORY AND THE LOCAL
GALAXY POPULATION}

\vskip 10pt
\noindent
A major aim of extragalactic astronomy is to combine the results of the previous two sections.
The equation describing the formation of stars in the Universe is
$$\Sigma_i \int_L \, L \, \phi_i(L) \, \Gamma_i \, {\rm d}L = \int_t {\rm SFR}(t) \, {\rm d}t,\eqno(7)$$
where $\Gamma_i$ is the mass-to-light ratio of stellar population $i$ ($\Gamma_i$ is derived from
stellar evolution and population synthesis models).  Having gotten $\phi_i(L)$  and SFR$(t)$, the
aim is then to evaluate this equation in detail, matching the contributions
to the sum in the first term to the relevant part of the integral in the second term.
The first stage will be to evaluate the integral on the right-hand side of Equation (7); note
that the sum on the left-hand side has already been computed in Section 3.5.
Integration of the Madau Plot described in Section 2 gives
$$\Omega_* \approx 0.0038 \,
\left(0.16 + 0.11 \, {{1+P_{150 \, {\rm nm}}}\over{4.7}} + 0.73 \, {{1+P_{280 \, {\rm nm}}}\over{2.7}}\right) \,
{{1}\over{1-f_{\rm IR}}}
.\eqno(8)$$
Here the $P$ values are contribution-weighted averages.
Given the numbers in Section 3.5, this would seem to provide evidence in favour of a small
value of $f_{\rm IR}$.

Below are listed some of the important observational and theoretical results that will
provide constraints on
the matching algorithm, along with descriptions of improvements that are expected
to happen over the next few years.

\vskip 10pt
\noindent
{\bf 4.1 Evolution of $\Omega_{*}$}  
\vskip 5pt

\noindent
In an important study Papovich et al.~[197] measured the evolution of
$\Omega_{*}$ by determining the stellar masses of an optically-selected
sample of field galaxies with spectroscopic or photometric
redshifts.  This was achieved by performing near-infrared
photometry and fitting the evolutionary
models described in Section 3.4 to broadband colours.
Because the wavelength baseline is so long, the stellar masses are well-determined
for the galaxies.  A more recent study using a larger sample was performed
by Dickinson et al.~[198], and their results are presented in Figure 11.

\includegraphics{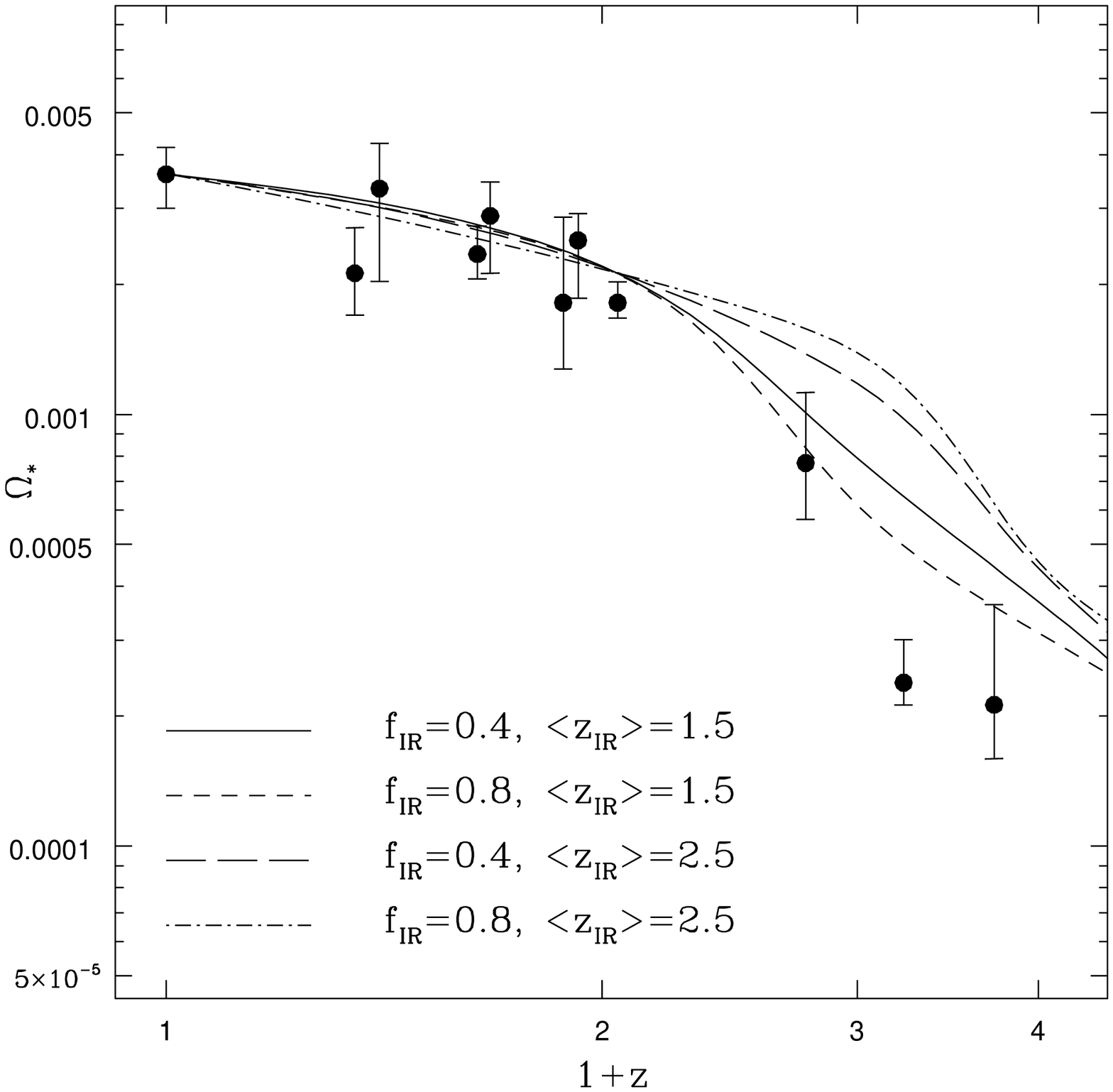}

\vskip 300pt
\noindent
{\bf Figure 11:} The evolution of $\Omega_{*}$ (adapted to our cosmology
and stellar IMF), the cosmic density in
stars.  The data are from Figure 7 of Dickinson et al.~[198] and come
from refs.~15, 47 and 199.
The lines represent the predictions of four groups of models described in
Figure 4, defined 
by $f_{\rm IR}$ and $<z_{\rm IR}>$. 

\vskip 20pt
These studies already provide tight constraints -- the error bars 
in Figure 11 are small.  It appears that about 85\% of the stars in
the Universe formed between $z=2$ and $z=1$.  Two immediate conclusions
are:
\vskip 1pt \noindent
1) another population of 
star-forming galaxies is required in addition
to optical galaxies.  This is because the shape 
of the dashed line in Figure 1 is such that its integral over time
between $z=2$ and $z=1$, while high, does not represent 85\% of the
total integral.  Given the discussion in Section 2, infrared galaxies would
seem to be a natural candidate for this extra population;
\vskip 1pt \noindent
2) the infrared Madau Plot must be heavily peaked at some redshift between
$z=2$ and $z=1$.  The four models shown in Figure 11 are all based on
Gaussians in redshift with widths $\sigma_z=0.5$, which are clearly too 
broad.    

Since the random errors are small, one particulary constructive
line of observational study will be to eliminate any possible systematic
errors.  The most serious systematic errors likely pertain to
source selection.  Much active star formation is hidden by dust 
and it is possible that many evolved stellar populations are also
obscured by dust,
particularly if dust dispersion timescales are long.
These stellar populations would be missing from optical surveys.
so multiwavelength selection may be required.
Also surface-brightness selection effects may be important, particularly
at high redshift where $(1+z)^4$ cosmological surface-brightness dimming
is important.
At low redshift this is less of a concern since the systems which dominate
the stellar mass in Figure 7 are high surface-brightness systems
e.~g.~elliptical galaxies.
 
\vskip 20pt
\noindent
{\bf 4.2 Stellar Population Age Matching} 
\vskip 5pt

\noindent
The construction of models describing the spectrophotometric evolution
of galaxies has become an active area of research.  These models are
discussed in Section 3.4 and have been very successful at explaining the
properties of nearby galaxies.

For higher redshift ($z>1$) galaxies, in which 
much of the cosmic star formation happens,
the situation is made more difficult by the fact that only broadband colours
and perhaps one or two emission line diagnostics are available for each galaxy.
This generally means that we
can get the instantaneous star formation rate reasonably precisely, as the
rest-frame blue and ultraviolet flux is heavily dominated by young, massive
OB stars.
If near-infrared colours are available, the total mass in stars may be determined
too, as in Section 4.1.  The star-formation history, however, is poorly
determined for the same reason as in Section 3.4 -- for all but the youngest
stars (and consequently populations of stars), the models predict that broadband
colours are a weak function of age.

Relative ages of populations, rather than absolute ages, are the most robust
outputs of the galaxy evolution models.  If we have a secure measurement of the
Madau Plot, these are sufficient to fully describe the match described by
Equation (7): both the left- and right-hand  
sides of this equation may be rewritten in differential form.

With the next generation of instruments, like those aboard the {\it Next Generation
Space Telescope}, the match between observation and theory will improve in two ways.
Firstly, we will be able to produce colour-magnitude diagrams for galaxies beyond
the Local Group to faint stellar luminosities.  This means that the galaxy
evolution models can be tested more precisely.  Secondly, observations of distant
galaxies will be made at much higher signal-to-noise, allowing a more detailed
comparison with the galaxy evolution models.

\vskip 20pt
\noindent
{\bf 4.3 Chemical and gas content evolution}
\vskip 5pt

\noindent  
In general, as galaxies evolve and form more stars, their gas content
decreases and their metal content increases.  Later generations of stars
will be more metal-rich as they will be made out of gas that is
increasingly polluted.  This is implicit in the models described in the
previous section.
Some of the metals, however, will not be retained by the galaxies and
will be ejected into the IGM by stellar winds and
supernovae, particularly from low-mass galaxies
with small gravitational potential wells.  This has a number of
observational consequences.

The most direct observable signatures are the abundances in Ly$\alpha$ forest
clouds or in the outer parts of damped
Ly$\alpha$ systems [200,201], 
thought to consist of baryonic material representative of the IGM.
At high redshift $z>2.5$, the chemical compositions of the clouds 
[202,203,204] can now be
determined e.~g.~using echelle spectroscopy on large telescopes.  But these 
high redshifts do not appear (from Figure 11) to be where most cosmic
star formation happens.  Measurements of abundances in clouds at low redshift
are difficult because they required high signal-to-noise ultraviolet
spectroscopy, which can be done only with the {\it Hubble Space Telescope}
(see e.~g.~the spectra in ref.~205) 
and even then are difficult to perform.  In the future, measurements such as these
will be possible with the {\it Next Generation     
Space Telescope} and will provide important constraints. 

\vfil \eject
\noindent
{\bf 4.4 Structural Parameter Matching}
\vskip 5pt

\noindent
If all that happens in the evolution of a galaxy is passive stellar evolution,
then the density of stars within galaxies
should be comparable to the gas densities in the star forming galaxies.

But secular evolution of stellar populations is important too.  Processes like
phase mixing and violent relaxation [206] occur in most galaxies,
but do not lead to huge density changes [207], when averaged over suitably large
volumes.  More dramatic changes like evaporation and core collapse (following
a gravothermal catastrophe) can happen, but the timescales on which 
these processes operate in galaxies is normally longer than a Hubble time.
The presence of a massive dark halo is important too, particular as regards making
galaxies stable to evaporation.

Stellar-kinematic mergers also play an important role in determining the final
stellar densities.  These are considered separately in the next section. 

Apart from studies of mergers, progress will be made in following areas;
\vskip 1pt \noindent
1) integral field units on large telescopes will permit spatially resolved spectroscopy
of star-forming galaxies.  This will permit us to determine the conditions at the
precise locations where most of the star formation is happening in these galaxies.
This will be particularly important at near-infrared wavelengths 
[208] since we may be
able to make measurements in infrared galaxies like the host of GRB 980703, at least in
the ones that are not too heavily obscured.  Additionally measurements of optical
galaxies will be possible, allowing a determination of $R$;
\vskip 1pt \noindent
2) as millimetre line receivers become more sensitive and as the number of
infrared galaxies known increases, studies of different transitions 
of different molecular species in the
infrared galaxies will become possible.
Given the discussion in Section 2.5, it might be expected that species like
HCN and CS, which probe extremely high densities, will not be detected in the
bulk of infrared galaxies, so a significant number of positive detections would 
imply that $f_{\rm IR}$ is low;  
\vskip 1pt \noindent
3) as spectrographs get more sensitive and are used on increasingly large
telescopes, absorption-line studies will become possible on increasingly
distant evolved stellar populations.  Constraints will come from such studies,
when performed on galaxies in the samples used in Section 4.1.  There will be two
types of constraints;
\vskip 1pt \noindent
3a) direct measurements of the kinematic of stars in high-redshift galaxies, and
\vskip 1pt \noindent
3b) very-high signal-to-noise measurements of the kinematics of stars in intermediate-redshift 
($0.1 < z < 1$) galaxies, like has already been done for a few $z=0$ elliptical
galaxies and bulges [209].
A template sample can then be constructed, and the
kinematics of stars in high-redshift galaxies determined by matching
to this template sample if the signal-to-noise of the high-redshift galaxy
spectra is low, perhaps with the help of non-spectroscopic indicators like colours and
morphologies.
 
\vskip 20pt
\noindent
{\bf 4.5 Mergers}
\vskip 5pt
 
\noindent
Galaxy mergers not only affect the galaxy luminosity by redistributing light to
systems of higher luminosity (with or without triggering new
star formation), but also the density structure of galaxies.  
In particular stars found at very low densities could have been ejected 
as debris from higher-density
regions during galaxy mergers, perhaps following the formation of tidal tails
(e.~g.~ref.~210).  These stars may fall back onto the remnant galaxy or
they may become unbound, depending on the ejection velocity and the parent galaxy
masses. 
 
The reason galaxy mergers happen is normally due to the collisions of dark matter
halos as structure builds up on large mass scales in hierarchical galaxy
formation scenarios like cold dark matter theory.  The halos drag their associated baryons,
which we observe as galaxies, with them, and the result is a galaxy merger. 
Galaxy mergers often have spectacular morphologies.

In summary, the final density structure of galaxies is determined by the joint effects of
the (gas) density at which the stars form, secular evolution, and the redistribution
of stars in stellar kinematic mergers.
Mergers are quite possibly the most important of the three, as they can lead to fairly
dramatic changes in the late stages of galaxy evolution.

Particular areas where the study of mergers is evolving rapidly are:
\vskip 1pt \noindent
1) observations of high-redshift mergers.  Larger optical telescopes mean improved
improved imaging capability in the outer parts of interacting galaxies where
dynamical clocks run slowest, the merger signature is strongest, but the signal-to-noise
is low;
\vskip 1pt \noindent
2) studies of low-redshift mergers.  This is driven by (1) increased sensitivity in
submillimetre and millimetre line and continuum (as well as
optical) imaging, and (2) increased computer
power.  Detailed numerical modelling of a large sample of local mergers, each observed
to a high level of precision, will provide an important template that 
can be compared to high-redshift galaxies, for which only limited measurements are
available;
\vskip 1pt \noindent
3) determination of the evolution of the luminosity function of evolved stellar
populations, identified as in Section 4.1.  This should provide a 
reasonably direct measurement of the evolution of the merger rate, so long as
contributions from newly-formed stars are removed.  Perhaps this will be difficult
if the fraction of mergers at high-redshift that are predominantly stellar-kinematic
and involve no new star formation is small.  If so, this will be important to know;
\vskip 1pt \noindent
4) measurement of the optical intergalactic light, that which contributes to the EBL but
is outside galaxies.  Techniques [102] are already in place to do this.
If it is assumed that the majority of stars seen between galaxies were flung out and
became unbound during mergers, this measurement provides a constraint on the integrated
(over time) merger rate.  If optical colours of this intergalactic light can also
be measure this provides (in conjunction with the galaxy evolution models) a constraint
on the redshift-dependance on the merger rate.
A possible source of contamination which needs to be subtracted from this measurement
is the contribution from galaxies that are not resolved because their surface
brightnesses are too low.  These are mainly low-luminosity galaxies, and most current
indications (e.~g.~ref.~16) are that this contribution is small.

\vskip 20pt
\noindent
{\bf 4.6 Joint Magorrian Relation Considerations}
\vskip 5pt

\noindent An identical analysis can be performed for the formation of supermassive black holes
in the centres of galaxies, many of which also seem to form in a dust-enshrouded
phase (e.~g.~ref.~211).  This process and the formation of stars must happen in
such a way so as to produce at the end the tight correlations observed between stellar mass and black
hole mass (e.~g.~ref.~39).  The joint consideration of star formation histories and 
supermassive black holes formation histories will provide constraints on the mapping
described by Equation (7).  It is worth discussing two extreme scenarios.

One possible scenario is that most of the mass build-up in AGNs happens during
mergers of two massive galaxies.
This is essentially the scenario proposed by Sanders et al.~[37,38] and described
at the end of Section 2.5.  The evidence on which this scenario is based is that the 
timescales for the build-up of mass in supermassive black holes are short 
[e.~g.~ref.~212] and that the host galaxies of {\it all} AGNs observed
(these are all at $z<0.5$) are massive systems.  This is true all along the
evolutionary sequence described in Section 2.5: both for
ULIGs that are probably AGN-powered (e.~g.~Markarian 231) and for optical
quasars (e.~g.~3C 273).  
As AGNs evolve along the sequence, their AGNs heat and then disrupt their
dust shrouds so that the most evolved luminous AGNs (radio-loud quasars [212]) 
reside in the most dynamically evolved stellar systems (e.g.~elliptical galaxies;
ref.~213).
The problem with this scenario is that most star formation appears to happen
between $z=2$ and $z=1$, whereas the quasar luminosity density
redshift distribution peaks at
$z>2$ [214], yet quasars are expected to happen in the very latest stages of
AGN evolution.

A second possible scenario is that most of the mass build-up in AGNs happens during
the early stages of galaxy assembly, before most stars formed.
The systems which later merged to form spheroids 
would already contain black holes at the time of the merger.
The Magorrian relation is then set by two physical processes:
1) spheroids are made during mergers 
[215] via physical processes like violent relaxation so that the Magorrian relation applies to
spheroids specifically and not disks [216], and
2) the proportion of gas ending up in the central black hole in the final
object is set by the proportion
of gas within a catchment volume that is capable of collapsing to very small radii
within one of the bound sub-units within the volume.   This proportion depends on
small-scale physical effects and is unlikely to vary significantly from one catchment
volume to another, since the volumes we need to average over to determine these
proportions are so big.
When the subsystems merge in this context, the stellar remnant (an elliptical
or spiral bulge) may then possess a binary black hole, depending on the
details of this merger.  Such a binary system is observed in both M31 [217]
and NGC 6240 [218], although the binary in NGC 6240, a rapidly evolving galaxy,
may eventually merge to form a single black hole.
The main problem with this scenario is that no local examples of galaxies with
massive black holes but few stars are known.  All local AGNs, including
low-luminosity ones, are in massive galaxies and these would in this context all
be secondary accretion events.
For this scenario to remain viable, we would need considerable fine tuning: the
entire process of gas collapse in subunits within a catchment volume
plus AGN build-up 
cannot be too efficient a process or the AGN luminosity density distribution [214] 
would peak at much higher redshifts or too inefficient a process or we would see
at least some local examples.

Maybe both scenarios are operating at some level.
In either case, a seed black hole [219] may be required to trigger the entire process.

The observational tests required here are conceptually simple: deep imaging of the
host galaxies of powerful AGNs is required.  
Paricularly important are measurements of AGNs at different evolutionary stages, perhaps
identified by their SEDs.  If the first scenario is correct, host galaxies of AGNs at
all evolutionary stages will be massive, evolved stellar systems and will show signs
of tidal debris in their outer parts (where dynamical clocks run slowest), even if
they look like elliptical galaxies in their centres.  
The morphologies of the host galaxies will become progressively more relaxed as the
AGNs grow.  If the second scenario is correct, host galaxies of AGNs at
all evolutionary stages (other than secondary
accretion events) will be gas-rich galaxies yet to form their stars.   
The existing stellar content of these systems will be small.

The constraints will improve in the future but already they are good enough to
provide perhaps the strongest constraint on a {\it low} value of
$f_{\rm IR}$.  Most galaxies that contribute to the infrared Madau plot have
redshifts $2 > z > 1$ (see Section 4.1).  Yet these are the galaxies which are
most likely to evolve into dense (Section 2.5) spheroidal (Section 3.5)
stellar systems, the ones that harbour black
holes.  It is the less dense optical galaxies that dominate the Madau Plot at
high redshift, but the most luminous AGNs appear to be in place by then.
Considerable fine tuning is then required to maintain a high $f_{\rm IR}$,
which is what is suggested by many of the other observations discussed in
Section 2.

\vskip 20pt
\noindent
{\bf 4.7 Number Counts}
\vskip 5pt

\noindent
Galaxies have SEDs characterized by $L_\nu$, which are
directly related to fluxes $f_\nu$ or apparent magnitudes
$m_\nu = -2.5 \log_{10} f_\nu$ + constant. 

For any model of the galaxy population evolution,
at each time or redshift each galaxy is
characterized by a particular SED.
Summing over the entire galaxy population,
$$n(f_\nu) = \int_V \int_{{{4 \pi d_L(z)^2 f_\nu}\over{(1+z)}}}  
\phi(L_\nu,z)
\, {\rm d}{L_\nu} \, {\rm d}{V(z)},\eqno(9)$$ 
where fluxes $f_\nu$ are related to magnitudes $m_\nu$ 
as above.  Here $\phi(L_\nu,z)$ is the redshift-dependent
luminosity-density function: the SED $L_\nu$ is related
to the bolometric luminosity $L$ by $L=\int_0^{\infty} L_{\nu}
\, {\rm d}{\nu}$ and the luminosity in any passband $T$
defined by the transmission function $T(\nu)$ as
$L_T = \int_0^{\infty} L_{\nu} \, T(\nu)\, {\rm d}{\nu} /
\int_0^{\infty}\, T(\nu)\, {\rm d}{\nu}$.
This must be consistent with optical [220] 
and near-infrared [221,222,223]  
number count data.

Normally the number counts are presented as binned
data, where the number of galaxies in each bin is
determined over a large surface area of sky (so
that uncertainties due to cosmic variance are
small).
The reasons a particular galaxy at a given redshift may
leave or enter a bin are
\vskip 1pt \noindent
1) the turning on of star formation in a galaxy.  This is 
normally a large effect.  If an infrared galaxy, the galaxy
can move up several bins for all $T$ in the infrared.  If
an optical galaxy, the galaxy
can move up (i.~e.~brightward)
several bins for all $T$ in the ultraviolet or
optical, unless it already has a substantial existing
stellar population (e.~g.~star formation that is suddenly 
turned on in local early-type or Sa galaxies), in which case
it only moves up a small number of bins; 
\vskip 1pt \noindent
2) stellar redistribution by mergers.  If two 
galaxies merge and the merger is purely stellar-kinematic
and not accompanied by new star formation, 
they both disappear from their existing
bins and reappear as a single unit is a higher bin.
This is normally a moderate effect, corresponding to a bin
shift of about 1 magnitude for an equal-mass merger.
The shift happens everywhere in the spectrum but is generally most
important at optical wavelengths, since this is where the SEDs of
existing stellar populations peak.  Conversely, if a galaxy is
broken into several smaller galaxies e.~g.~by a collision,
a single unit in a high-flux bin will reappear as several
smaller units in low-flux bins;
\vskip 1pt \noindent
3) as the stellar populations in galaxies evolve, they normally
do so in the sense of decreasing $L$ (fading) and a shift in
$L_{\nu}$ towards longer wavelengths (lower frequencies).
For such passively-evolving galaxies, there is therefore a slow
migration towards bins of fainter flux and decreasing
frequency;
\vskip 1pt \noindent
4) as dust in a galaxy clumps, disperses, or gets destroyed, more light
becomes visible at ultraviolet and optical wavelengths and the 
SED changes to reflect this.  This leads further to a trend of
increasing $n(m_\nu)$ towards lower frequencies.
This is particularly important if infrared galaxies at high redshift
evolve into (relatively) gas-poor stellar populations at low redshift;
\vskip 1pt \noindent
5) if AGN activity is turned on, a galaxy can move upward by several bins. 
Where in wavelength space this is most important depends on the
SED of the AGN  
(see Section 2.5), which in turn depends on its evolutionary state.
In principle this could affect number counts everywhere from the
radio to the X-ray, particularly at the highest flux levels.  However,
AGNs are unlikely to affect the bulk of the number counts since their
contribution to the EBL is small (see Section 2.2).

\vskip 20pt
\noindent
{\bf 4.8 The Milky Way Galaxy}
\vskip 5pt

\noindent
Historically measurements of stars within the Milky Way galaxy have provided 
the most rigorous constraints on its formation [224]. 
Over the next few years these measurements will increase in precision 
and stellar sample sizes will become much bigger.

Clearly our knowledge of the formation of the Milky Way will expand. The 
application to the wider galaxy formation problem is less clear to assess 
at this stage, but it seems a good idea to look for similarities 
between the Milky Way and field galaxies at all redshifts, as far as possible. 
This may well become a major observational industry over the next few 
years, as many more measurements of stars within the Milky Way are made. 
Both photometric and spectroscopic indicators will be useful.  Presumably 
the level of significance of all measurements will decrease with increasing 
distance so such a procedure will, no doubt, be piecemeal.

\vskip 20pt
\noindent
{\bf 5 SUMMARY}
\vskip 5pt

\noindent 
The formation of stars in galaxies is a well-defined problem and detailed
solutions will come from observational studies and modelling.  The
progress currently being made in both areas is rapid, but some results
are more secure than others in terms of their applicability to the
overall problem.
The aim of the current article is to place the results in this context.

Our current knowledge of even the basic form of the Madau Plot is
poor because the redshift distribution of the infrared Madau Plot is
unknown.  In fact, even the value of $f_{\rm IR}$ is unknown, although
various measurements, like the SCUBA 850-$\mu$m counts and the 
radio detections of optically-faint GRB host galaxies, lead us to suspect
that it is high.

Conversely the galaxy luminosity function at $z=0$ is now very well
determined.  The normalization is well known from the large number of
big redshift surveys currently in progress and the extreme faint end is
known from deep surveys of nearby diffuse groups of galaxies carried out
using large-format mosaic CCDs on 8 m telescopes.

Modelling of astrophysical processes like mergers and star
formation in galaxies is rapidly becoming more detailed because
1) increased computer power allows simulations with larger numbers of
particles and/or resolution elements, and 2) the observational constraints
on models are becoming tighter as new discoveries are made and the
signal-to-noise of existing observational results increases.  This is
similarly true of the stellar population evolutioon models described in
Section 3.4. 

Extragalactic astronomy is therefore a technologically-driven
field and much progress will happen over the next few years.

Many of the formalisms required to match the Madau Plot to the
local galaxy function have already been developed, as part of
semi-analytic galaxy formation models [225]. 
Conceptually semi-analytic models 
[23,226,227]
adopt quite a different  
approach to that described here.
They follow the gravitational growth of dark matter perturbations
and the collapse of baryons within them and attempt to model the
large number of physical processes that operate as this gas is
converted into stars.
The end product is the information listed in Section 3, and the
Madau Plot is output along the way.
Here we start with the Madau Plot, and 
deal with the various physical processes in a purely empirical
way, so are in effect only concerned with a limited part of the
galaxy formation process.  The semi-analytic models, on the
other hand, directly use recipes constructed from the laws of 
physics.  This approach has the advantage of putting the
entire galaxy formation process on a firm footing as regards the
laws of physics.  It has the disadvantage that many of the
more important physical processes,   
particularly feedback during star formation [228],
are so poorly constrained by observation that detailed modelling 
is not rigorous.  As the observational issues highlighted in this
article get resolved, detailed modelling of the physics is a natural
next step.

\vskip 20pt
\noindent
{\bf ACKNOWLEDGEMENTS} 
\vskip 5pt

\noindent
Figures 2 and 3 are reprinted from Physics Reports, Vol.~369, p.~111-176 "Submillimeter
Galaxies" by Blain et al., Copyright (2002) with permission from Elsevier.

\noindent
Discussions with my colleagues Andrew Blain,
Aaron Evans,
Jeff Goldader,
Simon Hodgkin,
Tae Sun Kim,
Priya Natarajan,
Bianca Poggianti,
Enrico Ramirez-Ruiz,
Martin Rees,
Dave Sanders,
Brent Tully, and
Mark Wilkinson are gratefully acknowledged.  
Also I thank Richard Sword for making Figure 9 and
Shireen Mohandes for helping to prepare the manuscript.

\vskip 20pt
\noindent
{\bf ADDENDUM} 
\vskip 5pt 

\noindent
In the few weeks since this article was submitted for publication,
a number of recent results have appeared in the literature that are
relevant to this work.  Three studies in particular are important in
the context of this review.

\vskip 10pt
\noindent
{\bf (i) A median redshift of 2.4 for galaxies bright at
submillimetre wavelengths -- Chapman et al.~2003, Nature}
\vskip 5pt

\noindent
Using the techniques outlined in Section 2.4.1.1, Chapman et al.~[A1] have
identified and measured the redshifts of a sample of 10 submillimetre sources.
They find a median redshift of 2.4.  

This is somewhat higher than the redshift of the bulk of cosmic star formation
required by the studies described in Section 4.1.  This could be reconciled
with a high $f_{\rm IR}$ and submillimetre background
if the bulk of the sources which dominate the
submillimetre background, the $S_{850}$ = 1 mJy sources, lie at lower 
redshifts than the sources Chapman et al.~observed.  There is some evidence
for this (see Fig.~A1).  
Interestingly, the two lowest-redshift sources of
have dust temperatures of only 16 K and 25 K, and consequently
low star-formation rates.  If these are typical of infrared
star-forming galaxies, then $f_{\rm IR}$ may be reasonably low  
even if the submillimetre background is high.

On the other hand, the lower-flux submillimetre sources may have
a similar redshift distribution to their higher-flux counterparts.
This could be reconciled with Figure (11) if either
\vskip 1pt
\noindent 1) the Chapman et al.~sample is contaminated by dust-enshroouded
AGNs.  This is suggested by the warm dust temperatures of the more energetic
sources in the sample and the concordance between the redshifts of the
SCUBA sources and of high-redshift quasars [214,A2] 
\vskip 1pt
\noindent 2) the SCUBA sources are forming stars that are later cannibalized 
by an AGN.  Each parcel of hydrogen is then responsible for creating two
sets of photons: one set when hydrogen is converted to helium in stars, and
one set when stellar mass is converted to black hole mass as the AGN grows.
This scenario has the advantage of generating a low $\Omega_*$ 
and local supermassive black hole mass density while maintaining a large
$I_{\rm opt}$ and $I_{\rm IR}$, all of which are supported by observation.  

\vskip 10pt
\noindent
{\bf (ii) Bimodality in the Clustering Properties
-- Budavari et al.~2003, ApJ}
\vskip 5pt

\noindent
Using photometric redshifts for a very large (2 million) sample of galaxies
from the Sloan Digital Sky Survey, Budavari et al.~[A3] noticed strong bimodality
of clustering properties when the galaxies were segregated by spectral type.
Evolved red galaxies have been known for some time to cluster more
strongly than star-forming galaxies but the new data shows that the transition
between the two regimes is very abrupt -- somewhat evolved, moderately red
galaxies cluster strongly like very evolved, very red galaxies, and
moderately blue galaxies cluster weakly like actively star-forming galaxies
that are very blue.

These results point towards two distinct formation environments for galaxies.
It may follow from this that there could exist two possible types of global
mechanism for converting gas into stars in galaxies -- perhaps this could
lead to two types of star forming galaxy: optical and infrared.

Following the evolution of the spectral-type subsamples defined by Budavari et al.~will
give a strong constraint on the match defined by Equation (7).
Tracking the galaxies backward in time allows us to estimate when their stars formed
out of gas and where on the Madau plot they lie.  Tracking them forward in time tells
us which stellar populations in the local Universe they correspond to and therefore
their contribution to the local luminosity function.
This constraint on the match defined by Equation (7) is a very indirect one,
but the significance
of the bimodality measured by Budavari et al.~is high enough that it will probably be
an important complement to the measurements desribed in Section 4.

\vfil \eject

\null 
\includegraphics{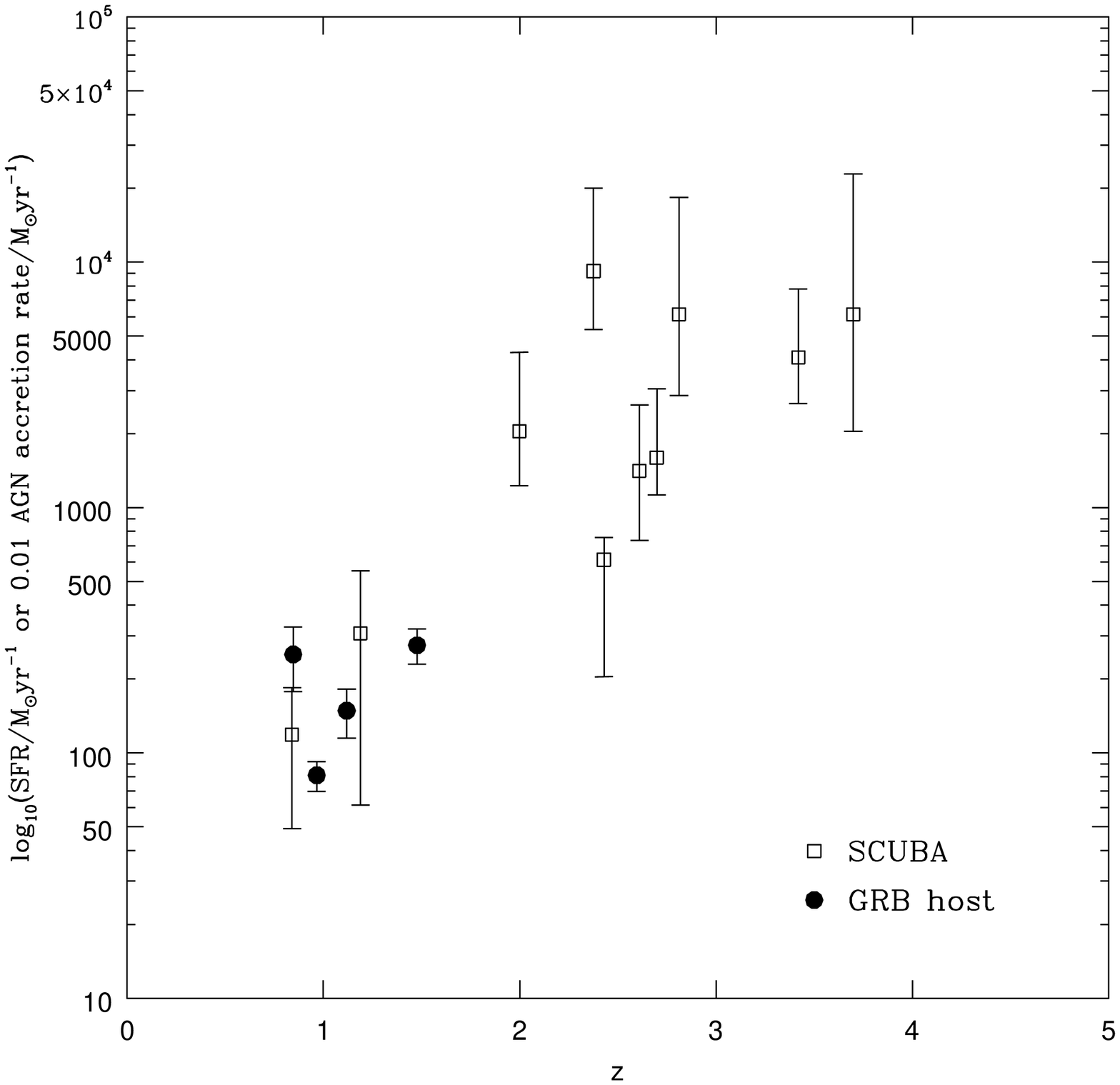}

\vskip 300pt
\noindent
{\bf Figure A1:} 
The redshifts and star-formation (or AGN mass-accretion) rates for the
SCUBA samples of Chapman et al.~(ref.~A1) and the hosts of the four
GRBs described in Section 2.4 that are infrared galaxies.

\vskip 30pt
\noindent
{\bf (iii) H$\alpha$ Spectroscopy of Galaxies at $z>2$ 
-- Erb et al.~2003, ApJ}
\vskip 5pt
\noindent
This work [A4] presents evidence that star-formation rate derived from nebular abundances is
somewhat higher relative to the star-formation rate derived from broadband
luminosities at $z=2$ than it is at $z=3$.

The redshift dependence of this effect appears to be moderated through the
galaxy luminosity.
The $z \sim 2$ galaxies studied by Erb et al.~have lower total luminosity and
star-formation rate than the $z \sim 3$ galaxies studied by Pettini et al.~[53]

If both samples of galaxies are typical of the ones which dominate the
cosmic star formation at  the relevent epoch, there are
important implications for how we determine the $P$ values in
Equation (8) -- these are redshift-averaged values weighted by the cosmic
star-formation happening at each redshift.  
Specifically $P_{280 \, {\rm nm}}$ may be higher relative to 
$P_{150 \, {\rm nm}}$ than we might expect from simple application  
of a Galactic extinction law.
The third term in the parentheses in Equation (8) then contributes more  
relative to the second term than we might expect and consequently the argument
made in Point 1) in Section 4.1 is weakened.
Indeed the considerable reradiated flux in
the $z \sim 2$ galaxies studied by Erb et al.~may be enough 
to turn these optical galaxies into the very galaxies postulated to exist there.

However, the Erb et al.~sample is small and therefore this kind of assertion
should be made with caution at this stage.

\vskip 20pt
\noindent
{\bf References}
\vskip 5pt

\noindent
[1] Blumenthal G.~R., Faber S.~M., Primack J.~R., Rees M.~J., 1984, 
Nature, 311, 517   

\noindent
[2] Bardeen J.~M., Bond J.~R., Kaiser N., Szalay A.~S., 1986, 
ApJ, 304, 15  

\noindent
[3] White S.~D.~M., Rees M.~J., 1978, MNRAS, 183, 341  

\noindent
[4] Madau P., Ferguson H.~C., Dickinson M.~E., Giavalisco M.,
Steidel C.~C., Fruchter A., 1996, MNRAS, 283, 1388

\noindent
[5] Hauser M.~G.~et al., 1998, ApJ, 508, 25

\noindent
[6] Schlegel D.~J., Finkbeiner D.~P., Davis M., 1998, ApJ, 500, 525

\noindent
[7] Fixsen D.~J.~et al., 1998, ApJ, 508, 123

\noindent
[8] Smail I., Ivison R.~J., Blain A.~W., 1997, ApJ, 490, L5 

\noindent
[9] Berger E., Kulkarni S.~R., Frail D.~A., 2001, ApJ, 560, 652

\noindent
[10] Berger E., Cowie L.~L., Kulkarni S.~R., Frail D.~A., Aussel H., 
Barger A.~J., 2003, ApJ, submitted (astro-ph/0210645) 

\noindent
[11] Frail D.~A.~et al., 2002, ApJ, 565, 829

\noindent
[12] Schechter P.~L., 1976, ApJ, 203, 297

\noindent
[13] Binggeli B., Sandage A., Tammann G.~A., 1988, ARA\&A, 26, 509

\noindent
[14] Blanton M.~R.~et al., 2001, AJ, 121, 2358  

\noindent
[15] Cole S.~et al., 2001, MNRAS, 326, 255

\noindent
[16] Trentham N., Tully R.~B., 2002, MNRAS, 335, 712  

\noindent
[17] Alcock C.~et al., 2000, ApJ, 542, 281 

\noindent
[18] Oppenheimer B.~R., Hambly N.~C., Digby A.~P., Hodgkin S.~T., Saumon D., 
2001, Science, 292, 698     

\noindent
[19] Fields B.~D., Freese K., Graf D.~S., 1998, New Astron, 3, 347 

\noindent
[20] Carroll S.~M., Press W.~H., Turner E.~L., 1992, ARAA, 30, 499

\noindent
[21] Lilly S.~J., Le F{\`e}vre O., Hammer F., Crampton D., 
1996, ApJ, 460, L1

\noindent
[22] Efstathiou G., Bridle S.~L., Lasenby A.~N., Hobson M.~P., Ellis R.~S.,
1999, MNRAS, 303, L47

\noindent
[23] Somerville R.~S, Primack J.~R., Faber S.~M., 2001, MNRAS, 320, 504

\noindent
[24] Leitherer C., Heckman T.~M., 1995, ApJS, 96, 9

\noindent
[25] Smail I., Ivison R.~J., Blain A.~W., Kneib J.-P.,
2002, MNRAS, 331, 495

\noindent
[26] Chapman S.~et al., 2000, MNRAS, 319, 318

\noindent
[27] Ivison R., Smail I., Le Borgne J.-F., Blain A.~W., Kneib J.-P.,
Bezecourt J., Kerr T.~H., Davies J.~K., 1998, MNRAS, 298, 583

\noindent 
[28] de Jong T., Clegg P.~E., Soifer B.~T., Rowan-Robinson M., Habing
H.~J., Aumann H.~H., Raimond E., 1984, ApJ, 278, L67

\noindent
[29] Djorgovski S.~et al., 1998, ApJ, 508, L17

\noindent
[30] Ramirez-Ruiz E., Trentham N., Blain A.~W., 2002, MNRAS, 329, 465

\noindent
[31] Blain A.~W., Kneib J.-P., Ivison R.~J., Smail I., 1999, ApJ, 512, L87

\noindent
[32] Blain A.~W., Smail I., Ivison R.~J., Kneib J.-P., 1999, MNRAS, 302, 632

\noindent
[33] Blain A.~W. 2001, in Tacconi~T., Lutz D. eds., Starburst Galaxies:
Near and Far. Springer, Berlin, p~303

\noindent
[34] Kroupa P., Tout C.~A., Gilmore G., 1993, MNRAS, 262, 545

\noindent
[35] Madau P., Pozzetti L., 2000, MNRAS, 312, L9 

\noindent
[36] Pozzetti L., Madau P., Zamorani G., Ferguson H.~C., Bruzual A.~G.,
1998, MNRAS, 298, 1133 

\noindent
[37] Sanders D.~B., Soifer B.~T., Elias J.~H., Madore B.~F., Matthews K.,
Neugebauer G., Scoville N.~Z., 1988, ApJ, 325, 74

\noindent
[38] Sanders D.~B., Soifer B.~T., Elias J.~H.,
Neugebauer G., Matthews K., 1988, ApJ, 328, L35

\noindent
[39] Magorrian J.~et al., 1998, AJ, 115, 2285

\noindent
[40] Blain A.~W., Smail I., Ivison R.~J., Kneib J.-P., Frayer D.~T.,
2002, Physics Reports, 369, 111

\noindent
[41] Gallego J., Zamorano J., Aragon-Salamanca A., Rego M., 1995, ApJ, 455, L1 

\noindent
[42] Tresse L., Maddox S.~J., Le F{\`{e}}vre O., Cuby J.-G., 2002, MNRAS, 337, 369 

\noindent
[43] Gallego J., Garc{\'{i}}a-Dab{\'{o}} C.~E., Zamorano J., Arag{\'{o}}n-Salamanca A.,
Rego M., 2002, ApJ, 570, L1 

\noindent
[44] Hicks E.~K.~S., Malkan M.~A.,  Teplitz H.~L., McCarthy P.~J., Yan L.,
2002, ApJ, 581, 205

\noindent
[45] Cowie L.~L., Songaila A., Barger A.~J., 1999, AJ, 118, 603 

\noindent
[46] Connolly A.~J., Szalay A.~S., Dickinson M., Subbarao M.~U., Brunner R.~J.,
1997, ApJ, 486, L11  

\noindent
[47] Cohen J.~G., 2002, ApJ, 567, 672

\noindent
[48] Steidel C.~C., Giavalisco M., Pettini M., Dickinson M., Adelberger K.~L., 1996,
ApJ, 462, L17  

\noindent
[49] Steidel C.~C., Adelberger K.~L., Giavalisco M., Dickinson M.,
Pettini M., 1999, ApJ, 519, 1

\noindent
[50] Flores H., Hammer F., Thuan T.~X., C{\'{e}}sarsky C., Desert F.~X., Omont A.,
Lilly S.~J., Eales S., Crampton D., Le F{\`{e}}vre O., 1999, ApJ, 1, 148

\noindent 
[51] Mathis J.~S., 1990, ARA{\&}A, 28, 37 

\noindent
[52] Calzetti D., 1997, AJ, 113, 162 

\noindent
[53] Pettini M., Shapley A.~E., Steidel C.~C., Cuby J.-G., Dickinson M.,
Moorwood A.~F.~W., Adelberger K.~L., Giavalisco M., 2001, ApJ, 554, 981

\noindent
[54] Weynmann R.J., Stern D., Bunker A., Spinrad H., Chaffee F.~H., Thompson R.~L.,
Storrie-Lombardi L.~J., 1998, ApJ, 505, L95

\noindent
[55] Spinrad H., Stern D., Bunker A., Dey A., Lanzetta K., Yahil A., Pascarelle S.,
Fern{\'{a}}ndez-Soto A., 1998, AJ, 116, 2617 

\noindent
[56] Hu E.~M., Cowie L.~L., McMahon R.~G., 1998, ApJ, 502, L99 

\noindent
[57] Hu E.~M., Cowie L.~L., McMahon R.~G., Capak P., Iwamuro F., Kneib J.-P., 
Maihara T., Motohara K., 2002, ApJ, 568, L75

\noindent
[58] Lanzetta K.~M., Yahata N., Pascarelle S., Chen H. -W., 
Fern{\'{a}}ndez-Soto A., 2002, ApJ, 570, 492

\noindent
[59] Sanders D.~B., Mirabel I.~F., 1996, ARA{\&}A, 34, 749

\noindent
[60] Trentham N., 2001, MNRAS, 323, 542 

\noindent
[61] Kaviani  A., Haehnelt M.~G., Kauffmann G., 2002, MNRAS submitted
(astro-ph/0207238)

\noindent
[62] Papadopoulos P., Ivison R., Carilli C., Lewis G., 2001, Nature, 409, 58

\noindent
[63] Holland W.~S.~et al., 1999, MNRAS, 303, 659 

\noindent
[64] Blain A.~W., Ivison R.~J., Smail I., 1998, MNRAS, 296, L29  

\noindent
[65] Barger A.~J., Cowie L.~L., Sanders D.~B., 1999, 518, L5

\noindent
[66] Wilner D.~J., Wright M.~C.~H., 1997, ApJ, 488, L67

\noindent
[67] Trentham N., Blain A.~W., Goldader J., 1999, MNRAS, 305, 61

\noindent
[68] Smail I., Ivison R.~J., Kneib J.-P., Cowie L.~L., Blain A.~W.,
Barger A.~J., Owen F.~N., Morrison G., 1999, MNRAS, 308, 1061

\noindent
[69] Gear W.~K., Lilly S.~J., Stevens J.~A., Clements D.~L., Webb T.~M., 
Eales S.~A., Dunne L., 2000, 316, L51

\noindent
[70] Dey A., Graham J.~R., Ivison R.~J., Smail I.,
Wright G.~S., Liu M.~C., 1999, ApJ, 519, 610 

\noindent
[71] Dunne L., Clements D.~L., Eales S.~A., 2000, MNRAS, 319, 813

\noindent
[72] Yun M.~S., Carilli C.~L., 2002, ApJ, 568, 88

\noindent
[73] Ivison R.~J.~et al., 2002, MNRAS, 337, 11

\noindent
[74] sirtf.caltech.edu/SSC/obs/overview.html

\noindent
[75] van Paradijs J., Kouveliotou C., Wijers R.~A.~M.~J., 2000, ARA{\&}A, 38, 379  

\noindent
[76] MacFadyen A., Woosley S., Heger A., 2001, ApJ, 550, 410

\noindent
[77] Piro L.~et al., 2000, Science, 290, 955

\noindent
[78] Amati L.~et al., 2000, Science, 290, 953

\noindent
[79] Bloom J.~S.~et al.~1999, Nature, 401, 453

\noindent
[80] http://www.mpe.mpg.de/
{\%}7ejcg/grbgen.html)

\noindent
[81] Sethi S., Bhargavi S.~G., 2001, A{\&}A, 376, 10 

\noindent
[82] Waxman E., Draine B.~T., 2000, ApJ, 537, 796

\noindent
[83] Galama T.~J., Wijers R.~A.~M.~J., 2001, ApJ, 49, L209

\noindent
[84] Venemans B.~P., Blain A.~W., 2001, MNRAS, 325, 1777

\noindent
[85] Fruchter A.~S., Krolik H. J., Rhoads J., 2001, ApJ, 563, 597

\noindent
[86] Sakamoto K., Scoville N.~Z., Yun M.~S., Crosas M., Genzel R.,
Tacconi L.~J., 1999, ApJ, 514, 68

\noindent
[87] Djorgovski S.~G., Frail D.~A., Kulkarni S.~R., Bloom J.~S.,
Odewahn S.~C., Diercks A., 2001, ApJ, 562, 654

\noindent
[88] Gorosabel J.~et al., 2003, A{\&}A, in press (astro-ph/0212334) 

\noindent 
[89] Elvis M., Wilkes B.~J., McDowell J.~C., Green R.~F., Bechtold J.,
Willner S.~P., Oey M.~S., Polomski E., Cutri R., 1994, ApJS, 95, 1 

\noindent
[90] Tran Q.~D., Lutz D., Genzel R., Rigopolou D., Spoon H.~W.~W., Sturm E.,
Gerin M., Hines D.~C., Morwood A.~F.~M., Sanders D.~B., Scoville N.,
Taniguchi Y., Ward M., 2001, ApJ, 552, 527

\noindent
[91] McMahon R.~G., Priddey R.~S., Omont A., Snellen I., Withington S.,
1999, MNRAS, 309, L1

\noindent
[92] Knudsen K.~K., van der Werf P., Jaffe W., 2001, in 
Lowenthal J., Hughes D.~H., eds,
Deep submillimeter surveys: implications for galaxy formation and
evolution, World Scientific, Singapore, p.~168   

\noindent
[93] Farrah D., Serjeant S., Efstathiou A., Rowan-Robinson M.,
Verma A., 2002, MNRAS, 335, 1163
 
\noindent
[94] Rowan-Robinson M., Efstathiou A., 1993, MNRAS, 263, 675 

\noindent
[95] Saunders W., Rowan-Robinson M., Lawrence A., Efstathiou G., Kaiser N.,
Ellis R.~S., Frenk C.~S., 1990, MNRAS, 242, 318

\noindent
[96] P{\'{e}}roux C., McMahon R.~G., Storrie-Lombardi L.~J., Irwin M.~J.,
2001, MNRAS, submitted (astro-ph/0107045) 

\noindent
[97] Lu L., Sargent W.~L.~W., Barlow T.~A., Churchill C.~W., Vogt S.~S.,
1996, ApJS, 107, 475

\noindent
[98] Rowan-Robinson M.~et al., 1997, MNRAS, 289, 490

\noindent
[99] Kennicutt R.~C., 1998, ARA{\&}A, 36, 189 

\noindent
[100] Blain A.~W., Jameson A., Smail I., Longair M.~S., Kneib J.-P.,
Ivison R.~J., 1999c, MNRAS, 309, 715  

\noindent
[101] Lagache G., Abergel A., Boulanger F., Desert F.-X., Puget J.-L., 1999,
A{\&}A, 344, 322  

\noindent
[102] Bernstein R.~A., Freedman W.~L., Madore B.~F., 2002, ApJ, 571, 56  

\noindent
[103]) Kormendy J., 1977, ApJ, 217, 406 

\noindent
[104] Kent S.~M., 1985, ApJS, 59, 115  

\noindent
[105] Simien F., de Vaucouleurs G., 1986, 302, 564   

\noindent
[106] Fukugita M., Hogan C.~J., Peebles P.~J.~E., 1998, ApJ, 1998, 503, 518

\noindent
[107] Toomre A., Toomre J., 1972, ApJ, 178, 623 
 
\noindent
[108] Bothun G.~D., Impey C.~D., Malin D.~F., Mould J.~R., 1987, AJ, 94, 23

\noindent
[109] Freeman K., Bland-Hawthorn J., 2002, AR{\&}A, 40, 487

\noindent
[110] Binney J., Merrifield M., 1997, Galactic Astronomy, Princeton University Press
Princeton

\noindent
[111] Odenkirchen M.~et al., 2001, ApJ, 548, L165

\noindent
[112] Hodgkin S.~T., Oppenheimer B.~R., Hambly N.~C., Jameson R.~F., Smartt S.~J.,
Steele I.~A., 2000, Nature, 403, 57

\noindent
[113] Reid I.~N., Sahu K.~C., Hawley S.~L., 2001, ApJ, 559, 942

\noindent
[114] Koopmans L.~V.~E., Blandford R.~D., 2001, in
Natarajan P., ed.,
The Shapes of Galaxies and their Halos, in press
(astro-ph/0106392)

\noindent
[115] Hansen B.~M.~S., 2003, ApJ, 582, 915

\noindent
[116] Alcock C.~et al., 2000, ApJ, 542, 281

\noindent
[117] Hansen B.~M.~S., 1999, ApJ, 517, L39 

\noindent
[118] Lynden-Bell D., Tout C.~A., 2001, ApJ, 558, 1

\noindent
[119] Gilmore G., 1999, in Spooner N.~J.~C.,  Kudryatsev V., eds., 
The Implications of Dark Matter.  World Scientific,
Singapore, p.~121

\noindent
[120] Alam J.-E., Raha S., Sinha B., 1999, ApJ, 513, 572

\noindent
[121] Derue F.~et al., 1999, A\&A, 351, 87 

\noindent
[122] Loveday J., Peterson B.~A., Efstathiou G., Maddox S.~J., 1992,
ApJ, 390, 338

\noindent
[123] Cowie L.~L., Songaila A., Hu E.~M., Cohen J.~G., 1996, AJ, 112, 839

\noindent
[124] Ellis R.~S., Colless M., Broadhurst T., Heyl J., Glazebrook K.,
1996, MNRAS, 280, 235

\noindent
[125] Lin H., Kirshner R.~P., Shectman S.~A., Landy S.~D., Oemler A.,
Tucker D.~L., Schechter P.~L., 1996, ApJ, 464, 60

\noindent
[126] Norberg P.~et al.~2002, MNRAS, 336, 907

\noindent
[127] Marzke R., Geller M.~J., Huchra J.~P., Corwin H.~G., 1994,
AJ, 108, 437

\noindent
[128] Huang J.-S., Cowie L.~L., Luppino G.~A., 1998, ApJ, 496, 31

\noindent
[129] Gilmore G., Howell D., 1998, The Stellar Initial Mass Function. 
ASP, San Francisco

\noindent
[130] Salpeter E.~E., 1955, ApJ, 121, 161

\noindent
[131] Scalo J.~M., 1986, Fundam.~Cosmic Phys., 11, 1

\noindent
[132] Gould A., Bahcall J.~N., Flynn C., 1996, ApJ, 465, 759

\noindent
[133] van der Marel R.~P., 1991, MNRAS, 253, 710

\noindent
[134] Faber S.~M., 1973, ApJ, 179, 731

\noindent
[135] Kormendy J., 1990, in Kron R.~G., ed., The Edwin Hubble Centennial Symposium:
The Evolution of the Universe of Galaxies.  Astronomical Society of the
Pacific, San Francisco, p.~33

\noindent
[136] Carter D., Inglis I., Ellis R.~S., Efstathiou G., Godwin J.~G., 1985, MNRAS, 212, 471

\noindent
[137] Disney M.~J., 1976, Nature, 263, 573

\noindent
[138] Binggeli B., 1994, in Meylan G., Prugneil P., ed., ESO
Conference and Workshop Proceedings No.~49: Dwarf Galaxies.
European Space Observatory, Munich, p.~13

\noindent
[139] Wilkinson M.~I., Kleyna J., Evans N.~W., Gilmore G., 2002, MNRAS, 330, 778

\noindent
[140] Armandroff T.~E., da Costa G., 1999, in Whitelock P., Cannon R.,
ed., IAU Symposium 192: The stellar content of Local Group galaxies.  ASP, San Francisco, p.~203

\noindent
[141] Whiting A.~B., Hau G.~K.~T., Irwin M.~J., 1999, AJ, 118, 2767

\noindent
[142] Tully R.~B., Verheijen M.~A.~W., Pierce M.~J., Huang J.$\,$-S.,
Wainscoat R.~J., 1996, AJ, 112, 2471

\noindent
[143] Trentham N., Tully R.~B., Verheijen M.~A.~W, 2001, MNRAS, 325, 385

\noindent
[144] Trentham N., Hodgkin S., 2002, MNRAS, 333, 423
 
\noindent
[145] Tully R. B., Somerville R. S., Trentham N., Verheijen M. A. W.,
2002, ApJ, 569, 573

\noindent
[146] Benson A.~J., Frenk C.~S., Baugh C.~M., Cole S., Lacey C.~G., 2002,
MNRAS, submitted (astro-ph/0210354)

\noindent
[147] Kormendy J., Djorgovski S., 1989, ARA{\&}A, 27, 235

\noindent
[148] Lauer T.~R., 1985, ApJ, 292, 104

\noindent
[149] Faber S.~M.~et al.~1997, AJ, 114, 1771

\noindent
[150] Bender R., Surma P., Doebereiner S., Moellenhoff C., Madejsky R., 1989, A\&A, 217, 35

\noindent
[151] Davies R.~L., Efstathiou G., Fall S.~M., Illingworth G., Schechter P.~L.,
1983, ApJ, 266, 41

\noindent
[152] Forbes D.~A., Franx M., Illingworth G.~D., Carollo C.~M., 1996, ApJ, 467, 126

\noindent
[153] van der Marel R.~P., 1999, AJ, 117, 744

\noindent
[154] Trentham N., Tully R.~B., Verheijen M.~A.~W, 2001, MNRAS, 325, 1275

\noindent
[155] Doublier V., Caulet A., Comte G.., 1999, A\&AS, 138, 213

\noindent
[156] Drinkwater M.~J., Gregg M.~D., 1998, MNRAS, 296, L15

\noindent
[157] Phillipps S.,. Drinkwater M.~J., Gregg M.~D., Jones J.~B., 2001, ApJ, 560, 201

\noindent
[158] Freeman K.~C., 1970, ApJ, 160, 811

\noindent
[159] Binggeli B., Cameron L.~M., 1991, A{\&}A, 252, 27

\noindent
[160] King I.~R., 1966, AJ, 71, 64

\noindent
[161] Harris W.~E.~et al., 1997, AJ, 114, 1030

\noindent
[162] Gnedin O., Lahav O., Rees M.~J., 2001 (astro-ph/0108034)

\noindent
[163] van der Marel R.~P., Gerssen J., Guhatakurta P., Peterson R.~C., Gebhardt K., 
2002, AJ, 124, 3255

\noindent
[164] Gerssen J., van der Marel R.~P., Gebhardt K.,
Guhatakurta P., Peterson R.~C., Pryor C., 2002, AJ, 124, 3270 

\noindent
[165] Mar{\'{i}}n-French A., Aparicio A., 2003, ApJ, 585, 714

\noindent
[166] Hodge P., 1994, in Munoz-Tunon C., Sanchez F.,
ed., The Formation and Evolution of Galaxies.  Cambridge University
Press, Cambridge, p.~1

\noindent
[167] Feltzing S., Gilmore G., Wyse R.~F.~G., 1999, ApJ, 516, 17

\noindent
[168] Wyse R.~F.~G., Gilmore G., Houdashelt M.~L., Feltzing S., Hebb L., Gallagher J.~S.,
Smecker-Hane T.~A., 2002, New Astron., 7, 395

\noindent
[169] Armandroff T.~E., da Costa G., 1991, AJ, 101, 1329

\noindent
[170] Bruzual A.~G., Charlot S., 1993, ApJ, 405, 538

\noindent
[171] Ferrini F., Poggianti B.~M., 1993, ApJ, 410, 44

\noindent
[172] Leiherer C.~et al.~1996, PASP, 108, 996

\noindent
[173] Fioc M., Rocca-Volmerange B., 1997, A\&A, 326, 950

\noindent
[174] Poggianti B.~M., Barbaro G., 1997, A\&A, 325, 1025 

\noindent
[175] Kodama T., Arimoto N., 1997, A\&A, 320, 41

\noindent
[176] Charlot S., Worthey G., Bressan A., 1996, ApJ, 457, 625 

\noindent
[177] Szokoly G.~P., Subbarao M.~U., Connolly A.~J., Mobasher B., 1998, ApJ, 492, 452

\noindent
[178] Thronson H.~A., Greenhouse M.~A., 1998, ApJ, 327, 671 

\noindent
[179] de Jong R.~S., 1996, A\&A, 313, 377
 
\noindent
[180] Walker T.~P., Steigman G., Kang H.-S., Schramm D.~M., Olive K.~M., 1991,
ApJ, 376, 51 

\noindent
[181] Smith M.~S., Kawano L.~H., Malaney R.~A., 1993, ApJS, 85, 219

\noindent
[182] Songaila A., Wampler E.~J.,  Cowie L.~L., 1997, Nature, 385, 137

\noindent
[183] O'Meara J.~M., Tytler D., Kirkman D., Suzuki N., Prochaska J.~X.,
Lubin D., Wolfe A.~M., 2001, ApJ, 552, 718

\noindent
[184] Ruhl J.~E.~et al., 2002 (astro-ph/0212229)

\noindent
[185] Benoit A.~et al., 2003, A\&A, 399, L25 

\noindent
[186] Weinberg D., Miralda-Escud{\'{e}} J., Hernquist L., Katz N., 1997, ApJ, 490, 564

\noindent
[187] Perlmutter S.~et al., 1999, ApJ, 517, 565

\noindent
[188] White S.~D.~M., Navarro J.~F., Evrard A.~E., Frenk C.~S., 1993,
Nature, 366, 429

\noindent
[189] Evans N.~W., Wilkinson M.~I., 2000, MNRAS, 316, 929

\noindent
[190] Trentham N., Moeller O., Ramirez-Ruiz E., 2001, MNRAS, 322, 658

\noindent
[191] Press W.~H., Schechter P., 1974, ApJ, 187, 425

\noindent
[192] Efstathiou G., Bond J.~R., White S.~D.~M, 1992, MNRAS, 258, 1p

\noindent
[193] Wilkinson M.~I., Evans N.~W., 1999, MNRAS, 310, 645

\noindent
[194] Kleyna J.~T., Wilkinson M.~I., Evans N.~W., Gilmore G., 2001, ApJ, 563, L115

\noindent
[195] Gerhard O.~E., Spergel D.~N., 1992, ApJ, 389, L9

\noindent
[196] van den Bergh S., 1992, A\&A, 264, 75

\noindent
[197] Papovich C., Dickinson M., Ferguson H.~C., 2001, ApJ, 559, 520 
 
\noindent
[198] Dickinson M., Papovich C., Ferguson H.~C., Budavari T., 2003, 
ApJ, in press (astro-ph/0212242) 

\noindent
[199] Brinchmann J., Ellis R.~S., 2000, ApJ, 536, L77 

\noindent
[200] Barlow T.~A., Tytler D., 1998, AJ, 115, 1725

\noindent
[201] Pettini M., Ellison S.~L., Schaye J., Songaila A., Steidel C., Ferrara A.,
2001, Ap\&SSS, 277, 555

\noindent
[202] Ellison S.~L., Songaila A., Schaye J., Pettini M., 2000, AJ, 120, 1175

\noindent
[203] Misawa T., Tytler D., Iye M., Storrie-Lombardi L.~J., Suzuki N., Wolfe A.~M.,
2002, AJ, 123, 1847

\noindent
[204] Telfer R.~C., Kriss G.~A., Zheng W., Davidsen A.~F., Tytler D., 2002, ApJ, 579, 500

\noindent
[205] Webb J.~K., Carswell R.~F., Lanzetta K.~M., Ferlet R., Lemoine M., Vidal-Madjar A.,
Bowen D.~V., 1997, Nature, 388, 250

\noindent
[206] Lynden-Bell D., 1967, MNRAS, 136, 101

\noindent
[207] Fulton E., Barnes J.~E., 2001, Ap{\&}SS, 2001, 276, 851

\noindent
[208] http://www.ast.cam.ac.uk/$^{\sim}$optics/cirpass/index.html

\noindent
Mihos J.~C., Dubinski J., Hernquist L., 1998, ApJ, 494, 183

\noindent
[209] Dejonghe H., Habing J., 1993, IAU Symposium 153: Galactic Bulges.~Kluwer
Academic Publishers, Dordrecht

\noindent
[210] Mihos J.~C., Dubinski J., Hernquist L., 1998, ApJ, 494, 183

\noindent
[211] Haehnelt M.~G., Natarajan P., Rees, 1998, MNRAS, 300, 817

\noindent
[212] Lacy M., Ridgway S., Trentham N., 2002, in Biretta J., Koekemoer A. M.,
Perlman E. S., O'Dea C. P., eds, Life Cycles of Radio Galaxies,  
New Astronomy Reviews, 46, 211 

\noindent
[213] Kukula M.~J., Dunlop J.~S., McLure R.~J., Miller L., Percival W.~J.,
Baum S.~A., O'Dea C.~P., 2001, MNRAS, 326, 1533

\noindent
[214] Boyle B.~J., Terlevich R.~J., 1998, MNRAS, 293, L49

\noindent
[215] Schweizer F., 1986, Science, 231, 227  

\noindent
[216] Gebhardt K.~et al., 2001, AJ, 122, 2469

\noindent
[217] Kormendy J., Bender R., M31 binary

\noindent
[218] Komossa S., Burwitz V., Hasinger G., Predehl, P.; Kaastra, J. S.; Ikebe, Y.,
2003, ApJ, 582, L15    

\noindent
[219] Gnedin O., 1999, BAAS, 65.04

\noindent
[220] Metcalfe N., Shanks T., Campos A., McCracken H.~J.,
Fong R., 2001, MNRAS, 323, 795

\noindent
[221] Gardner J.~P., Cowie L.~L., Wainscoat R.~J., 1993,
ApJ, 415, L9

\noindent
[222] Cowie L.~L., Gardner J.~P., Hu E.~M., Songaila A., Hodapp K.-W.,
Wainscoat R.~J., 1994, ApJ, 434, 114

\noindent
[223] Huang J.-S., Cowie L.~L., Gardner J.~P., Hu E.~M., Songaila A.,
Wainscoat R.~J., 1997, ApJ, 476, 12 

\noindent
[224] Eggen O.~J., Lynden-Bell D., Sandage A.~R., 1962, ApJ, 136, 748

\noindent
[225] White S.~D.~M, Kauffmann G., in Munoz-Tunon C., Sanchez F., eds.,
The Formation and Evolution of Galaxies.~Cambridge University Press,
Cambridge, p.~455

\noindent
[226] Cole S., Lacey C.~G., Baugh C.~M., Frenk C.~S., 2000, MNRAS, 319, 168

\noindent
[227] Mathis H., Lemson G., Springel V., Kauffmann G., White S.~D.~M,
Eldar A., Dekel A., 2002, MNRAS, 333, 739 

\noindent
[228] Efstathiou G., 2000, MNRAS, 317, 697  

\noindent
[A1] Chapman S.~C., Blain A.~W., Ivison R.~J., Smail I.~R.,
2003, Nature, 422, 695

\noindent
[A2] Boyle B.~J.~et al., MNRAS, 317, 1014

\noindent
[A3] Budavari T.~et al., 2003, ApJ, in press (astro-ph/0305603)

\noindent
[A4] Erb D.~K., Shapley A.~E., Steidel C.~C., Pettini M.,
Adelberger K.~L., Hunt M.~P., Moorwood A.~F., Cuby J.-G., 2003,
ApJ, in press (astro-ph/0303392)

\end